\documentclass[11pt]{article}

\usepackage{jheppub}
\usepackage{amsmath}
\usepackage{bbm}
\usepackage{url}

\def\beq{\begin{equation}}
\def\eq{\end{equation}}
\def\eeq{\end{equation}}

\newcommand{\beqa}{\begin{eqnarray}}
\newcommand{\eeqa}{\end{eqnarray}}

\newcommand{\hc}{\mbox{h.c.}}
\newcommand{\lp}{\left(}
\newcommand{\rp}{\right)}

\def\centeron#1#2{{\setbox0=\hbox{#1}\setbox1=\hbox{#2}\ifdim
\wd1>\wd0\kern.5\wd1\kern-.5\wd0\fi
\copy0\kern-.5\wd0\kern-.5\wd1\copy1\ifdim\wd0>\wd1
\kern.5\wd0\kern-.5\wd1\fi}}
\def\ltap{\;\centeron{\raise.35ex\hbox{$<$}}{\lower.65ex\hbox{$\sim$}}\;}
\def\gtap{\;\centeron{\raise.35ex\hbox{$>$}}{\lower.65ex\hbox{$\sim$}}\;}

\def\MET{{\not \!  \! E}_T}

\def\chii0{\chi_i^0}
\def\chij0{\chi_j^0}

\def\GTLT{ \mathop{}_{<}^{>} }

\def\foursqr#1#2{{\vcenter{\vbox{
 \hrule height.#2pt
 \hbox{\vrule width.#2pt height#1pt \kern#1pt
 \vrule width.#2pt}
 \hrule height.#2pt
 \hrule height.#2pt
 \hbox{\vrule width.#2pt height#1pt \kern#1pt
 \vrule width.#2pt}
 \hrule height.#2pt
     \hrule height.#2pt
 \hbox{\vrule width.#2pt height#1pt \kern#1pt
 \vrule width.#2pt}
 \hrule height.#2pt
     \hrule height.#2pt
 \hbox{\vrule width.#2pt height#1pt \kern#1pt
 \vrule width.#2pt}
 \hrule height.#2pt}}}}
\def\psqr#1#2{{\vcenter{\vbox{\hrule height.#2pt
 \hbox{\vrule width.#2pt height#1pt \kern#1pt
 \vrule width.#2pt}
 \hrule height.#2pt \hrule height.#2pt
 \hbox{\vrule width.#2pt height#1pt \kern#1pt
 \vrule width.#2pt}
 \hrule height.#2pt}}}}
\def\sqr#1#2{{\vcenter{\vbox{\hrule height.#2pt
 \hbox{\vrule width.#2pt height#1pt \kern#1pt
 \vrule width.#2pt}
 \hrule height.#2pt}}}}

\def\figin{\epsfcheck\figin}\def\figins{\epsfcheck\figins}
\def\epsfcheck{\ifx\epsfbox\UnDeFiNeD
\message{(NO epsf.tex, FIGURES WILL BE IGNORED)}
\gdef\figin##1{\vskip2in}\gdef\figins##1{\hskip.5in}
\else\message{(FIGURES WILL BE INCLUDED)}%
\gdef\figin##1{##1}\gdef\figins##1{##1}\fi}
\def\DefWarn#1{}
\def\figinsert{\goodbreak\midinsert}
\def\ifig#1#2#3{\DefWarn#1\xdef#1{fig.~\the\figno}
\writedef{#1\leftbracket fig.\noexpand~\the\figno}%
\figinsert\figin{\centerline{#3}}\medskip\centerline{\vbox{\baselineskip12pt
\advance\hsize by -1truein\noindent\footnotefont{\bf
Fig.~\the\figno:\ } \it#2}}
\bigskip\endinsert\global\advance\figno by1}

\def\fig#1#2#3#4{\vskip 0.5cm \begingroup \midinsert \centerline{
\psfig{file=#1,width=#2}} \vskip 0.4cm
\global\advance\figno by 1
\centerline{\vbox{\baselineskip=12pt \noindent Figure \the\figno: #3}}
\endinsert \endgroup {\xdef#4{\the\figno}} }

\def\figcrop#1#2#3#4#5#6#7#8{\vskip 0.5cm \begingroup \midinsert \centerline{
\psfig{file=#1,width=#2,bbllx=#3,bblly=#4,bburx=#5,bbury=#6}} \vskip 0.4cm
\global\advance\figno by 1
\centerline{\vbox{\baselineskip=12pt \noindent Figure \the\figno: #7}}
\endinsert \endgroup {\xdef#8{\the\figno}} \vskip .5cm}

\def\figlabel#1{\xdef#1{\the\figno}}
\def\encadremath#1{\vbox{\hrule\hbox{\vrule\kern8pt\vbox{\kern8pt
\hbox{$\displaystyle #1$}\kern8pt}
\kern8pt\vrule}\hrule}}
\def\underarrow#1{\vbox{\ialign{##\crcr$\hfil\displaystyle
 {#1}\hfil$\crcr\noalign{\kern1pt\nointerlineskip}$\longrightarrow$\crcr}}}

\title{Multi-Lepton Signals of Multiple Higgs Bosons}

\author[a,b]{Nathaniel Craig,}

\author[a]{Jared A. Evans,}

\author[a]{Richard Gray,}

\author[c]{Can Kilic,}

\author[a]{Michael Park,}

\author[a]{\\ Sunil Somalwar,}

\author[a]{Scott Thomas}

\affiliation[a]{Department of Physics, Rutgers University \\
Piscataway, NJ 08854 }
\affiliation[b]{ School of Natural Sciences, Institute for Advanced Study \\
Princeton, NJ 08540}
\affiliation[c]{Theory Group, Department of Physics and Texas Cosmology Center, \\
The University of Texas at Austin \\
Austin, TX 78712}

\preprint{RU-NHETC-2012-20, UTTG-12-12, TCC-012-12}

\abstract{ {
We identify and investigate novel multi-lepton signatures of extended Higgs sectors 
at the LHC in the guise of CP- and flavor-conserving two-Higgs-doublet models (2HDMs). 
Rather than designing individual searches tailored to specific 2HDM
signals, we employ the combination of many exclusive multi-lepton search 
channels to probe the collective signal from the totality of 
production and decay processes. 
Multi-lepton signals of 2HDMs can arise from a variety of sources, including 
Standard Model-like production of the CP-even scalars, $h$ and $H$, through 
gluon-fusion with $h,H \to ZZ^{(*)}$, or associated 
production with vector bosons or top quarks, with 
$h,H \to WW^{(*)}, ZZ^{(*)},\tau \tau$.  
Additional sources include
gluon-fusion production of the heavy 
CP-even scalar with cascade decays through the light CP-even scalar, 
the CP-odd scalar, $A$, or the charged scalar, $H^\pm$, 
such as $H \to hh$, $H \to AA$, $H \to H^+ H^-$, $H \to ZA$, 
with $A \to Zh, \tau \tau$, $H^\pm \to Wh$, 
and $h \to WW^*, ZZ^*,\tau \tau$.
Altogether, the combined multi-lepton signal may greatly exceed that of the 
Standard Model Higgs boson and provides a sensitive probe of extended 
Higgs sectors over a wide range of parameters. 
As a proof of principle, we use a factorized mapping procedure between model 
parameters and signatures 
to determine multi-lepton sensitivities
in four different flavor conserving 2HDM parameter spaces
by simulating the acceptance times efficiency in 20 exclusive 
multi-lepton channels for 222 independent production 
and decay topologies
that arise for four benchmark  
2HDM spectra within each parameter space.  
A comparison of these sensitivities 
with the results of a multi-lepton search conducted by the CMS collaboration 
using 5 fb$^{-1}$ of data collected from 7 TeV $pp$ collisions
yields
new limits in some regions of 2HDM parameter space that have 
not previously been covered by other types of direct experimental searches.
}
}

\begin{document}

\maketitle


\section{Introduction}

Probing the mechanism of electroweak symmetry breaking (EWSB) is one of the primary objectives of the Large Hadron Collider (LHC). Fulfilling this goal includes characterization of
the Standard Model-like Higgs boson corresponding to excitation of 
the scalar condensate responsible for EWSB \cite{125HiggsCMS, 125HiggsATLAS}.  
Yet it also extends much more broadly to include the 
search for additional Higgs states that could be a window into the underlying physics of EWSB.

Two Higgs doublet models (2HDMs) offer a canonical framework for extended electroweak symmetry breaking.  
Indeed, in many extensions of the minimal Standard Model (SM), supersymmetric or otherwise, 
the Higgs sector is extended to two scalar doublets \cite{2HDMoriginal}.  
It is therefore worthwhile to study the generic features of the 2HDM scenario independent of the specific 
underlying model, 
purely as an effective theory for extended EWSB. 
The phenomenology of 2HDMs is rich, as five physical Higgs sector particles 
remain after EWSB: two neutral CP-even scalars, $h$, $H$; one neutral 
CP-odd pseudoscalar, $A$; and two charged scalars, $H^+$ and $H^-$.  
All of these states could have masses at or below the TeV scale, in a regime accessible to the LHC. 
The parameter space of the 2HDM scenario is large enough to accommodate a wide diversity of  
modifications to the production and decay modes of the lightest Higgs boson, as well as to provide non-negligible production mechanisms for the heavier Higgs states that may decay directly to SM final states, or through cascades that yield multiple Higgs 
states. 

Much of the study of 2HDM phenomenology to date has been devoted to the specific setup that arises in minimal supersymmetric models \cite{SUSYHiggssearches}, which occupies 
a restricted subset of possible 2HDM signals. 
Even more general 2HDM studies  \cite{Gunion:1989we,2HDMreview,2HDMsearches} have largely focused on the direct production and decays of scalars in SM-like channels, or on specific cascade decays between scalars. 
In this work, we wish to pursue a more inclusive objective: the sensitivity of the LHC 
to the sum total of production and decay modes available in a given 2HDM, 
including both direct decays of scalars and all kinematically available scalar cascades. 
Such an approach exploits the large multiplicity of signals arising from 
production and decay of the various states in an extended EWSB sector.

Searches for final states involving three or more leptons  are well matched 
to this objective, since both direct scalar decays and scalar cascades populate multi-lepton final states with low Standard Model backgrounds. 
The CMS multi-lepton search strategy 
\cite{CMSMulti, CMSMulti5} is particularly well-suited in this respect, since its power lies in the combination of numerous exclusive channels. 
While the sensitivity to new physics in any
individual channel alone is not necessarily significant, 
the exclusive combination across multiple channels can provide considerable sensitivity. 
This is particularly effective in the search for extended EWSB sectors such as 2HDMs, where multi-lepton final states may arise from many different production and decay processes
that would evade detection by searches narrowly 
focused on  kinematics or resonantly-produced final states of specific topologies.  
With a potentially sizable multiplicity of rare multi-lepton signatures, 
an extended Higgs sector therefore provides an excellent case study for the sort
of new physics that could first be discovered in an exclusive multi-channel multi-lepton search at the LHC. 

Multi-lepton searches are already sensitive to Standard Model Higgs production  \cite{us}, 
as well as the production of a SM-like Higgs in rare decay 
modes of states with large production cross sections \cite{tch}. 
This suggests that these studies may be particularly amenable to searching for evidence of extended Higgs sectors. 
Theories with two Higgs doublets enjoy all of the multi-lepton final states available to the Standard Model Higgs, albeit with modified cross sections, as well as the multi-lepton final states of additional scalars and cascade decays between scalars  that often feature on-shell $W$ and $Z$ bosons in the final state. 
These additional particles give rise to numerous new production mechanisms for multi-lepton final states.

The goal of this paper is to perform a detailed survey of the multi-lepton signals that arise in some
representative 2HDM parameter spaces.
In particular, we will consider four different CP- and flavor-conserving 
2HDM benchmark mass spectra that
have qualitatively distinct production and decay channels. 
For each mass spectrum, we will consider each of the four discrete types of 
2HDM tree-level Yukawa couplings between the Higgs doublets and the 
SM fermions that are guaranteed to be free of tree-level flavor changing neutral currents (FCNCs).  
A study of the sensitivity to the myriad rare production and decay processes over a grid of points 
in the parameter spaces defining these sixteen representative 2HDMs 
using standard simulation techniques, while in principle straightforward, 
is computationally prohibitive.  
So instead we employ a factorized mapping procedure to go between model 
parameters and signatures \cite{sunil}.
In this procedure the acceptance times efficiency for each individual 
production and decay topology 
is independently 
determined from monte carlo simulation, 
assuming unit values for all branching ratios in the decay topology.  
The production cross section and branching ratios are then calculated externally as functions of 
model parameters. 
The total cross section times branching ratio into any given final state at any 
point in parameter space is then 
given by a sum over the production cross section times acceptance and 
efficiency for each topology times a product of the branching ratios at 
that parameter space point.  
For the study here, we simulate the acceptance times efficiency 
in 20 exclusive multi-lepton channels for 222 independent 
production and decay topologies that arise in the four benchmark 
2HDM spectra.  
For each benchmark spectrum we combine the 20 exclusive multi-lepton 
channels to obtain an overall sensitivity as a function of 
two-dimensional mixing angle parameter spaces that characterize each of the four discrete types 
of flavor conserving 2HDMs.   
With this, we identify regions of 2HDM parameter space that are excluded by 
the existing CMS multi-lepton search \cite{CMSMulti5}, 
as well as those regions where future multi-lepton searches at the LHC will have 
sensitivity.  
  

Beyond requiring CP-conservation and 
no direct tree-level flavor violation in the Higgs sector, 
we will not address constraints imposed by low energy 
precision flavor measurements on the parameter space of 2HDMs (see \cite{2HDMreview} and references therein, and \cite{Fajfer:2012jt} for a very recent analysis). 
In general, contributions to loop-induced flavor changing processes, such as $B\to X_s \gamma$, may be reduced by destructive interference among different loops, so that new physics outside of our low-energy effective theory can relax flavor bounds on the 2HDM sector.  Additionally, with the assumptions employed here, 
flavor constraints are driven by the mass of the charged Higgs, which typically does not play a significant role in the production of multi-lepton final states. 
For the benchmark spectra we consider, the charged Higgs may generally be decoupled in mass without substantially altering the phenomenology.  More generally, we emphasize that our benchmark spectra are intended to qualitatively illustrate the relevant topologies for producing multi-lepton final states. Various scalar masses may be raised to accommodate flavor physics 
without changing the qualitative multi-lepton signatures, though of course particular numeric limits will be altered.

The outline of the paper is as follows: In section \ref{sec:theory}, we will briefly review the relevant 
aspects of 2HDMs and define the parameter space within which we will conduct our survey. 
In section \ref{sec:pheno}, we will give an overview of the most interesting production and decay channels for 2HDM collider phenomenology which result in multi-lepton final states.  
Additionally, we select benchmark spectra that have a representative set of multi-lepton 
production and decay topologies. 
Section \ref{sec:search} is devoted to summarizing the multi-lepton search strategy and the simulation methods we use. The results of our study are displayed in section \ref{sec:results} 
where we identify the regions of parameter space that are excluded on the basis of the 
existing CMS multi-lepton search with 5 fb$^{-1}$ of 7 TeV proton-proton collisions \cite{CMSMulti5}
 as well as those regions to which future searches will have sensitivity.  
 In section \ref{sec:conclusions} we suggest some refinements 
to future 
multi-lepton searches that could enhance the sensitivity to extended Higgs sectors.


\section{Two Higgs Doublet Models}
\label{sec:theory}

The physically relevant parameter space specifying the most general 
2HDM is large (for a review of general 2HDMs see, for example, 
\cite{Gunion:1989we} and \cite{2HDMreview}). 
The goal here is not to consider the most general theory, but rather 
to define a manageable parameter space in which to characterize 
multi-lepton signals.  
The couplings of physical Higgs states 
that are relevant to the production and decay topologies considered 
below include those of a single Higgs boson to two fermions 
or two gauge bosons, couplings of two Higgs bosons to a single gauge boson, 
and couplings of three Higgs bosons. 
Other higher multiplicity couplings do not appear in the simplest topologies.  

For simplicity we consider CP-conserving 2HDMs that are automatically 
free of tree-level flavor changing neutral currents. 
With these assumptions, the renormalizable couplings of a single physical 
Higgs boson to pairs of fermions or gauge bosons, and of two Higgs bosons to a gauge boson, 
 are completely 
specified in terms of two mixing angles, as detailed below. 
With a mild restriction to renormalizable potentials of a  certain 
class described below, couplings involving three Higgs bosons are 
specified in terms of Higgs masses and these same mixing angles. 

The absence of tree-level 
flavor changing neutral currents in multi-Higgs theories 
is guaranteed by the Glashow-Weinberg condition 
\cite{Glashow:1976nt} which postulates that all fermions of a given 
gauge representation receive mass through renormalizable
Yukawa couplings to a single Higgs doublet.  
With this condition, tree-level couplings of neutral Higgs bosons are 
diagonal in the mass basis.  
In the case of two Higgs doublets with Yukawa couplings 
\beq
-V_{\rm yukawa} = \sum_{i=1,2} \left( Q \tilde{H}_i y_i^u \bar{u}
  +  Q H_i y_i^d \bar{d} +  L H_i y_i^e \bar{e} +\hc \right)
\label{eqn:yukawa}
\eeq
the Glashow-Weinberg condition 
is satisfied by precisely four discrete 
types of 2HDMs distinguished by the possible assignments of 
fermion couplings with either $y_1^F=0$ or $y_2^F=0$ for each of $F=u,d,e$. 
Under this restriction, we can always denote the Higgs doublet that couples to the up-type quarks as $H_u$. Having fixed this, we have two binary choices for whether the down-type quarks and the 
leptons in (\ref{eqn:yukawa}) couple to $H_u$ or $H_d$. Of these four possibilities, ``Type I'' is commonly referred to as the fermi-phobic Higgs model in the limit of zero mixing, as all fermions couple to one doublet and the scalar modes of the second doublet couple to vector bosons only. ``Type II'' is MSSM-like, since this is the only choice of charge assignments consistent with a holomorphic superpotential.  ``Type III'' is often referred to as ``lepton-specific,'' since it assigns one Higgs doublet solely to leptons.  Finally, ``Type IV'' is also known as ``flipped,'' since the leptons have a ``flipped'' coupling relative to Type II. These possible couplings are illustrated in Table \ref{tab:2hdm}. We will restrict ourselves to these four choices as they exhaust all possibilities where tree-level FCNCs are automatically forbidden.

\begin{table}[htdp]
\begin{center}
\begin{tabular}{|c|c|c|c|c|} \hline
& 2HDM I & 2HDM II & 2HDM III & 2HDM IV \\ \hline
$u$ & $H_u$ & $H_u$ & $H_u$ & $H_u$ \\
$d$ & $H_u$ & $H_d$ & $H_u$ & $H_d$ \\
$e$ & $H_u$ & $H_d$ & $H_d$ & $H_u$  \\ \hline
\end{tabular}
\caption{The four discrete types of 2HDM $H_u$ and $H_d$ Yukawa couplings 
to right-handed quarks and leptons 
that satisfy the Glashow-Weinberg condition. 
By convention 
$H_{u}$ is taken to couple 
to right handed up-type quarks, and the assignments of the remaining couplings are indicated.
\label{tab:2hdm}}
\end{center}
\end{table}%

For any of the CP-conserving 2HDMs satisfying the Glashow-Weinberg condition, the coefficient of the 
couplings of a single physical Higgs boson to fermion pairs through the Yukawa couplings 
(\ref{eqn:yukawa}) depend on the fermion mass, the ratio of the Higgs expectation 
values, conventionally defined as 
$\tan \beta \equiv \langle H_u \rangle / \langle H_d \rangle$, 
and the mixing angle $\alpha$ that diagonalizes the $2\times2$ neutral scalar 
$h-H$
mass squared matrix. 
The parametric dependences  of these couplings on $\alpha$ and $\beta$ 
relative to coupling of the Standard Model Higgs boson with a single Higgs 
doublet are given in Table \ref{tab:couplings}. 
The parametric dependence of the 
couplings of the charged scalar, $H^{\pm}$, are
the same as those of the pseudo-scalar, $A$.  

The renormalizable 
couplings of a single physical Higgs boson to two gauge bosons 
are fixed by gauge invariance in terms of the mixing angles in any CP-conserving 
2HDM as 
\beq
g_{hVV}= \sin(\beta-\alpha) g_V 
\hspace{8mm}  g_{HVV} = \cos(\beta-\alpha) g_V  
\hspace{8mm} g_{AVV}=0
\hspace{8mm}
g_{H^\pm W^\mp Z}=0
\eeq
where for $V=W,Z$ the Standard Model Higgs couplings are 
$g_W = g$ and $g_Z = g/ \cos \theta_W$, where $g$ is the $SU(2)_L$ 
gauge coupling and $\theta_W$ the weak mixing angle. 
The renormalizable couplings of two physical Higgs bosons to a single gauge boson are 
likewise fixed in any CP-conserving 2HDM as 
\begin{eqnarray}
g_{hZA}=& \frac 12 g_Z \cos(\beta-\alpha) \hspace{12mm} 
g_{HZA} & = -\frac 12   g_Z \sin(\beta-\alpha) \hspace{11mm} 
   \nonumber 
   \\
g_{hW^\mp H^\pm}=& \mp \frac i2 g \cos(\beta-\alpha) \hspace{9mm} 
g_{HW^\mp H^\pm} & = \pm \frac i2  g \sin(\beta-\alpha)   \hspace{9mm} 
g_{AW^\mp H^\pm} = \frac 12 g
\label{eqn:hhv}
\end{eqnarray}
None of these couplings involve additional assumptions about the form of the full non-renormalizable scalar potential, 
beyond CP conservation.


\begin{table}[h]
\begin{center}
\begin{tabular}{|c|c|c|c|c|} \hline
$y_{\rm 2HDM} / y_{\rm SM}$ & 2HDM I & 2HDM II & 2HDM III & 2HDM IV \\ \hline
$hVV$ & $\sin(\beta - \alpha)$ &  $\sin(\beta - \alpha)$ &  $\sin(\beta - \alpha)$ &  $\sin(\beta - \alpha)$ \\
$h Q u $ & ${\cos \alpha}/{\sin \beta}$ & ${\cos \alpha}/{\sin \beta}$ & ${\cos \alpha}/{\sin \beta}$& ${\cos \alpha}/{\sin \beta}$  \\
$h Q d$ & ${\cos \alpha}/{\sin \beta}$ & $- {\sin \alpha}/{\cos \beta}$ & ${\cos \alpha}/{\sin \beta}$& $- {\sin \alpha}/{\cos \beta}$  \\
$h L e$ & ${\cos \alpha}/{\sin \beta}$ & $- {\sin \alpha}/{\cos \beta}$ & $- {\sin \alpha}/{\cos \beta}$& ${\cos \alpha}/{\sin \beta}$  \\ \hline
$HVV$ & $\cos(\beta - \alpha)$ & $\cos(\beta - \alpha)$ & $\cos(\beta - \alpha)$& $\cos(\beta - \alpha)$   \\
$H Q u$ & ${\sin \alpha}/{\sin \beta}$ & ${\sin \alpha}/{\sin \beta}$ & ${\sin \alpha}/{\sin \beta}$& ${\sin \alpha}/{\sin \beta}$  \\
$H Q d$ & ${\sin \alpha}/{\sin \beta}$ & ${\cos \alpha}/{\cos \beta}$ & ${\sin \alpha}/{\sin \beta}$&${\cos \alpha}/{\cos \beta}$  \\
$H L e$ & ${\sin \alpha}/{\sin \beta}$ & ${\cos \alpha}/{\cos \beta}$ &${\cos \alpha}/{\cos \beta}$& ${\sin \alpha}/{\sin \beta}$  \\\hline
$AVV$ & 0 & 0 & 0 & 0 \\
$AQu$ & $\cot \beta$ & $\cot \beta$ & $\cot \beta$& $\cot \beta$   \\
$AQd$ & $-\cot \beta$ & $\tan \beta$& $- \cot \beta$ & $\tan \beta$   \\
$ALe$ & $- \cot \beta$ & $\tan \beta$  & $\tan \beta$& $- \cot \beta$  \\ \hline
\end{tabular}
\caption{Tree-level couplings of the neutral Higgs bosons
to up- and down-type quarks, leptons, and massive gauge bosons
in the four types of 2HDM models relative to the SM Higgs boson couplings as functions of $\alpha$ and $\beta$.
The coefficients of the couplings of the charged scalar $H^{\pm}$, are
the same as those of the pseudo-scalar, $A$\label{tab:couplings}}
\end{center}
\end{table}%

The couplings between three physical Higgs bosons 
depends on details of the Higgs scalar potential. 
Specifying these therefore requires additional assumptions to completely specify the
branching ratios that appear in some of the decay topologies discussed below. 
The main goal here is to present multi-lepton sensitivities to 2HDMs 
in relatively simple, manageable parameter spaces.  
A straightforward condition that fulfills this requirement is to consider 
2HDM Higgs potentials that, in additional to being CP-conserving, 
are renormalizable and restricted by a (discrete) Peccei-Quinn 
symmetry that forbids terms with an odd number of 
 $H_u$ or $H_d$ fields. 
The most general potential of this type is given by 
\beqa
V_{\rm scalar} &=& m_{u}^2 H_u^\dagger H_u + m_{d}^2 H_d^\dagger H_d  
   + \frac 12 \lambda_1 ( H_u^\dagger H_u )^2 + \frac 12 \lambda_2 ( H_d^\dagger H_d )^2 +
  \lambda_3  ( H_u^\dagger H_u ) ( H_d^\dagger H_d )  \nonumber \\
   & &  + ~ \lambda_4 ( H_u^\dagger H_d )( H_d^\dagger H_u ) 
    + \left[ \frac 12 \lambda_5 ( H_u^\dagger H_d )^2 + \hc \right]
 \label{eqn:potential}
\eeqa
This potential has seven free parameters, 
which may be exchanged for the overall Higgs expectation value, 
the four physical masses $m_h, m_H, m_A$, and $m_{H^\pm}$, 
and the two mixing angles, $\alpha$ and $\beta.$ 
So all the Higgs boson couplings in a renormalizable 2HDM with the potential (\ref{eqn:potential})
are, for a given mass spectrum, specified entirely in terms of the mixing angles 
$\alpha$ and $\beta$. 
The couplings of three physical Higgs bosons from the potential 
(\ref{eqn:potential}) that are relevant to the production and decay 
topologies studied below are 
\begin{eqnarray}
g_{Hhh} &=& \frac 1v (m_H^2 +2m_h^2)\cos(\beta-\alpha) 
    (\sin2\alpha /  \sin2\beta)  
    \nonumber
    \\
g_{HAA} &=& \frac 1v  \lp m_H^2 \lp \cos\beta \cot\beta \sin\alpha + \sin\beta \tan\beta \cos\alpha\rp + 2m_A^2\cos(\beta-\alpha) \rp 
   \nonumber
    \\
g_{HH^+H^-} &=& \frac 1v  \lp m_H^2 \lp \cos\beta \cot\beta \sin\alpha + \sin\beta \tan\beta \cos\alpha\rp + 2m_{H^\pm}^2\cos(\beta-\alpha) \rp  
\label{tri-couplings}
\end{eqnarray}
We emphasize that the choice of the potential (\ref{eqn:potential})
 is illustrative to allow a simple presentation in terms of 
 a two-dimensional parameter space of mixing angles for a given physical spectrum. 
 Although there is additional parametric freedom available in the most general CP-conserving
2HDM potential, the phenomenology is qualitatively similar. 
The only important generalization in the production and decay 
topologies studied below for the most general CP- and flavor-conserving 
2HDMs as compared  with the assumptions outlined here is that 
the partial decay widths of the CP-even heavy Higgs boson, $H$, to pairs 
of lighter Higgs bosons become free parameters, rather than being specified 
in terms of $\alpha$ and $\beta$ through the couplings (\ref{tri-couplings}).


\section{Multi-lepton Signals of Two Higgs Doublet Models}
\label{sec:pheno}

The wide range of possibilities for Higgs boson mass spectrum hierarchies and branching ratios 
in 2HDMs yields a diversity of 
production and decay channels that are relevant for multi-lepton signatures at the LHC.
Multi-lepton final states become especially important 
when the decay of one Higgs scalar to a pair 
of Higgs scalars or a Higgs scalar and a vector boson 
is possible.  
Of course, the availability of these inter-scalar decays comes at a price, 
as the decaying Higgs must be sufficiently heavy for the decay modes to be 
kinematically open, so that the production cross section is reduced. 
Performing a full multi-dimensional scan of the mass spectra 
of 2HDMs is not only computationally untenable, but also unnecessary for our purposes; 
most of the salient features may be illustrated by exploring a few 
benchmark scenarios in which all the relevant types of cascade decays are realized. 
We will focus on four such mass spectra with various orderings of the 
scalar mass spectrum, fixing the lightest CP-even Higgs mass at 125~GeV in each case.

The various 2HDM production and decay topologies that give rise to 
multi-lepton signatures fall into two broad categories: 
those resulting from the direct production and decay of an individual scalar,
and those resulting from cascades involving more than one scalar. 
The first category includes the resonant four-lepton signals of the Standard Model-like Higgs 
 $h$, from gluon fusion 
and vector boson fusion production followed by $h \to ZZ^*$
with $Z^{(*)} \! \to \ell \ell$. 
Other resonant and 
non-resonant multi-lepton signals arise from 
quark--anti-quark fusion production of $Wh, Zh,$ along with 
$t t h$ associated production with $t \to Wb$, all 
followed by $h \to WW^*, ZZ^*,\tau \tau$ with leptonic decays of 
(some of the) $W \to \ell \nu$, $Z^{(*)} \! \to \ell \ell$ and $\tau \to \ell \nu \nu$.
These modes were studied in depth in \cite{us} to obtain multi-lepton 
limits on the Standard Model Higgs and simple variations. 
The same modes of production and decay are available to the heavy 
CP-even scalar, $H$, albeit with reduced production cross sections due to 
its larger mass and 
mixing suppression of some of its couplings.
While the branching fractions of these modes depend 
on the parameters of the theory, their existence is robust and common 
to all benchmark spectra we consider. 
In contrast, the sole 
multi-lepton mode involving direct production of 
the pseudoscalar, $A$, without cascade decays through other scalars 
is $t  t A$  associated production followed by $t \to Wb$ and 
$A \to \tau \tau$ with 
leptonic decays of (some of the) $W \to \ell \nu$ and $\tau \to \ell \nu \nu$.  
And there are no multi-lepton signals resulting from
direct production of the charged Higgs, $H^{\pm}$,
without cascade decays through other scalars.

Scalar cascades add a variety of new multi-lepton processes, 
including production and decay modes that contribute
to some of the same final states that arise from 
a Standard Model Higgs boson.
Processes of this type include gluon fusion production of $A$ with $A \to Zh,ZH$
followed by $h,H \to WW^*, ZZ^*,\tau \tau$ with 
(some of the) $W \to \ell \nu$, $Z^{(*)} \! \to \ell \ell$, and $\tau \to \ell \nu \nu$.
Another example of this type is 
gluon fusion and vector boson fusion production of $H$ with $H \to AA,hh$ 
followed by $A \to \tau \tau$ or 
$h \to bb, WW^*, ZZ^* , \tau \tau $ with 
(some of the) $W \to \ell \nu$, $Z^{(*)} \! \to \ell \ell$ and $\tau \to \ell \nu \nu$.
With only a single Higgs doublet, direct Standard Model 
di-Higgs production is a very rare process, but 
resonant heavy Higgs production and decay into these final states can be 
up to two orders of magnitude larger in 2HDMs.  

Scalar cascade decays of the heavy Higgs scalar, $H$, 
can also contribute to entirely new multi-lepton final 
states that do not arise with a single Higgs doublet.  
These include gluon fusion and vector boson fusion production of $H$ with
$H \to AA, H^+ H^-, ZA, W H^\pm$ with 
$A \to bb, Zh, \tau \tau$, and $H^{\pm} \to tb, \tau \nu, Wh$ with  
$t \to Wb$ followed by 
$h \to bb, WW^*, ZZ^* , \tau \tau $ with 
(some of the) $W \to \ell \nu$, $Z^{(*)} \! \to \ell \ell$ and $\tau \to \ell \nu \nu$.
These processes can give final states with up to
six $W$ and/or $Z$ bosons. 
Similar processes in this same category 
include gluon fusion production of $A$ with 
$A \to ZH$ followed by $H \to hh$ with $h \to bb, WW^*, ZZ^* , \tau \tau $
 with  
(some of the) $W \to \ell \nu$, $Z^{(*)} \! \to \ell \ell$ and $\tau \to \ell \nu \nu$.
These processes can give final states with up to
five $W$ and/or $Z$ bosons. 

Direct di-Higgs production of non-Standard Model-like 
Higgs bosons either with or without scalar cascade decay processes  
can also give rise to multi-lepton final states that do not arise 
with a single Higgs doublet.  
These include quark--anti-quark fusion production of 
$Ah, AH, H^{\pm}A$ followed by 
$H \to WW^*, ZZ^*,\tau \tau, AA$,  
and $H^{\pm} \to tb, \tau \nu, Wh, WA$ with $t \to Wb$,  
and $A \to bb, \tau \tau$, all with 
$h,H \to WW^*, ZZ^*,\tau \tau$
with 
(some of the) 
$W \to \ell \nu$, $Z^{(*)} \to \! \ell \ell$ and $\tau \to \ell \nu \nu$. 
The existence of some of these processes is sensitive to  
mass hierarchies in the Higgs spectrum; other production 
and decay processes of this type can  arise 
depending on mass orderings.  

Additional multi-lepton final states not associated with a single Higgs doublet
can arise from production of non-Standard Model-like Higgs bosons 
in association with top quarks.  
These include 
$ttH, ttA$, and $tbH^{\pm}$ associated production with $t \to Wb$ followed by  
$H \to AA$, and $H^{\pm} \to Wh, WA$, and $A \to Zh, bb, \tau \tau$, 
all with 
$h,H \to WW^*, ZZ^*,\tau \tau$
with
(some of the) $W \to \ell \nu$, $Z^{(*)} \! \to \ell \ell$ and $\tau \to \ell \nu \nu$. 
While the production and decay processes listed here and above 
do not completely exhaust all possibilities for contributions to 
multi-lepton signatures in every conceivable  2HDM mass spectrum, 
they do include the leading topologies for a very wide range of mass hierarchies.

All of the production and decay processes outlined above are 
represented in one or more of the benchmark Higgs mass spectra 
described below.  
The value of the scalar masses chosen for each benchmark spectrum are shown in Table \ref{tab:specs}. 
In the benchmark spectra 1-3, for simplicity the pseudoscalar and the charged Higgs 
are taken to form an isotriplet with degenerate masses. 
In spectrum 4, this simplifying assumption is relaxed, with 
the pseudoscalar Higgs taken to be the lightest scalar.   
For all four 2HDM spectra, the light, CP-even scalar, $h$, 
has no available decay modes beyond those of a Standard Model Higgs boson, 
although the branching fractions may significantly differ from the SM values.   

The simplest benchmark spectrum is that with all the heavy non-Standard Model like Higgs 
bosons decoupled.  
In this case the remaining Standard Model Higgs boson can be produced in gluon fusion, 
vector boson fusion, and in assocation with vector bosons and top quarks, 
and 
it can decay to $h \to WW^*, ZZ^*, \tau \tau$.  
The leading topologies that contribute to multi-lepton signatures 
from these production and decay channels
are given in Table
\ref{tab:sm-table}.
These topologies are associated to the Standard Model-like 
Higgs boson in all 2HDMs.
The important additional production and decay channels that 
contribute to 
multi-lepton signatures 
(beyond those of the Standard Model-like Higgs boson)
in each of our four 2HDM benchmark spectra are as follows: 

\begin{table}
\begin{center}
\begin{tabular}{|cccccc|}
\hline
  & SM & Spectrum 1 & Spectrum 2 & Spectrum 3 & Spectrum 4 \\
 &  (GeV) & (GeV) &  (GeV) &  (GeV) & (GeV) \\
\hline 
$h$            & 125 & 125  & 125  & 125 & 125 \\
$H$            & $-$ & 300  & 140  & 500  &  200   \\
$A$            & $-$ & 500  & 250  & 230   &  80   \\
$H^{\pm}$ & $-$  & 500  & 250  & 230   &  250    \\
\hline
\end{tabular}
\end{center}
\caption{Higgs boson masses in the SM Benchmark and our four 2HDM Benchmark Spectra. \label{tab:specs}}
\end{table}

{\bf Benchmark spectrum 1:} The heavy neutral Higgs, $H$, 
is produced mainly through gluon fusion and vector boson fusion, and 
can decay through the same channels as a heavy Standard Model Higgs, 
plus the new kinematically allowed decay $H \to hh$. 
The pseudoscalar, $A$, is produced mainly through 
gluon fusion 
and can decay by $A \to Zh, ZH$. 
The charged Higgs, $H^{\pm}$, does not play an important role in this spectrum. 
The complete list of topologies that contribute to multi-lepton signatures 
from these production and decay channels, along with those from the Standard Model-like 
Higgs boson,  
are given in Table \ref{spec1}.

{\bf Benchmark spectrum 2:} This spectrum 
 is qualitatively similar to the first, but with $H \to hh$ 
 no longer kinematically allowed.  
 Production of the Heavy Higgs, $H$,  can proceed through 
 gluon fusion, vector boson fusion, and in association 
 with vector bosons and top quarks, with decays to Standard Model channels. 
Production of the pseudoscalar, $A$, 
through gluon fusion production and in association with top quarks 
with $A \to Zh, ZH, \tau \tau$ 
 is much greater than in spectrum 1 due to the lower $A$ mass.
The charged Higgs, $H^{\pm}$, can also be produced in association 
with a top quark, and can decay by $H^{\pm} \to Wh$.  
The complete list of topologies that contribute to multi-lepton signatures 
from these production and decay channels, along with those from the Standard Model-like 
Higgs boson,   
are given in Table \ref{spec2}.

{\bf Benchmark spectrum 3:} This spectrum is the most rich in the multiplicity of 
multi-lepton final states, 
as the decay channels $H \to hh, AA, H^+ H^-,AZ$ are all kinematically open, 
in addition to the Standard Model decay channels.  
The heavy Higgs, $H$, can be produced in gluon fusion and vector boson fusion.  
The pseudoscalar, $A$, is produced in gluon fusion, as well as from decays
of the $H$, with decays $A \to Zh,\tau \tau$. 
The charged Higgs, $H^{\pm}$, can be produced in association with a top quark, 
or from decay of $H$ with decays $H^{\pm} \to \tau \nu, Wh$.  
This spectrum includes topologies with 
sequential cascade decays through up to three Higgs scalars.  
The complete list of topologies that contribute to multi-lepton signatures 
from all these production and decay channels, along with those from the Standard Model-like 
Higgs boson,   
are given in Table \ref{spec3}. 

{\bf Benchmark spectrum 4:} This spectrum breaks the degeneracy 
between the pseudoscalar, $A$, and the charged Higgs, $H^{\pm}$, 
 in order to highlight the role of a light pseudoscalar. 
Quark--anti-quark fusion production of $A$ with the scalar Higgses, $H,h$ or  
charged Higgs, $H^{\pm}$, is significant, with decays $A \to bb, \tau \tau$ and
$H^{\pm} \to \tau \nu, Wh, WA$ as well as $H \to AA$, in addition to the Standard Model 
decay channels.   
The later decay yields a topology with three pseudoscalar Higgses in the 
final state. 
The pseudoscalar, $A$, as well as $H$ and $H^{\pm}$, can also be produced in 
association with top quarks.  
The heavy Higgs, $H$, can also be produced in gluon fusion and 
vector boson fusion. 
 The very small partial width for the decay $h\to AA^*$ in this spectrum will be ignored. 
The complete list of topologies that contribute to multi-lepton signatures 
from all these production and decay channels, along with those from the Standard Model-like 
Higgs boson,  
are given in Table \ref{spec4}.

All 233 production and decay topologies listed 
in Tables  \ref{tab:sm-table} - \ref{spec4} 
were individually simulated in our studies of multi-lepton signatures of the Standard Model Higgs 
and our four 2HDM spectra benchmarks. 
Certain channels for the 2HDM benchmarks were omitted for the sake of conciseness. 
In general, channels were omitted if the production cross section times fixed 
Standard Model branching ratios to multi-lepton final states was much less 
than 1 fb even in the most promising regions of parameter space. 
For nominal simplicity, for the 2HDM benchmarks, we omitted associated 
production channels for $h$ with $h \to ZZ^*$, having found in \cite{us} that 
with the integrated luminosity considered here, these channels 
did not contribute significantly to even low-background search channels. 
However, with significantly more integrated luminosity these channels would 
begin to contribute to the sensitivity.

\begin{table}[h]
\begin{center}
{\small
\begin{tabular}{|l|l|} \hline
Production & Decay \\ \hline
$gg \to h$ & $h \to 4 \ell$ \\
${\rm VBF} \to h$ & $h \to 4 \ell$ \\
$q \bar q \to Wh$ & $Wh \to WWW, WZZ, W\tau\tau$ \\
$q \bar q \to Zh$ & $Zh \to ZWW, ZZZ, Z\tau\tau$ \\
$t \bar t h$ & $t \bar t h \to t \bar t WW , t \bar t ZZ, t \bar t \tau \tau$ \\
\hline
\end{tabular}
\caption{The 11 independent production and decay topologies 
simulated for the Standard Model Higgs Boson with $m_h = 125$ GeV. 
The Higgs boson branching ratios are factored out of each topology. 
All top-quark, $\tau$-lepton, and $W$- and $Z$ bosons branching ratios are Standard Model. 
\label{tab:sm-table}}
}
\end{center}
\end{table}

\begin{table}[h]
\begin{center}
{\small
\begin{tabular}{|l|l|} \hline
Production & Decay \\ \hline
$gg \to h$ & $h \to 4 \ell$ \\
${\rm VBF} \to h$ & $h \to 4 \ell$ \\
$gg \to H$ & $ H \to 4 \ell$ \\
		& $H \to hh \to 4W, WW \tau \tau, 4 \tau, ZZb \bar b, ZZ WW, 4Z, ZZ \tau \tau$ \\
${\rm VBF} \to H$ & $ H \to 4 \ell$ \\
		& $H \to hh \to 4W, WW \tau \tau, 4 \tau, ZZb \bar b, ZZ WW, 4Z, ZZ \tau \tau$ \\
$gg \to A$ & $A \to Zh \to ZWW, Z\tau \tau, ZZZ$ \\
		&  $A \to ZH \to ZWW, Z\tau\tau, ZZZ$\\
		& $A \to ZH \to Zhh \to ZWWWW, ZWW\tau\tau, Z\tau\tau\tau\tau, ZZZb \bar b, ZZZ WW,
		5Z, ZZZ \tau \tau$ \\
$q \bar q \to Wh$ & $Wh \to WWW, W\tau\tau$ \\
$q \bar q \to Zh$ & $Zh \to ZWW, Z\tau\tau$ \\
$t \bar t h$ & $t \bar t h \to t \bar t WW , t \bar t \tau \tau$ \\
\hline
\end{tabular}
\caption{
The 37 independent production and decay topologies simulated for the 2HDM Benchmark Spectrum 1 with 
$m_h = 125$ GeV, 
$m_H = 300$ GeV, 
$m_A = m_{H^\pm} =  500$ GeV.
All Higgs boson branching ratios are factored out of each topology. 
All top-quark, $b$-quark, $\tau$-lepton, and $W$- and $Z$-boson branching ratios are Standard Model. 
\label{spec1}}
}
\end{center}
\end{table}%

\begin{table}[h]
\begin{center}
{\small
\begin{tabular}{|l|l|} \hline
Production & Decay \\ \hline
$gg \to h$ & $h \to 4 \ell$ \\
${\rm VBF} \to h$ & $h \to 4 \ell$ \\
$gg \to H$ & $ H \to 4 \ell$ \\
${\rm VBF} \to H$ & $H \to 4 \ell$ \\
$gg \to A$ & $A \to Zh \to ZWW, Z\tau \tau,  ZZZ$ \\
		&  $A \to ZH \to ZWW, Z\tau\tau, ZZZ$\\
$q \bar q \to Wh$ & $Wh \to WWW, W\tau\tau$ \\
$q \bar q \to Zh$ & $Zh \to ZWW, Z\tau\tau$ \\
$q \bar q \to WH$ & $WH \to WWW, W\tau\tau$ \\
$q \bar q \to ZH$ & $ZH \to ZWW, Z\tau\tau$ \\
$t \bar t h$ & $t \bar t h \to t \bar t WW , t \bar t \tau \tau$ \\
$t \bar t H$ & $t \bar t H \to t \bar t WW , t \bar t \tau \tau$ \\
$t \bar t A$ & $t \bar t A \to t \bar t \tau \tau$ \\
		& $t \bar t A \to t \bar t Zh \to t \bar t Z WW, t \bar t Z \tau \tau, t \bar t Z b \bar b,  t\bar t ZZZ$ \\
		& $t \bar t A \to t \bar t ZH \to t \bar t Z WW, t \bar t Z \tau \tau, t \bar t Z b \bar b,  t\bar t ZZZ$ \\
$t b  H^\pm$ & $t b   H^\pm \to   t b W h\to t  b WWW, t b  W \tau\tau, t b  W ZZ$\\
\hline
\end{tabular}
\caption{
The 34 independent production and decay topologies simulated for the 2HDM Benchmark Spectrum 2 with 
$m_h = 125$ GeV, 
$m_H = 140$ GeV, 
$m_A = m_{H^\pm} =  250$ GeV.
All Higgs boson branching ratios are factored out of each topology. 
All top-quark, $b$-quark, $\tau$-lepton, and $W$- and $Z$-boson branching ratios are Standard Model. 
\label{spec2}}
}
\end{center}
\end{table}%

\begin{table}[h]
\begin{center}
{\small
\begin{tabular}{|l|l|} \hline
Production & Decay \\ \hline
$gg \to h$ & $h \to 4 \ell$ \\
${\rm VBF} \to h$ & $h \to 4 \ell$ \\
$gg \to H$ & $ H \to 4 \ell$ \\
		& $H \to hh \to 4W, WW \tau \tau, 4 \tau, ZZb \bar b, ZZ WW, 4Z, ZZ \tau \tau$ \\
		& $H \to AA \to 4 \tau$ \\
		& $H \to AA \to \tau \tau Zh \to \tau \tau ZWW, \tau \tau Z \tau \tau, \tau \tau Z b \bar b,  \tau\tau ZZZ$ \\
		& $H \to AA \to Zh Zh \to ZZ WW WW, ZZ WW \tau \tau,  ZZ WW b\bar b, ZZ \tau \tau b \bar b, ZZ \tau \tau \tau \tau$ \\
		& $H \to AA \to Zh Zh \to ZZ b \bar b b \bar b,  ZZ ZZ b \bar b, ZZ ZZ \tau\tau, ZZ ZZ W W, 6Z$ \\
		& $H \to H^+ H^- \to Wh Wh \to WW WW WW, WW WW \tau \tau, WW WW b \bar b, WW \tau \tau \tau \tau$ \\
		& $H \to H^+ H^- \to Wh Wh \to WW \tau \tau b \bar b, WW ZZ b\bar b, WW WW ZZ, WW ZZ ZZ, WW ZZ \tau\tau$ \\
		&$H \to H^+ H^- \to   \tau \nu Wh \to \tau \nu W WW, \tau \nu W\tau \tau, \tau \nu WZZ$ \\
		&$H \to H^+ H^- \to   t b Wh \to t b W WW, t b W\tau \tau,  t b WZZ$ \\
		&$H \to ZA  \to Z\tau\tau$ \\
		&$H \to ZA  \to  ZZh \to ZZ\tau\tau, ZZWW, ZZb\bar b, ZZZZ$ \\
		&$H \to WH^\pm  \to  WWh \to WW\tau\tau, WWWW, WWZZ$ \\
${\rm VBF} \to H$ & $ H \to 4 \ell$ \\
		& $H \to hh \to 4W, WW \tau \tau, 4 \tau, ZZb \bar b, ZZ WW, 4Z, ZZ \tau \tau$ \\
		& $H \to AA \to 4 \tau$ \\
		& $H \to AA \to \tau \tau Zh \to \tau \tau ZWW, \tau \tau Z \tau \tau, \tau \tau Z b \bar b,  \tau\tau ZZZ$ \\
		& $H \to AA \to Zh Zh \to ZZ WW WW, ZZ WW \tau \tau,  ZZ WW b\bar b, ZZ \tau \tau b \bar b, ZZ \tau \tau \tau \tau$ \\
		& $H \to AA \to Zh Zh \to ZZ b \bar b b \bar b,  ZZ ZZ b \bar b, ZZ ZZ \tau\tau, ZZ ZZ W W, 6Z$ \\
		& $H \to H^+ H^- \to Wh Wh \to WW WW WW, WW WW \tau \tau, WW WW b \bar b, WW \tau \tau \tau \tau$ \\
		& $H \to H^+ H^- \to Wh Wh \to WW \tau \tau b \bar b, WW ZZ b\bar b, WW WW ZZ, WW ZZ ZZ, WW ZZ \tau\tau$ \\
		&$H \to H^+ H^- \to   \tau \nu Wh \to \tau \nu W WW, \tau \nu W\tau \tau, \tau \nu WZZ$ \\
		&$H \to H^+ H^- \to   t b Wh \to t b W WW, t b W\tau \tau,  t b WZZ$ \\
		&$H \to ZA  \to Z\tau\tau$ \\
		&$H \to ZA  \to  ZZh \to ZZ\tau\tau, ZZWW, ZZb\bar b, ZZZZ$ \\
		&$H \to W H^\pm  \to  WWh \to WW\tau\tau, WWWW, WWZZ$ \\
$gg \to A$ & $A \to Zh \to ZWW, Z\tau \tau, ZZZ$ \\
$q \bar q \to Wh$ & $Wh \to WWW, W\tau\tau$ \\
$q \bar q \to Zh$ & $Zh \to ZWW, Z\tau\tau$ \\

$t \bar t h$ & $t \bar t h \to t \bar t WW , t \bar t \tau \tau$ \\
$t \bar t A$ & $t \bar t A \to t \bar t \tau \tau$ \\
		& $t \bar t A \to t \bar t Zh \to t \bar t Z WW, t \bar t Z \tau \tau, t \bar t Z b \bar{b},  t\bar t ZZZ$ \\
$t b  H^\pm$ & $ t  b H \to  t b W h\to t b W WW, t b W \tau\tau,t b WZZ$\\
\hline
\end{tabular}
\caption{
The 111 independent production and decay topologies simulated for the 2HDM Benchmark Spectrum 3 with 
$m_h = 125$ GeV, 
$m_H = 500$ GeV, 
$m_A = m_{H^\pm} =  230$ GeV.
All Higgs boson branching ratios are factored out of each topology. 
All top-quark, $b$-quark, $\tau$-lepton, and $W$- and $Z$-boson branching ratios are Standard Model. 
\label{spec3}}
}
\end{center}
\end{table}%

\begin{table}[h]
\begin{center}
{\small
\begin{tabular}{|l|l|} \hline
Production & Decay \\ \hline
$gg \to h$ & $h \to 4 \ell$ \\
${\rm VBF} \to h$ & $h \to 4 \ell$ \\
$gg \to H$ & $ H \to 4 \ell$ \\
		& $H \to AA \to 4 \tau$ \\
${\rm VBF} \to H$ & $H \to 4 \ell$ \\
		& $H \to AA \to 4 \tau$ \\
$q \bar q \to Wh$ & $Wh \to WWW, W\tau\tau$ \\
$q \bar q \to Zh$ & $Zh \to ZWW, Z\tau\tau$ \\
$t \bar t h$ & $t \bar t h \to t \bar t WW , t \bar t \tau \tau$ \\
$t \bar t H$ & $t \bar t H \to t \bar t WW , t \bar t \tau \tau$ \\
		& $t \bar t H \to t \bar t AA \to t \bar t \tau \tau \tau \tau, t \bar t \tau \tau bb$ \\
$t \bar t A$ & $t \bar t A \to t \bar t \tau \tau$ \\
$t b  H^\pm$ & $t b H^\pm \to t b  Wh\to t b WWW,t b W \tau\tau,t b  W ZZ$\\
                          & $t b  H^\pm \to t b  W A\to t b W \tau\tau$\\                          
$q \bar q \to H^{\pm} A$ & $H^\pm A \to 	Wh b \bar b \to W WW b \bar b, W \tau \tau b \bar b,  W ZZ b \bar b $	   \\
				& $H^\pm A \to 	Wh \tau \tau \to W WW \tau \tau, W \tau \tau \tau \tau, W b \bar b \tau \tau,  W ZZ \tau \tau$	   \\
				& $H^\pm A \to \tau \nu \tau \tau, t \bar b \tau \tau $ \\
				& $H^\pm A \to WA A \to W \tau \tau \tau \tau, W \tau \tau b \bar b$ \\
$q \bar q \to Ah$ & $Ah \to \tau \tau WW, \tau \tau \tau \tau, \tau \tau ZZ$ \\
$q \bar q \to AH$ & $AH \to \tau \tau WW, \tau \tau \tau \tau, \tau \tau ZZ$ \\
			& $AH \to AAA \to 6 \tau, \tau \tau \tau \tau b \bar b$ \\
		\hline
\end{tabular}
\caption{
The 40 independent production and decay topologies simulated for the 2HDM Benchmark Spectrum 4 with 
$m_h = 125$ GeV, 
$m_H = 200$ GeV, 
$m_A = 80 $ GeV, 
$m_{H^\pm} =  250$ GeV.
All Higgs boson branching ratios are factored out of each topology. 
All top-quark, $b$-quark, $\tau$-lepton, and $W$- and $Z$-boson branching ratios are Standard Model. 
\label{spec4}}
}
\end{center}
\end{table}%


\section{Search Strategy and Simulation Tools}
\label{sec:search}

In principle, it might be 
possible to design a multi-lepton search with sensitivity 
specifically tailored to certain features of 
the signatures that arise from some of the production and decay topologies of 
2HDMs. 
However, designing such a dedicated search would 
require a detailed understanding of backgrounds in many channels 
that is well beyond the scope of a theory-level study. 
Instead, as done previously in a study of the multi-lepton signatures of the 
Standard Model Higgs boson \cite{us}, 
we will adopt the selection cuts and background estimates 
of an existing CMS multi-lepton analysis \cite{CMSMulti, CMSMulti5} 
to demonstrate the efficacy of a 2HDM multi-lepton search. 
In the conclusions, we will comment briefly on how a focussed search could be  
further optimized to maximize sensitivity to multi-lepton final states arising from an extended scalar sector.

Although the CMS analysis includes hadronically decaying $\tau$-leptons, 
for simplicity of simulation, 
we will consider only strictly leptonic $\ell = e, \mu$ final states (of course, still including leptonic $\tau$ decays).  Additionally, we treat all hadronic taus as having failed selection criteria, thus being identified as jets.  Because of this, some events (mainly those involving $4\tau$ final states) will be categorized differently than in the CMS analysis. For instance, an event with three $e/\mu$ and one hadronic $\tau$ that the CMS analysis would have included in a $4\ell$ (with $1\tau$) bin, will instead be included in a $3\ell$ bin in our analysis, potentially with higher $H_T$ due to the additional energy of the hadronic $\tau$-lepton. 
While this is a deviation from the exact procedure of the CMS analysis, it goes in the conservative direction, as the $4\ell$ with $1\tau$ bins have significantly smaller backgrounds than the $3\ell$ with $0\tau$ bins. Thus, if we could implement a satisfactory modeling of hadronic $\tau$ identification in our study, we would expect our bounds to become stronger in regions of parameter space where $4\tau$ final states are driving the limits. For other final states such as $H\to hh\to 4W$, the impact of this effect on our signal is at the few percent level or less.

\subsection{Signal channels}\label{subsec:channels}

The prompt irreducible Standard Model backgrounds to multi-lepton searches 
are small and arise predominantly through leptonic decays of $W$ and $Z$ bosons. 
Such backgrounds may therefore be reduced by demanding significant hadronic activity and/or missing energy in the events. Hadronic activity can be quantified by the variable $H_T$, defined as the scalar sum of the transverse energies of all jets passing the preselection cuts. The missing transverse energy (MET) is the magnitude of the vector sum of the momenta of all particles in the event.

In order to make use of $H_T$ and MET, the CMS 
analysis of \cite{CMSMulti, CMSMulti5} divides events with $H_T > 200$ (MET $>50$) GeV 
into a high $H_T$ (MET) category, and those 
with $H_T < 200$ (MET $< 50$) GeV into a low $H_T$ (MET) category. 
The HIGH $H_T$ and HIGH MET requirements (individually or in combination) 
lead to a significant reduction in Standard Model 
backgrounds.\footnote{In the CMS study, a separate binning is also considered using $S_T$, a variable defined to be the scalar sum of MET, $H_T$, and leptonic $p_T$  \cite{CMSMulti}.  
For simplicity, we will not make use of $S_T$ here.}

 Another useful observable in reducing backgrounds is the presence of 
 $Z$ candidates, specifically the existence of an opposite-sign same-flavor 
 (OSSF) lepton pair with an invariant mass between $75-105$ GeV. 
 Events are thus further subdivided, and assigned a No $Z$ channel if no such pair exists. 
 It is also useful to characterize events according to whether 
 they may contain {\it off-shell} $\gamma^*$/$Z^*$ candidates, given by the number of OSSF lepton pairs. 
 Thus, for instance, three-lepton events are assigned to the DY0 (no possible Drell-Yan pairs) 
 or DY1 category (one OSSF pair). 
 The full combination of 3 and 4 lepton events results in 20 possible 
 categories of $H_T$ high/low; MET high/low; $Z$/no $Z$; and DY0/DY1. 
 The 20 channels are presented in Table~\ref{tab:SM}. 
 For each of the $3 \ell$ and $4 \ell$ categories, channels are listed from top to bottom in
approximately
descending order of backgrounds, or equivalently
ascending order of sensitivity, with the last such channel at the bottom
dominated by Standard Model backgrounds. 
Events are entered in the table exclusive-hierarchically from the top to the bottom. 
This ensures that 
 each event appears only once in the table, and in the lowest 
 possible background channel consistent with its characteristics.
Although the backgrounds in the individual channels vary over a wide range,  
all 20 channels are used to compute sensitivity limits.


\subsection{Simulation}

For simulating signal processes, we have used MadGraph v4 \cite{Maltoni:2002qb,Alwall:2007st}. In order to simulate a general 2HDM in MadGraph, we treat the 2HDM as a simplified model using a modified version of the \texttt{2HDM4TC} model file \cite{2HDM4TC}.  Cascade decays were performed in BRIDGE \cite{Meade:2007js}. Subsequent showering and hadronization effects were simulated using Pythia \cite{Sjostrand:2006za}. 
Detector effects and object reconstruction was 
simulated using PGS \cite{PGS} with the isolation algorithm for muons and taus modified to more accurately reflect the procedure used by the CMS collaboration. In particular, we introduce a new output variable 
called \texttt{trkiso} for each muon \cite{Gray:2011us}. The variable \texttt{trkiso} is defined to be the sum $p_T$ of all tracks, ECAL, and HCAL deposits within an annulus of inner radius 0.03 and outer radius 0.3 in $\Delta R$ surrounding a given muon. Isolation requires that for each muon, $I$=\texttt{trkiso}/$p_T$ of the muon be less than 0.15.  The efficiencies of PGS detector effects were normalized by simulating the mSUGRA benchmark studied in \cite{CMSMulti} and comparing the signal in 3$\ell$ and $4 \ell$ channels. To match efficiencies with the CMS study, we applied a lepton ID efficiency correction of 0.87 per lepton to our signal events. As discussed earlier, we applied preselection and analysis cuts  in accordance with those in \cite{CMSMulti}.

In order to assess the multi-lepton signatures of the 2HDMs studied here we employ a factorized mapping 
procedure \cite{sunil} to go between model parameters and signatures.  
In this procedure the acceptance times efficiency is independently determined in each of the 20 exclusive multi-lepton channels
by monte carlo simulation 
of each individual production and decay topology in each of the four 2HDM mass spectra 
as well as for the individual topologies of the Standard Model Higgs boson. 
The cross section times branching ratio times acceptance and efficiency in any of the 20 exclusive channels 
at any point in parameter space in a given mass spectrum is then given by a sum over the production cross 
section times acceptance and efficiency for each topology of that spectrum, times a product
of the branching ratios that appear in each topology 
\beq
\sigma \! \cdot \! {\rm Br} \! \cdot \! {\cal A} (pp \to f) = 
\sum_{t}\sigma(pp \to t) 
{\cal A}( pp \to t \to f) 
\prod_{a} {\rm Br}_{a}(t \to f) 
\label{factorized} 
\eq
where $f$ is a given exclusive final state channel, 
$t$ labels the topology, and $a$ the branching ratios of the 
decays in the $t$-th topology. 
Dependence on the parameter space characterized by $\alpha$ and 
$\beta$ enters only through the 
production cross sections and decay branching ratios. 
The factorized terms in (\ref{factorized}) are determined as follows: 
%
%
\begin{itemize}
{\item {\bf Acceptance times Efficiency:} 
For each individual production and decay topology listed in Tables \ref{tab:sm-table} - \ref{spec4}, 
the acceptance times detector efficiency into each of the 20 exclusive multi-lepton channels 
listed in Table \ref{tab:SM} was simulated with the monte carlo tools 
described above. 
The acceptance times efficiency of each topology was calculated assuming unit branching ratios 
for all Higgs boson decays but with Standard Model values for 
decays of $W$ and $Z$ bosons, and top quarks and $\tau$-leptons.  
A total of 50,000 events were simulated for each topology to ensure good statistical 
coverage of all the exclusive multi-lepton channels. 
 }
{\item {\bf Cross Sections:} 
For the case of the Standard Model Higgs boson, the NLO production cross sections 
for gluon fusion, vector boson fusion, and production in association with a vector 
boson or top quarks 
are taken from the LHC Higgs Cross Section Group \cite{LHCHiggsCrossSectionWorkingGroup:2011ti}. 
For the 2HDM spectra the ratio of LO production partial widths in each 
production channel for $h$ and $H$ relative to a Standard Model Higgs boson of 
the same mass are calculated analytically 
from the couplings presented in section \ref{sec:theory} 
as functions of the mixing parameters $\alpha$ and $\beta$. 
The NLO Standard Model Higgs production cross sections in each production 
channel are then rescaled 
by these factors to obtain an estimate for the 
NLO cross sections; for instance the $\alpha, \beta$ dependent 
cross section for gluon fusion production of $H$ is taken to be 
\beq
\sigma_{\rm NLO}(gg \to H)|_{\alpha,\beta} = \sigma_{\rm NLO}(gg \to h_{\rm SM}) ~
  \frac{\Gamma_{\rm LO}(H \to gg){\big |}_{\alpha,\beta}}{\Gamma_{\rm LO}(h_{\rm SM} \to gg)}
\eq
The same procedure of normalizing to Standard Model Higgs boson 
NLO cross sections through the $\alpha$ and $\beta$ dependent 
ratios of LO production partial widths 
is used for production of $A$ by gluon fusion or in association 
with top quarks. 
This is expected to be a good approximation since 
the fractional size of NLO corrections in these cases should not 
be strongly dependent on the parity of the Higgs scalar.  
For the modes that involve production of two Higgs bosons, 
or of the charged Higgs in association with a top quark, 
the LO cross sections are calculated using  Madgraph v4 
with a conservative $K$-factor of $K=1.2$ applied.  
These cross sections are calculated for a single canonical value of 
$\alpha$ and $\beta$ and then rescaled analytically 
using the couplings in section \ref{sec:theory} to obtain 
the cross sections at general values.   
}
%
{\item {\bf Higgs Bosons Branching Ratios:} 
For the case of the Standard Model Higgs boson, the NLO partial 
decay widths and branching ratios 
are taken from the LHC Higgs Cross Section Group \cite{LHCHiggsCrossSectionWorkingGroup:2011ti}. 
For the 2HDM spectra the ratio of LO partial decay widths for $h$ relative to a Standard 
Model Higgs boson of the same mass are calculated analytically as functions 
of the mixing parameters $\alpha$ and $\beta$ using 
the couplings presented in section  \ref{sec:theory}. 
The NLO Standard Model Higgs boson partial decay widths are then rescaled 
by these factors to obtain  estimates for the 
NLO partial widths; for instance the $\alpha, \beta$ dependent 
partial width for the light scalar $h$ to $b\bar{b}$ is taken to be 
\beq
\Gamma_{\rm NLO}(h \to b \bar b)|_{\alpha,\beta} = \Gamma_{\rm NLO} (h_{\rm SM} \to b \bar b) ~ 
\frac{\Gamma_{\rm LO}(h \to b \bar b){\big |}_{\alpha,\beta}}{\Gamma_{\rm LO}(h_{\rm SM} \to b \bar b)}
\eq
The same procedure of normalizing to Standard Model Higgs boson 
NLO partial decay widths through the ratio of LO decay widths 
is used for the $H$ and $A$ decay modes listed in Table \ref{higgs-decays}
that are in common with the $h$ decay modes.  
This estimate is used since, just as for a production cross section, 
the fractional size of NLO corrections to decay widths in these cases should not 
be strongly dependent on the parity of the Higgs scalar.  
For the remainder of the $H$ and $A$ decay modes listed in Table \ref{higgs-decays}
that are kinematically open in a given spectrum, 
as well as the $H^{\pm}$ decay modes given in the Table that are open, 
the LO decay widths are calculated analytically \cite{Djouadi:1995gv} 
as a function of  
$\alpha$ and $\beta$ using the couplings in section \ref{sec:theory}. 
Except for the charged Higgs decays to quarks, none of these
decay modes involve strongly interacting particles, so LO widths 
should be a good approximation in this case. 
The partial widths for all the open decay modes of each Higgs 
scalar in Table \ref{higgs-decays} are then used to calculate the 
$\alpha$ and $\beta$ dependent total 
widths and branching ratios in each mass spectrum.  
}
\end{itemize}
\begin{table}[h]
\begin{center}
\begin{tabular}{|c|l|l|} \hline
Higgs Boson  & ~~~ Decay Modes\\ \hline
$h$ & $bb,cc, \tau \tau, WW^*, ZZ^*, gg, \gamma \gamma, Z \gamma$ \\
$H$ & $tt, bb, cc, \tau \tau, WW^{(*)}, ZZ^{(*)}, 
      hh, AA, H^+ H^-, ZA, WH^{\pm}, 
     gg, \gamma \gamma, Z \gamma$ \\
$A$ & $tt, bb, cc, \tau \tau, Zh, ZH,  
     gg, \gamma \gamma, Z \gamma$ \\
$H^{\pm}$ & $tb, ts, cs, \tau \nu, WA, Wh, WH$ \\
\hline
\end{tabular}
\caption{Decay modes of the Higgs boson scalars used in branching ratio calculations.  
Partial widths of the kinematically open decay modes are calculated in 
each benchmark spectrum
as a function of the mixing parameters $\alpha$ and $\beta$  
to determine the total width and 
individual branching ratios.  
\label{higgs-decays}}\end{center}
\end{table}
Using this factorized mapping procedure, 
each of the 20 exclusive multi-lepton channels 
for a given benchmark spectrum 
over the entire $\alpha, \beta$ plane in all four 2HDM types 
is covered by 
a single set of monte carlo samples for the production and decay topologies. 

In some cases, particularly in Spectrum 3, the total widths of some scalars (particularly $H$) increase drastically in certain regions of parameter space, typically due to enhanced scalar couplings.  Our simulation and normalization techniques, however, treat all particles in the narrow width approximation and assume the validity of perturbation theory in the scalar couplings.  In the regions of parameter space where scalar widths grow large, one expects higher-order effects to modify the limits; in this respect the limits we find in high-width regions should be viewed as rough estimates subject to potentially large corrections beyond the scope of our approach.

\section{Results}
\label{sec:results}

In this section, we present the results of the analysis outlined above using the CMS multi-lepton search  based on $5~{\rm fb}^{-1}$ of 7 TeV proton-proton collisions at the LHC \cite{CMSMulti5}. 
We first consider the sensitivity of 
the CMS multi-lepton search 
to a Standard Model Higgs boson near 125 GeV before presenting limits in the full 2HDM parameter space for our four benchmark spectra.

For each benchmark, we briefly discuss the major processes that contribute to multi-lepton final states, including direct production and decay of individual scalars as well as cascades among scalars. 
We also illustrate many of the partial widths and $\sigma \cdot {\rm Br}$'s for key scalar cascades, which helps to capture the qualitative shape of the multi-lepton limits in the space of $(\sin \alpha, \tan \beta)$. In many cases, the signals of Type I and Type III 2HDM (and separately Type II and Type IV 2HDM) are often similar, up to final states involving $\tau$-leptons. 
These similarities arise because in each case the quark couplings are identical for the pairs of 2HDM types, so in particular the scaling of the $h \to b \bar{b}$ partial widths that often govern the total width (as well as the $h t \bar t$ couplings that governs the gluon fusion production rate) are identical. The only substantial distinction arises in standard channels with $\tau$ final states, since the lepton couplings differ among these pairs of 2HDM types.

In each case, we show the regions of parameter space excluded by the 5 fb$^{-1}$ CMS 
multi-lepton search. In regions not yet excluded, we show the 95\% CL limits on the production cross section times branching ratio in multiples of the theory cross section times branching ratio for the benchmark spectrum and 2HDM type.  To compute our 95\% CL limits, we used a Bayesian likelihood function 
assuming poisson distributions for each of the 20 channels with a flat prior for the signal.  
We treated the magnitude of the backgrounds in each exclusive channel 
as nuisance parameters with distributions given by a truncated positive 
definite Gaussian distribution with 
width equal to the background uncertainty.
The number of signal events in each exclusive channel 
for a given $\alpha$ and $\beta$ 
was obtained from the cross section times branching times acceptance and efficiency
in each channel 
times the integrated luminosity.  
For simplicity, we assumed there was no error on the signal. 
To generate the expected limits, a large number of background-only pseudo-experiments were used in place of data.   

For comparison, we also show regions where the heavy, CP-even scalar, $H$, is currently excluded by standard Higgs searches at 7 TeV \cite{125HiggsCMS} at roughly the same luminosity of the multi-lepton search. For Spectra 1, 3, and 4 we use the combined CMS Higgs limit at 5  fb$^{-1}$ of 7 TeV collisions, which is driven by $ZZ$ and $WW$ final states. For Spectrum 2, where $m_H = 140$ GeV, we use the $WW \to 2 \ell 2 \nu$ CMS Higgs limit at 5  fb$^{-1}$ of 7 TeV collisions, which dominates the exclusion limit at this mass. We also consider direct limits on the pseudoscalar $A$ and the charged Higgses $H^\pm$, but these do not impact the parameter space explored here. For the pseudoscalar, the best current CMS limits come from MSSM Higgs searches for $b\bar{b} A$ associated production with $A \to \tau \tau$ \cite{Chatrchyan:2012vp}. For a Type II 2HDM, the current exclusion is relevant only for $\tan \beta > 10$, and in all other 2HDM types the $\sigma \cdot {\rm Br}$ for $b\bar{b} A$ associated production 
with $A \to \tau \tau$ is smaller than in the Type II case. 
Searches for di-tau resonances \cite{Chatrchyan:2012hd} do not lead to meaningful limits. 
Finally,  searches for charged Higgses such as \cite{Chatrchyan:2012cw} 
are sensitive only to $H^\pm$ production in decays of the top quark, 
which are not relevant for the benchmark spectra considered here.

\subsection{Standard Model Higgs}

\begin{table}[h]
\begin{center}
{\small
\begin{tabular}{|lllccc|}
\hline
  & & &   Observed & Expected & SM Higgs   \\ 
  & & & & & Signal \\  \hline
    & & & & & \\
       4 Leptons & & & & & \\
   & & & & & \\
~~$^\dagger$MET HIGH & HT HIGH & No Z   & 0 & 0.018 $\pm$ 0.005 & 0.03  \\
~~$^\dagger$MET HIGH & HT HIGH &~~~~~Z        & 0 & 0.22 $\pm$ 0.05 & 0.01 \\
~~$^\dagger$MET HIGH & HT LOW & No Z   & 1 & 0.20 $\pm$ 0.07 & 0.06 \\
~~$^\dagger$MET HIGH & HT LOW &~~~~~Z         & 1 & 0.79 $\pm$ 0.21  & 0.22 \\
~~$^\dagger$MET LOW & HT HIGH & No Z    & 0  & 0.006 $\pm$ 0.001 & 0.01 \\
~~$^\dagger$MET LOW & HT HIGH &~~~~~Z         & 1& 0.83 $\pm$ 0.33 & 0.01 \\
~~$^\dagger$MET LOW & HT LOW & No Z     & 1 & 2.6 $\pm$ 1.1 & 0.36 \\
~~$^\dagger$MET LOW & HT LOW &~~~~~Z          & 33& 37 $\pm$ 15 & 1.2 \\

  & & & & & \\
  3 Leptons & & & & &\\
  & & & & &  \\
~~$^\dagger$MET HIGH & HT HIGH & DY0              & 2 & 1.5 $\pm$ 0.5 & 0.15 \\
~~$^\dagger$MET HIGH & HT LOW & DY0               & 7 & 6.6 $\pm$ 2.3  & 0.67\\
~~$^\dagger$MET LOW & HT HIGH & DY0                & 1 & 1.2 $\pm$ 0.7  & 0.04 \\
~~$^\dagger$MET LOW & HT LOW & DY0                & 14 & 11.7 $\pm$ 3.6 & 0.63 \\
~~$^\dagger$MET HIGH & HT HIGH & DY1 No Z      & 8 & 5.0 $\pm$ 1.3 & 0.38  \\
~~$^\dagger$MET HIGH & HT HIGH & DY1~~~~~~Z   & 20 & 18.9 $\pm$ 6.4 & 0.19 \\
~~$^\dagger$MET HIGH & HT LOW & DY1 No Z       & 30 & 27.0 $\pm$ 7.6  & 1.8 \\
~~~~\!MET HIGH & HT LOW & DY1~~~~~~Z    & 141 &  134 $\pm$ 50 & 1.6 \\
~~$^\dagger$MET LOW & HT HIGH & DY1 No Z       & 11 & 4.5 $\pm$ 1.5 & 0.13 \\
~~$^\dagger$MET LOW & HT HIGH & DY1~~~~~~Z   & 15 & 19.2 $\pm$ 4.8 & 0.09  \\
~~~~\!MET LOW & HT LOW & DY1 No Z          &  123  & 144 $\pm$ 36  & 1.8 \\
~~~~\!MET LOW & HT LOW & DY1~~~~~~Z    & 657 & 764 $\pm$ 183 & 4.3  \\
  & & & & & \\
    \hline
\end{tabular}
\caption{
Observed and expected number of events in various exclusive multi-lepton channels 
from the CMS multi-lepton search with 
5 fb$^{-1}$ of 7 TeV proton-proton collisions \cite{CMSMulti5}, 
along with expected  number
of Standard Model Higgs boson signal events for $m_h = 125$ GeV 
after acceptance and efficiency. 
HIGH and LOW for MET and HT indicate $\MET \GTLT $ 50 GeV and $H_T \GTLT  200$ GeV respectively.
DY0 $\equiv \ell^{\prime \pm} \ell^{\mp} \ell^{\mp}$, DY1 $\equiv \ell^{\pm} \ell^+ \ell^-, \ell^{\prime \pm} \ell^+ \ell^- $,
for $\ell = e, \mu$.
No Z and Z indicate $|m_{\ell \ell} - m_Z| \GTLT 15$ GeV for any opposite sign same flavor pair.
The channels with moderate to good sensitivity to multi-lepton Higgs boson signals 
are indicated with daggers. 
\label{tab:SM}}
}
\end{center}
\end{table}

We begin by briefly considering the multi-lepton signals of a Standard Model Higgs boson. 
This is useful both as an update to the multi-lepton Higgs search proposed in \cite{us} and as a way of understanding certain aspects of the 2HDM multi-lepton signals. 
In the alignment limit defined by $\sin(\beta - \alpha) = 1$ the 
Higgs expectation values and physical CP-even $h$ eigenstate are aligned, 
and the tree-level couplings of $h$ 
are identical to those of the 
Standard Model Higgs boson.  
So in the alignment limit, a 2HDM has an irreducible contribution to multi-lepton signatures 
that is equal to that of the Standard Model Higgs boson, with additional 
contributions coming from the heavier Higgs bosons.   
The decoupling limit is a special case of the alignment limit in which the heavy Higgs scalars
are decoupled with large masses. 
In this respect the Standard Model Higgs multi-lepton signals represents a lower bound 
over a sub-space of the 2HDM parameter space, and a limit of the 
general spectrum space. 

\begin{table}[h]
\begin{center}
\begin{tabular}{|cccc|}
\hline
$m_h$ & 120 GeV & 125 GeV & 130 GeV \\ \hline
Observed & 5.4 & 4.9 & 3.5  \\
Expected & 4.2  & 3.8  & 2.8   \\
\hline
\end{tabular}
\caption{Observed and expected 95\% CL limits 
from the CMS multi-lepton search with 5 fb$^{-1}$ of 7 TeV proton-proton collisions \cite{CMSMulti5}  
on the Higgs boson production cross section times branching ratio in multiples 
of that for Standard Model Higgs 
multi-lepton production and decay topologies listed in Table \ref{tab:sm-table}
with Standard Model branching ratios. 
Limits are obtained from an exclusive combination of the observed and expected number 
of events in all the multi-lepton channels presented in Table \ref{tab:SM}. 
\label{tab:SMlimits}}
\end{center}
\end{table}

For the Standard Model Higgs, we consider the resonant channels $gg \to h \to ZZ^* \to 4 \ell$ and $q \bar q \to h \to ZZ^* \to 4 \ell$; the non-resonant channels $gg \to h \to ZZ^* \to 2 \ell 2 \tau$ and $q \bar q \to h \to ZZ^* \to 2 \ell 2 \tau$; and the associated production channels $Zh, Wh,$ and $t \bar{t} h$ with $h \to ZZ^*$, $WW^*$, and $\tau \tau$, all with many possible states yielding multi-lepton signatures.  The combined signal expectations for a Higgs at 125 GeV in each of the 20 exclusive multi-lepton channels are shown in Table \ref{tab:SM}.
As 3$\ell$ bins require exactly 3 leptons and $4\ell$ bins require $\geq4$ leptons, each event appears in the table only once.
Although limits may be placed on the signal from
any individual channel in the multi-lepton search, the greatest sensitivity comes from combining all exclusive channels. Combining all multi-lepton channels, we find that the 5 fb$^{-1}$
multi-lepton CMS results \cite{CMSMulti5} yield the expected and observed limits for a Standard Model Higgs at $m_h = 120, 125,$ and $130$ GeV shown in Table \ref{tab:SMlimits}. The dominant decay modes and exclusive channels contributing to these limits were discussed in detail in \cite{us}.

The multi-lepton signals of $h$ remain important in the general 2HDM parameter space, both through Standard Model production of $h$ and the production of $h$ in scalar cascades. The variation in these signals as a function of $\sin \alpha$ and $\tan \beta$ for the four types of 2HDM was studied in detail in \cite{Craig:2012vn}; in what follows, we will often refer to these results to understand the parametric changes in the multi-lepton limit across the 2HDM parameter space.

\subsection{Spectrum 1}

Now let us turn to the multi-lepton signals and limits of our 2HDM benchmark spectra. The multi-lepton limits on the first benchmark spectrum for all four types of 2HDM are shown in Figure \ref{fig:s1ex}.  
Limits in this and the following figures were 
    obtained from an exclusive combination of the observed and expected number of events in all the 
    multi-lepton channels presented in Table \ref{tab:SM}
      on an evenly-spaced grid in 
    $-1 \leq \sin \alpha \leq 0$ and $1 \leq \tan \beta \leq 10$ with spacing 
    $\Delta(\sin \alpha) = 0.1$ and $\Delta(\tan \beta) = 1$; 
    contours were determined by numerical interpolation between these points. 
    
\begin{figure}[h]
\centering
\includegraphics[width=2in]{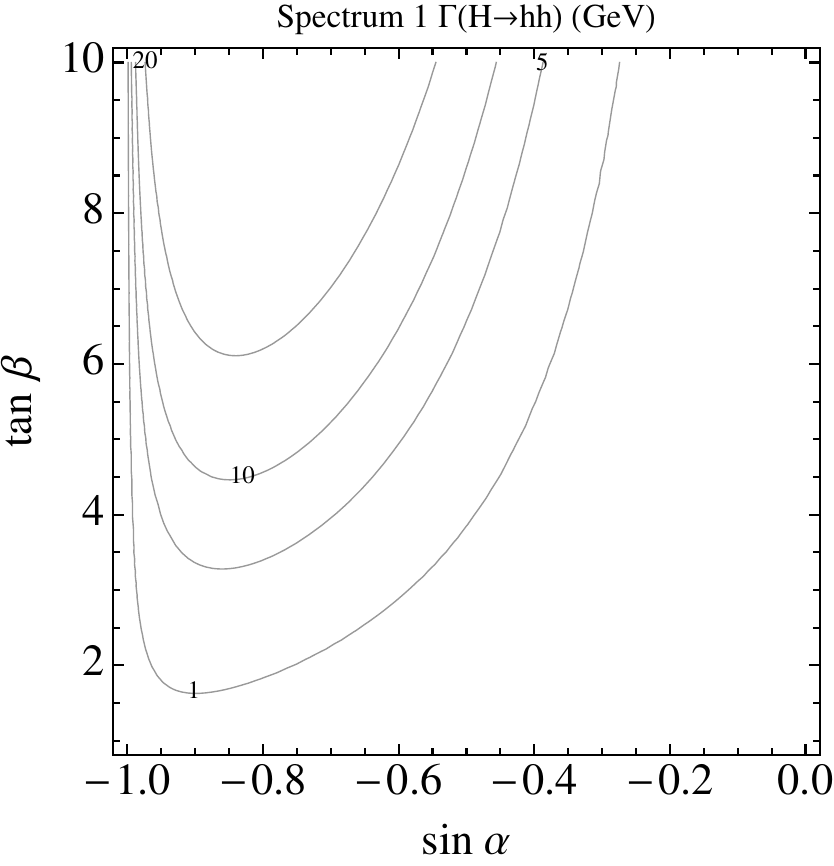}
\includegraphics[width=2in]{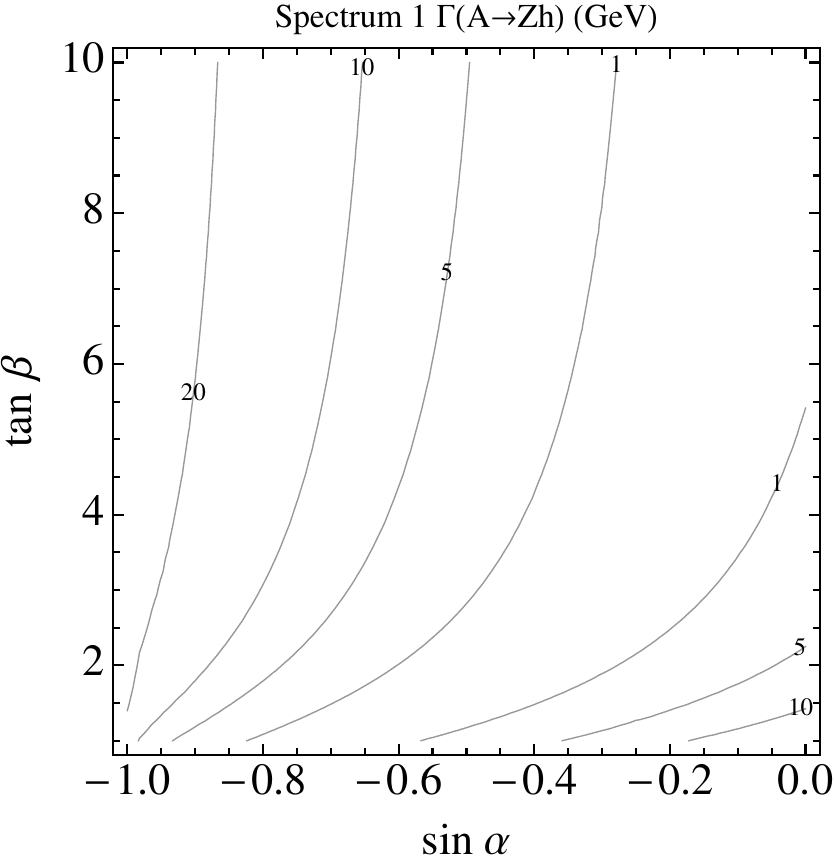}
\includegraphics[width=2in]{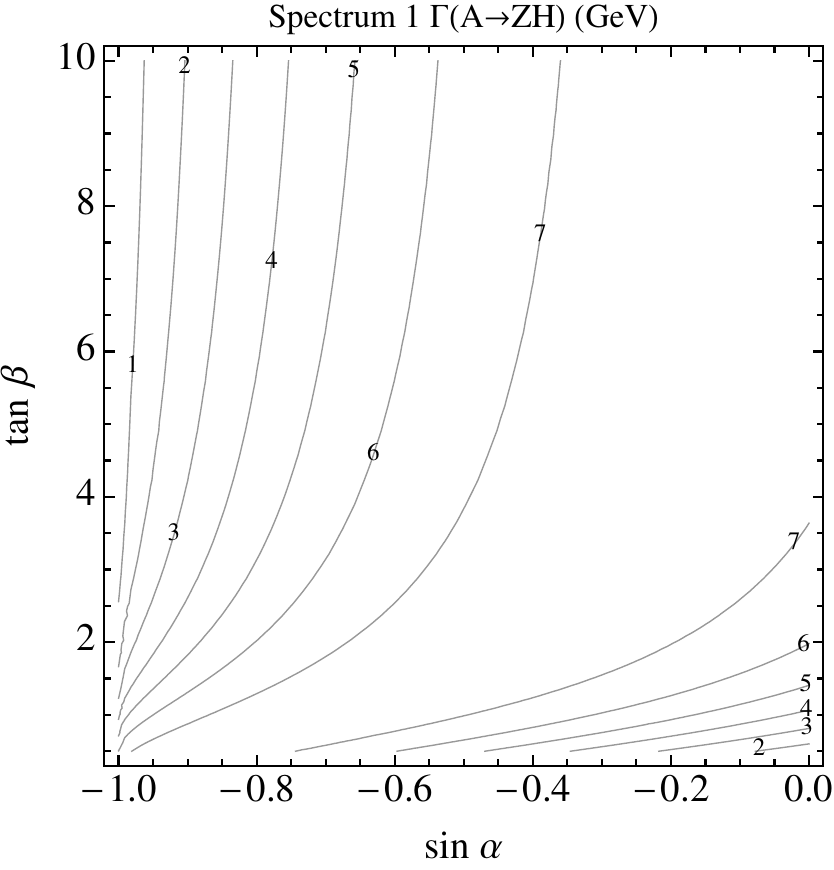}
\includegraphics[width=2in]{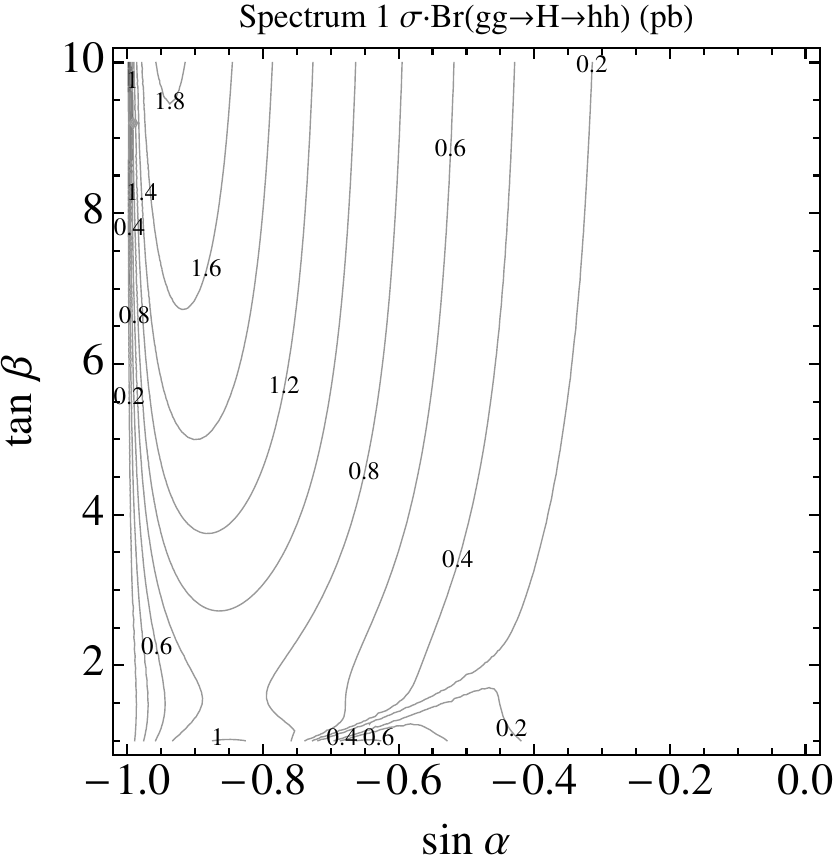}
\includegraphics[width=2in]{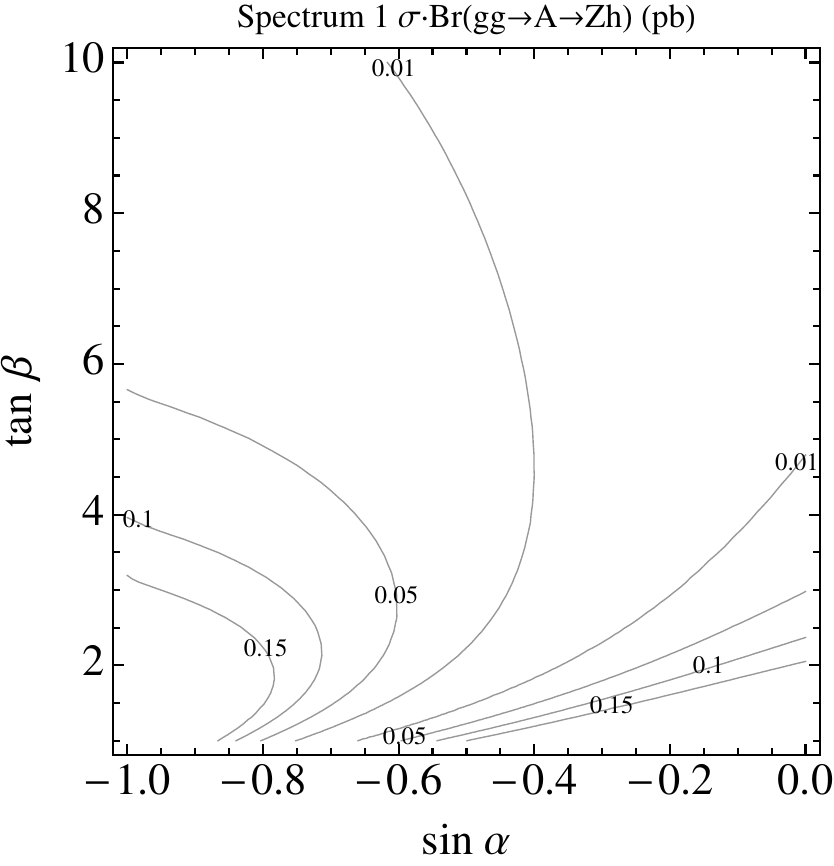}
\includegraphics[width=2in]{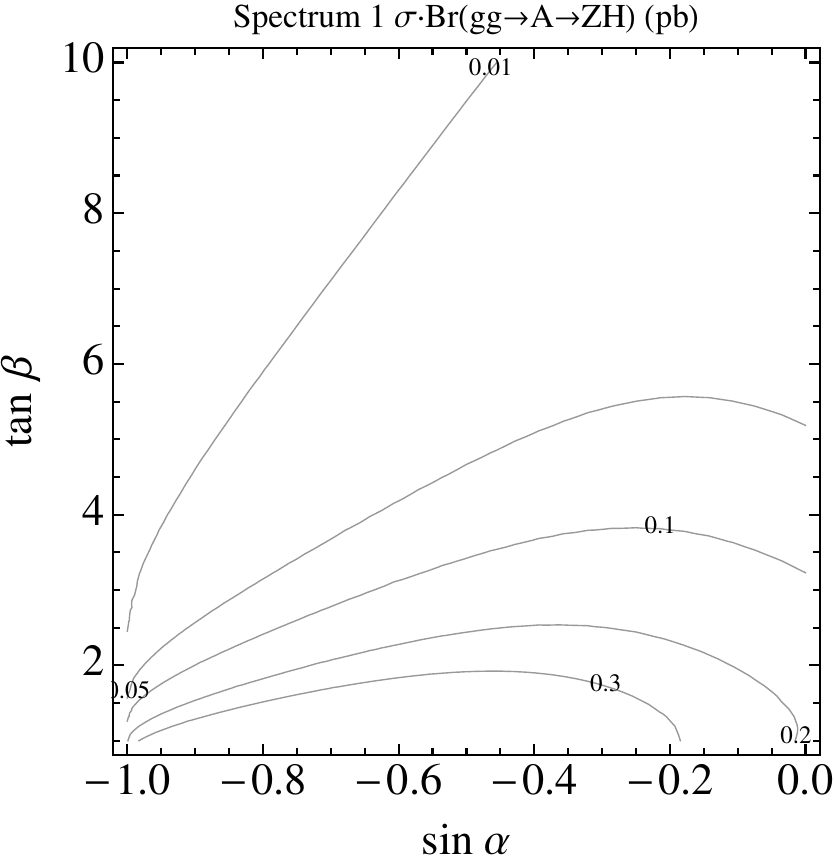}
\caption{2HDM Benchmark Spectrum 1 partial widths $\Gamma(H \to hh)$, $\Gamma(A \to Zh)$, 
and $\Gamma(A \to ZH)$
in units of GeV, 
and cross section times branching ratios 
$\sigma \cdot {\rm Br}(gg \to H \to hh)$, $\sigma \cdot {\rm Br}(gg \to A \to Zh)$, 
and $\sigma \cdot {\rm Br}(gg \to A \to ZH)$ in units of pb, all for Type I couplings. 
These partial widths and $\sigma \cdot {\rm Br}$s 
are qualitatively similar for the other types of 2HDM couplings; 
the production cross sections $\sigma(gg \to H, A)$ 
are moderately enhanced at large $\tan \beta$ for Type II and Type IV 2HDM due to the contribution from bottom loops.}
\label{fig:s1widths}
\end{figure}

In addition to the Standard Model-like production and decays of scalars to SM final states, the first benchmark spectrum also features the inter-scalar decays $H \to hh$, $A\to Zh$, and $A \to ZH$. The partial widths for these three inter-scalar decays (which are independent of the 2HDM type) and the $\sigma \cdot {\rm Br}$ for the dominant processes $gg \to H \to hh,$ $gg \to A \to Zh$, and $gg \to A \to ZH$ (which depend weakly on the 2HDM type; here, we display those of a Type I 2HDM) are shown in Figure \ref{fig:s1widths}; their parametric behavior as a function of $\sin \alpha$ and $\tan \beta$ helps to explain many of the detailed features of the exclusion limits in Figure \ref{fig:s1ex}.

\begin{figure}[h]
   \centering
   \includegraphics[width=2.8in]{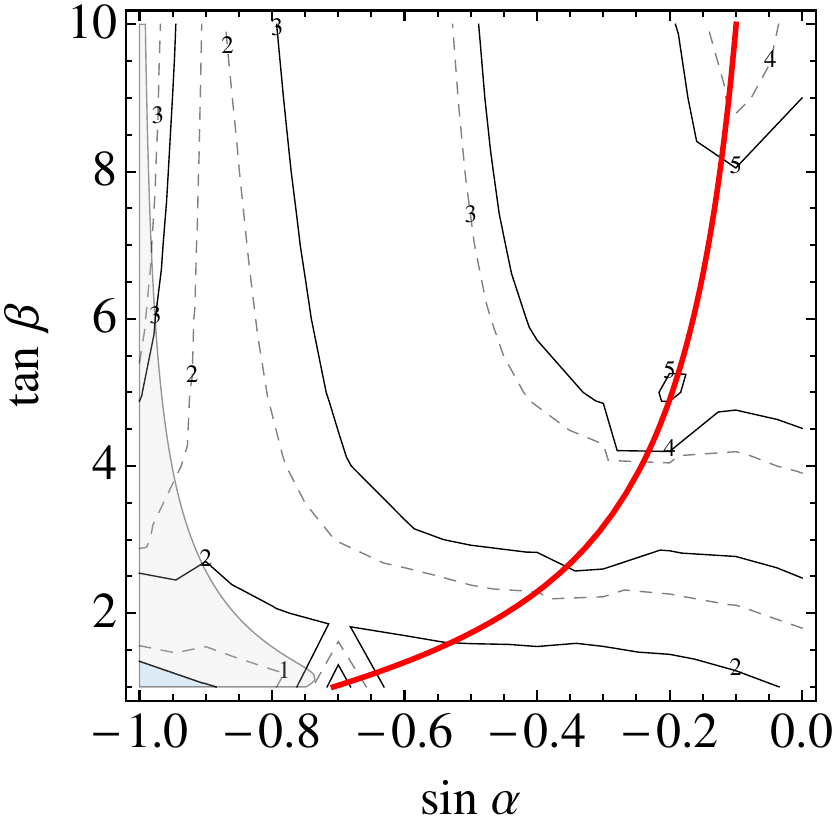}
   \includegraphics[width=2.8in]{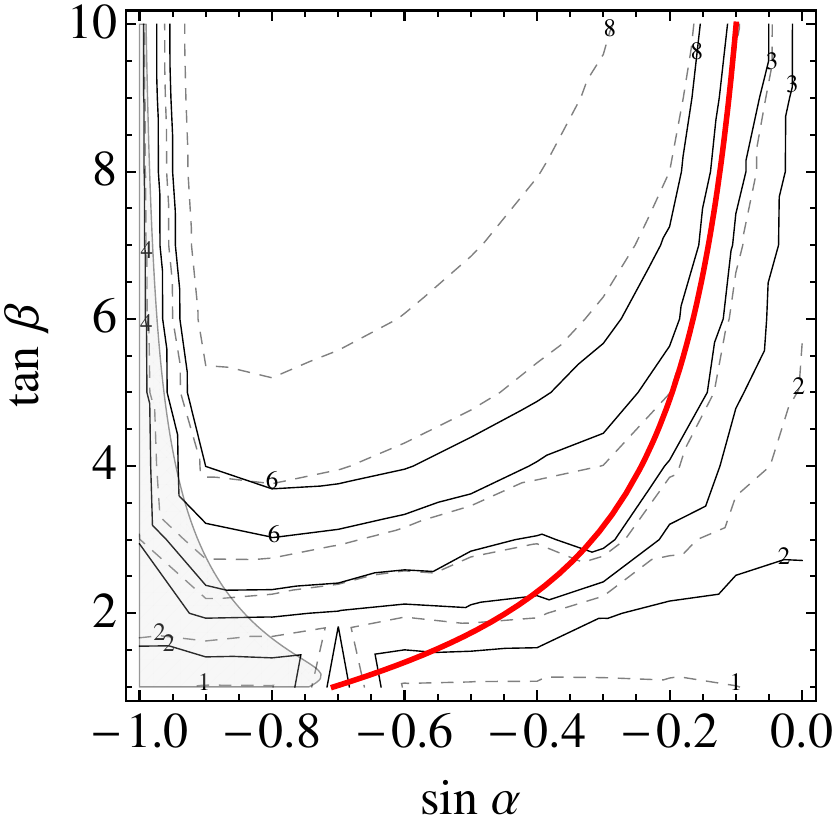}
   \includegraphics[width=2.8in]{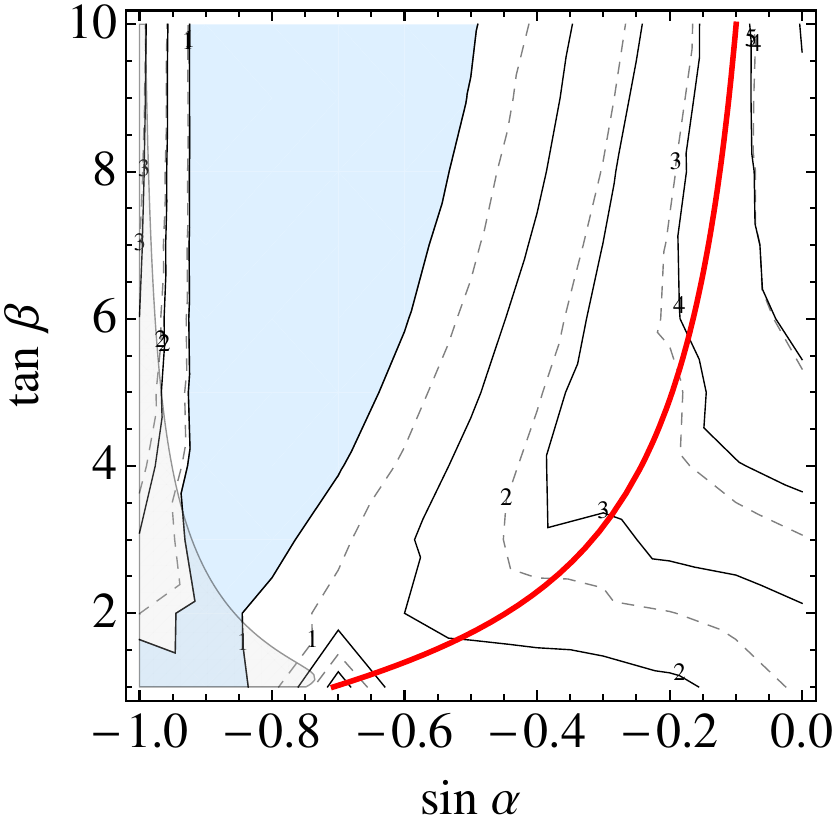}
   \includegraphics[width=2.8in]{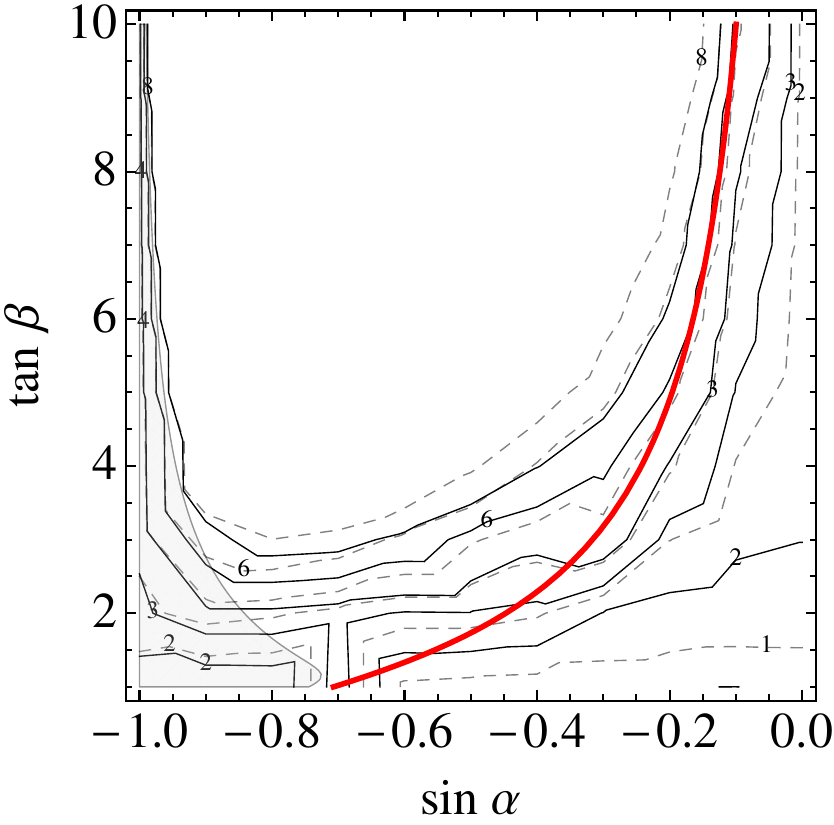}
   \caption{Multi-lepton limits 
   from the CMS multi-lepton search with 5 fb$^{-1}$ of 7 TeV proton-proton collisions \cite{CMSMulti5}   
   for the production and decay topologies of 
    Benchmark Spectrum 1 given in Table \ref{spec1}, 
    for Type I (top left), Type II (top right), Type III (bottom left), and Type IV (bottom right) couplings as a function 
    of $\sin \alpha$ and $\tan \beta$.  
        Limits were 
    obtained from an exclusive combination of the observed and expected number of events in all the 
    multi-lepton channels presented in Table \ref{tab:SM}.
        The solid and dashed lines correspond to the observed and expected 95\% CL limits 
    on the production cross section times branching ratio in multiples of the theory cross 
    section times branching ratio for the benchmark spectrum and 2HDM type. 
   The blue shaded regions denote excluded parameter space. 
   The solid red line denotes the alignment limit $\sin(\beta - \alpha) = 1$. 
   The gray shaded region corresponds to areas of parameter space where vector 
   decays of the heavy CP-even Higgs, $H \to VV$, are excluded at 95\% CL by the 
   SM Higgs searches at 7 TeV \cite{125HiggsCMS}.  }
   \label{fig:s1ex}
\end{figure}

The partial width, $\Gamma(H \to hh)$, has a complicated dependence on $\alpha, \beta$, but is greatest when $\tan \beta$ is large and $\sin \alpha \simeq - 0.85$. This process only contributes significantly to multi-lepton limits in 2HDM types for which the multi-lepton decays of $h$ are unsuppressed in the same region where ${\rm Br}(H \to hh)$ is large. The partial width, $\Gamma(A \to Zh) \propto \cos^2(\beta - \alpha)$, is largest away from the alignment limit, while the partial width, $\Gamma(A \to ZH) \propto \sin^2(\beta -\alpha)$, is largest in the alignment limit. 
In both cases, the multi-lepton limits are strongest for 2HDM types 
where the multi-lepton decays of $h$ and $H$
 are significant when ${\rm Br}(A \to Zh)$ and ${\rm Br}(A \to ZH)$ are respectively large.

On the production side, the dominant production cross section for $H$, $\sigma(gg \to H)$, is largest at small $\tan \beta$ and $\sin \alpha\to -1$, while the dominant cross section for $A$, $\sigma(gg \to A)$, is independent of $\sin \alpha$ (since the pseudoscalar couplings to fermions, and hence gluons, depend only on $\tan \beta$) and increases as $\tan \beta \to 0$. These production cross sections and scalar partial widths are largely independent of the 2HDM type; the gluon fusion rates for Type II and Type IV 2HDM increase slightly at large $\tan \beta$ due to the sizable bottom quark coupling.

The threefold combination of production rates, inter-scalar decay widths, 
and multi-lepton widths of scalars determines the shape of limits in the 
plane of $\sin \alpha$ and $\tan \beta$. These vary among different 2HDM types, 
though similarities between Type I \& III and between Type II \& IV make 
it worthwhile to discuss these two sets together.

\subsubsection*{Types I \& III }

In the Type I 2HDM, the multi-lepton signals of the SM-like Higgs, $h$, generally decrease as we move away from the alignment limit (in large part because the coupling to vectors is suppressed, reducing both the $Vh$ associated production rate and the branching ratios, ${\rm Br}(h \to VV^*)$; for an extended discussion, see \cite{Craig:2012vn}), but are not a strong function of $\sin \alpha$ and $\tan \beta$; only near $\sin \alpha \to -1$ are the $\sigma \cdot {\rm Br}$ for the conventional multi-lepton channels of $h$ significantly diminished. However, the SM-like multi-lepton signals of $h$ are typically never enhanced as we move away from the alignment limit (the exception being a mild enhancement of VBF and $Vh$ associated production with $h \to VV^*$ at small $\tan \beta$ and $\sin \alpha \to -1$; see \cite{Craig:2012vn} for more detail).   In the region where the multi-lepton signals of $h$ are diminished, the conventional multi-lepton signals of $H$ are correspondingly enhanced since the $HVV$ coupling is complementary to the $hVV$ coupling. While for $m_H = 300$ GeV, the production cross section for $H$ is somewhat smaller than that of $h$, it nonetheless contributes significantly to multi-lepton limits near $\sin \alpha \to -1$ through primarily SM-like production and decay modes. Note that the direct decays of the pseudoscalar $A$ never result in more than two leptons, so the pseudoscalar contributes to the multi-lepton signal only through scalar cascades and $t \bar{t} A$ associated production.

In addition to the conventional SM-like production and decay modes of $h$ and $H$, we must also consider the various production channels involving inter-scalar decays. The $\sigma \cdot {\rm Br}(gg \to H \to hh)$ is largest at large $\tan \beta$ and $\sin \alpha \sim - 0.8$ where $g_{Hhh}$ is largest.  The parametric behavior of this $\sigma \cdot {\rm Br}$, along with the fact that the multi-lepton final states of $h$ in a Type I 2HDM are only mildly suppressed when $\sigma \cdot {\rm Br}(gg \to H \to hh)$ is significant, largely explains the strengthening of the multi-lepton limit around $\sin \alpha \sim -0.85$.

For the pseudoscalar, $\sigma \cdot {\rm Br}(gg \to A \to Zh)$ is large away from the  alignment limit, but decreases at large $\tan \beta$ due to the falling gluon fusion rate for $A$.  Similarly, $\sigma \cdot {\rm Br}(gg \to A \to ZH)$ is large only at low $\tan \beta$, since the branching ratio for $A \to ZH$ is large along the alignment line but the gluon fusion rate for $A$ again decreases at large $\tan \beta$. Thus, both $\sigma \cdot {\rm Br}(gg \to A \to Zh)$ and $\sigma \cdot {\rm Br}(gg \to A \to ZH)$ contribute to limit-setting at small $\tan \beta$, essentially independent of $\sin \alpha$, while $\sigma \cdot {\rm Br}(gg \to A \to Zh)$ also contributes at larger $\tan \beta$ for $\sin \alpha \lesssim - 0.5$.

\begin{figure}[h]
   \centering
   \includegraphics[width=3in]{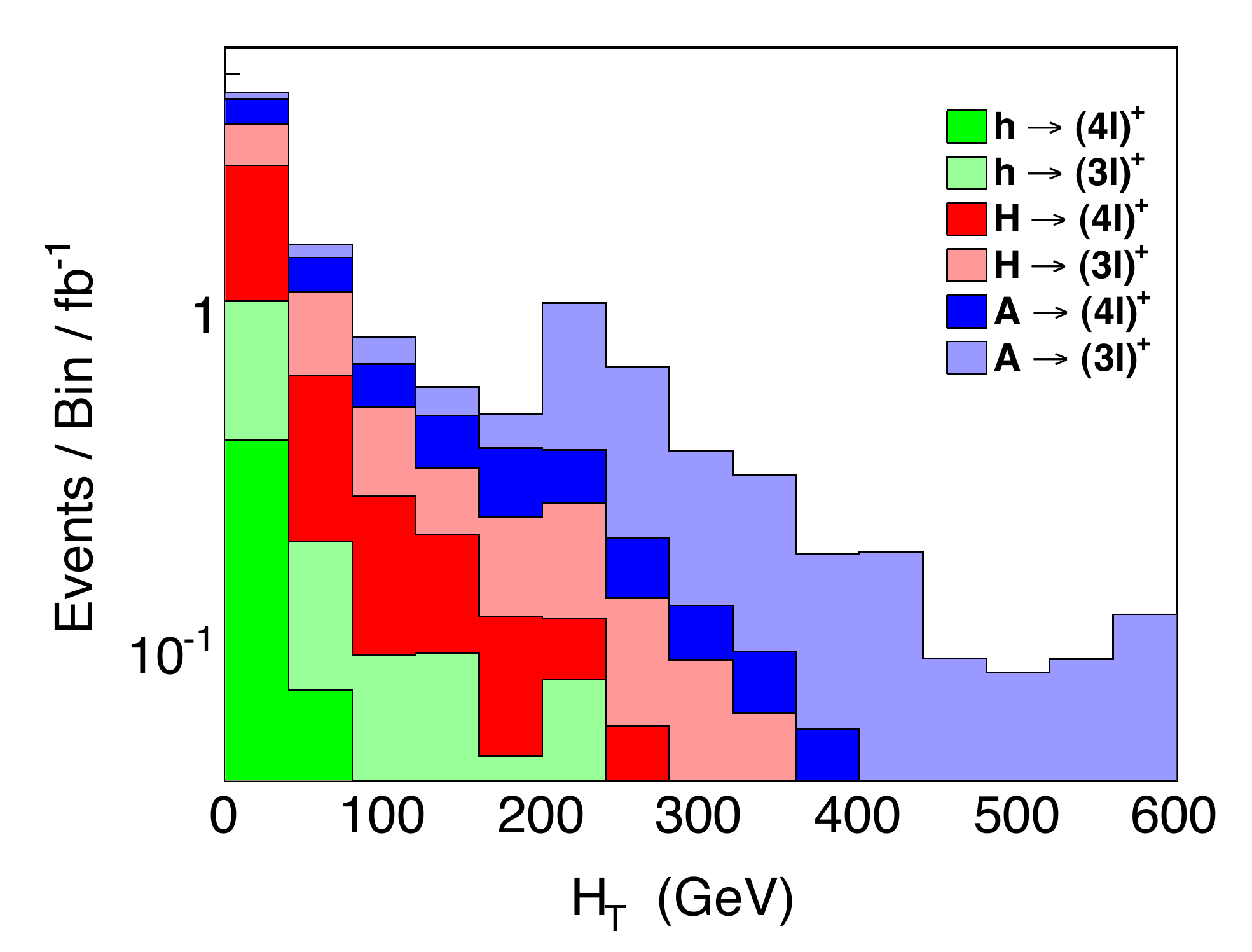}
   \includegraphics[width=3in]{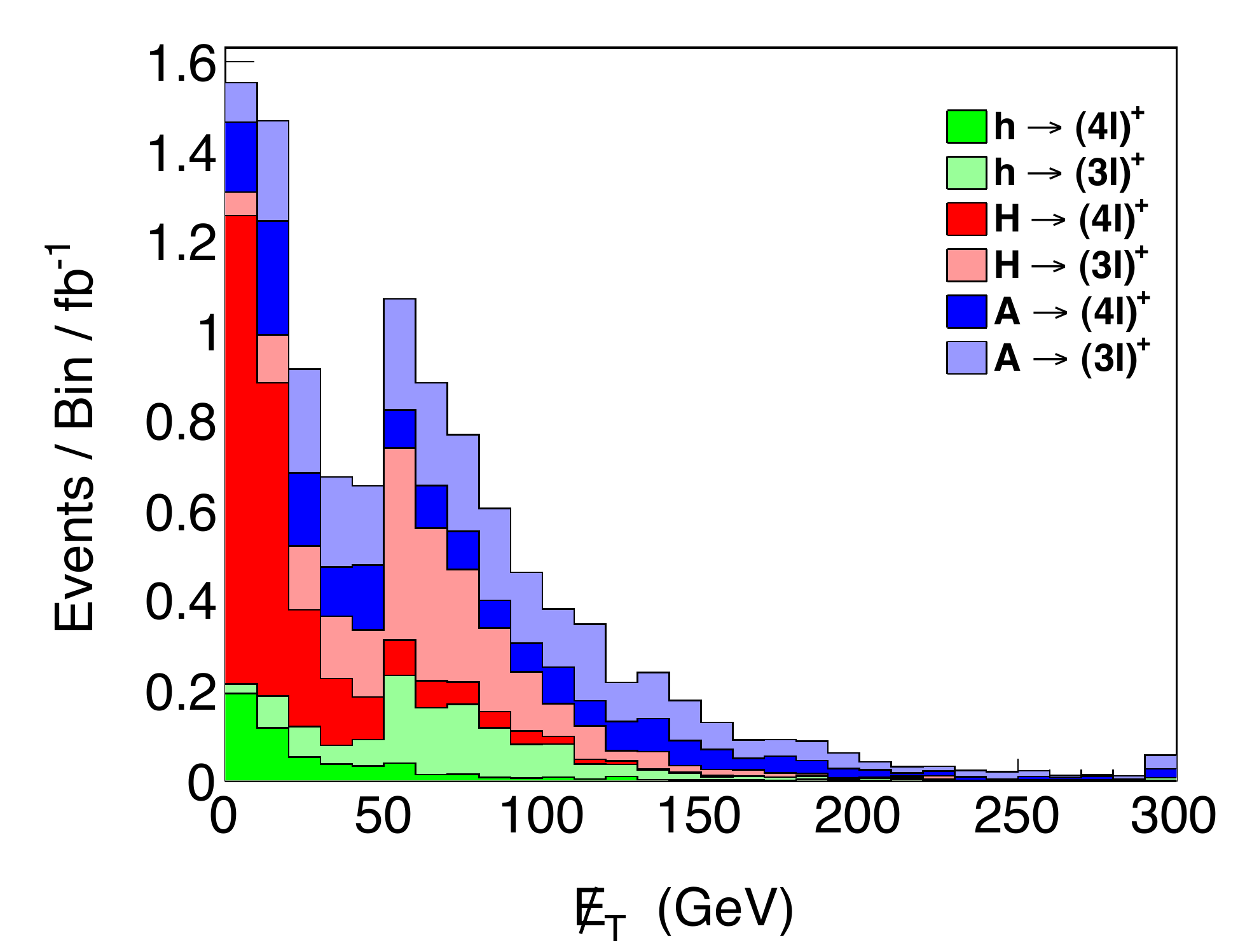}
   \caption{The 2HDM signal transverse
   hadronic energy distribution (left) and missing transverse energy distribution (right) 
   after acceptance and efficiency for 7 TeV proton-proton collisions arising from the production 
   and decay topologies of Benchmark Spectrum 1
   given in Table \ref{spec1} with $m_h = 125$ GeV, $m_H = 300$ GeV, 
   $m_{H^\pm} =  m_A = 500$ GeV,
   for Type I 2HDM couplings with  $\sin \alpha = -0.9$ and 
   $ \tan \beta = 1.0$. 
   Signal events correspond 
   to those falling in the exclusive three- or four-lepton channels
   labelled with 
   a dagger in Table \ref{tab:SM} that have moderate to good sensitivity.
   The colors indicate the initial type of Higgs boson produced.  
   For each color, the lighter shade corresponds to three-lepton channels, while the darker shade corresponds to
   four-lepton channels. 
   The bin size is 40 GeV for $H_T$ and 10 GeV for $\MET$, and in both cases the highest bin includes overflow.   }
   \label{fig:s1t1}
\end{figure}

All three scalar decays contribute to setting the strongest limits at small $\tan \beta$ (relatively insensitive to $\sin \alpha$), while $\sigma \cdot {\rm Br}(gg \to H \to hh)$ predominantly explains the limits at large $\tan \beta$ around $\sin \alpha \sim -0.85$. The additional contributions from scalar cascades are exemplified in Figure \ref{fig:s1t1}, which illustrates the $H_T$ and MET distributions for the sum of multi-lepton events at the point $(\sin \alpha = -0.9, \tan \beta = 1.0)$, distinguished by the initial scalar produced in each multi-lepton event.

The multi-lepton signals in the Type III, or ``lepton-specific,'' model are similar to those of the Type I model, since the couplings of the Higgs scalars to quarks and vectors are identical for these 2HDM types. The exception is a significant improvement in the limits around $-0.9 \lesssim \sin \alpha \lesssim -0.6$ relative to the Type I 2HDM. Here, the branching ratio, ${\rm Br}(h \to \tau \tau)$, is substantially increased over the SM rate and contributes both through SM-like associated production of $h$ and production of $H \to hh$ with one or both $h$ decaying to $\tau \tau $. 
Indeed, processes such as $Vh$ associated production with $h \to \tau \tau$ are as much as ten times larger than the SM rate, with $\sigma \cdot {\rm Br} (Wh \to W \tau \tau)$ as large as several hundred fb. Scalar cascades involving $\tau$s are even more important, with $\sigma \cdot {\rm Br}(gg \to H \to hh \to 4 \tau)$ as large as several pb. The enhancement of $\Gamma(h \to \tau \tau)$ renders this the 2HDM type most amenable to detection by the multi-lepton search, and, in fact, a large region of parameter space is already excluded by the CMS multi-lepton search with 5 fb$^{-1}$. While some of this region is already excluded by conventional searches for $h \to \tau \tau$, there exist regions not constrained by current searches where the dominant multi-lepton limit comes from scalar cascades.

\subsubsection*{Types II \& IV}

A very important difference in the phenomenology of the Type II \& IV 2HDM compared to the preceding description of the Type I \& III phenomenology is that the down-type quarks now couple to $H_d$ rather than $H_u$, thus the partial width of $h\to b\bar{b}$ has an entirely different parametric dependence. 
Since this decay mode dominates in the SM-like alignment limit, its variation sharply affects the Br's of 
all other decay modes as well. For instance, the multi-lepton 
signals of the SM-like Higgs $h$ change rapidly as we move 
away from the alignment limit, decreasing sharply with increasing $\tan\beta$ above the $\sin(\beta - \alpha) = 1$ line due to the rapidly increasing partial width, $\Gamma(h \to b \bar b)$, 
and rising rapidly below $\sin(\beta - \alpha) = 1$ as $\Gamma(h \to b \bar b)$ drops. 
Thus at large $\tan \beta$ above the alignment line, the multi-lepton signals of $h$ diminish rapidly, weakening the limit both from SM-like production of $h$ and from new associated production, such as $H \to hh$. 
The only exception are multi-lepton signals involving $h \to \tau \tau$, 
since $\Gamma(h \to \tau \tau)/ \Gamma(h \to b \bar b)$ is fixed in a Type II 2HDM.  
On the other hand, below the alignment line there is an overall enhancement of multi-lepton decays involving $h\to VV^*$ since the partial width $\Gamma(h \to b \bar b)$ drops, leading to an increase in the purely SM-like multi-lepton production and decay modes of $h$. As $\sin \alpha \to -1$, the direct multi-lepton decays of $H$ somewhat compensate for the loss of $h$ signals, but there is a wide region of large $\tan \beta$ and moderate $\sin \alpha$ where neither $h$ nor $H$ decays significantly to multi-lepton final states; this is clearly displayed by the weak limits in the range $-0.9 \lesssim \sin \alpha \lesssim -0.2$.

Scalar cascade decays do not significantly help to constrain a Type II 2HDM. While the $\sigma \cdot {\rm Br}(gg \to H \to hh)$ is parametrically similar to the Type I 2HDM, in a Type II 2HDM the SM-like Higgs $h$ decays predominantly to $b \bar b$ in this region, so this channel does not contribute substantially to multi-lepton limits (except for the rare $hh \to 4\tau$). Likewise, the contributions from $\sigma \cdot {\rm Br}(gg \to A \to Zh)$ at large $\tan \beta$ lead to multi-lepton signals only through $h \to \tau \tau$.

\begin{figure}[h]
   \centering
   \includegraphics[width=3in]{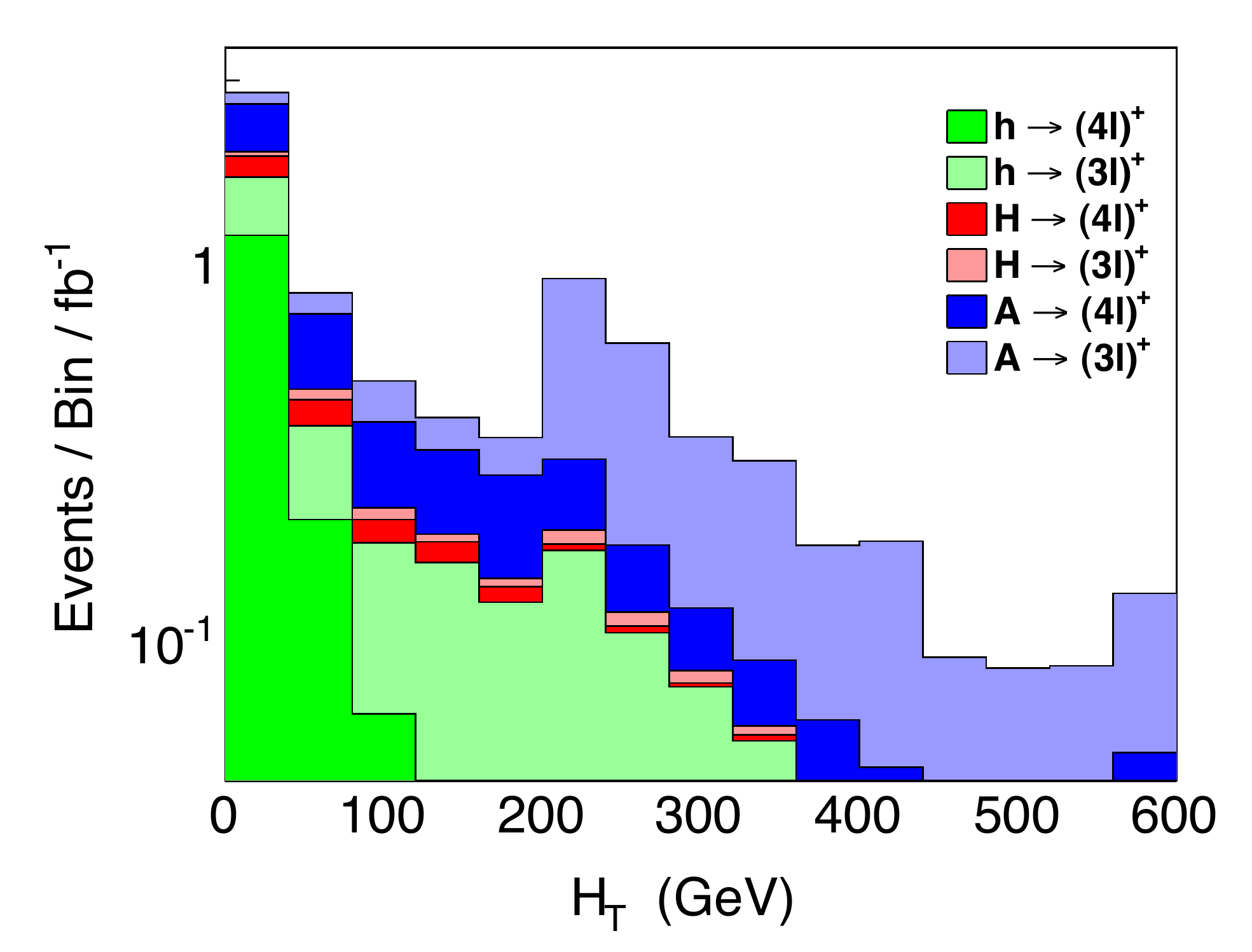}
   \includegraphics[width=3in]{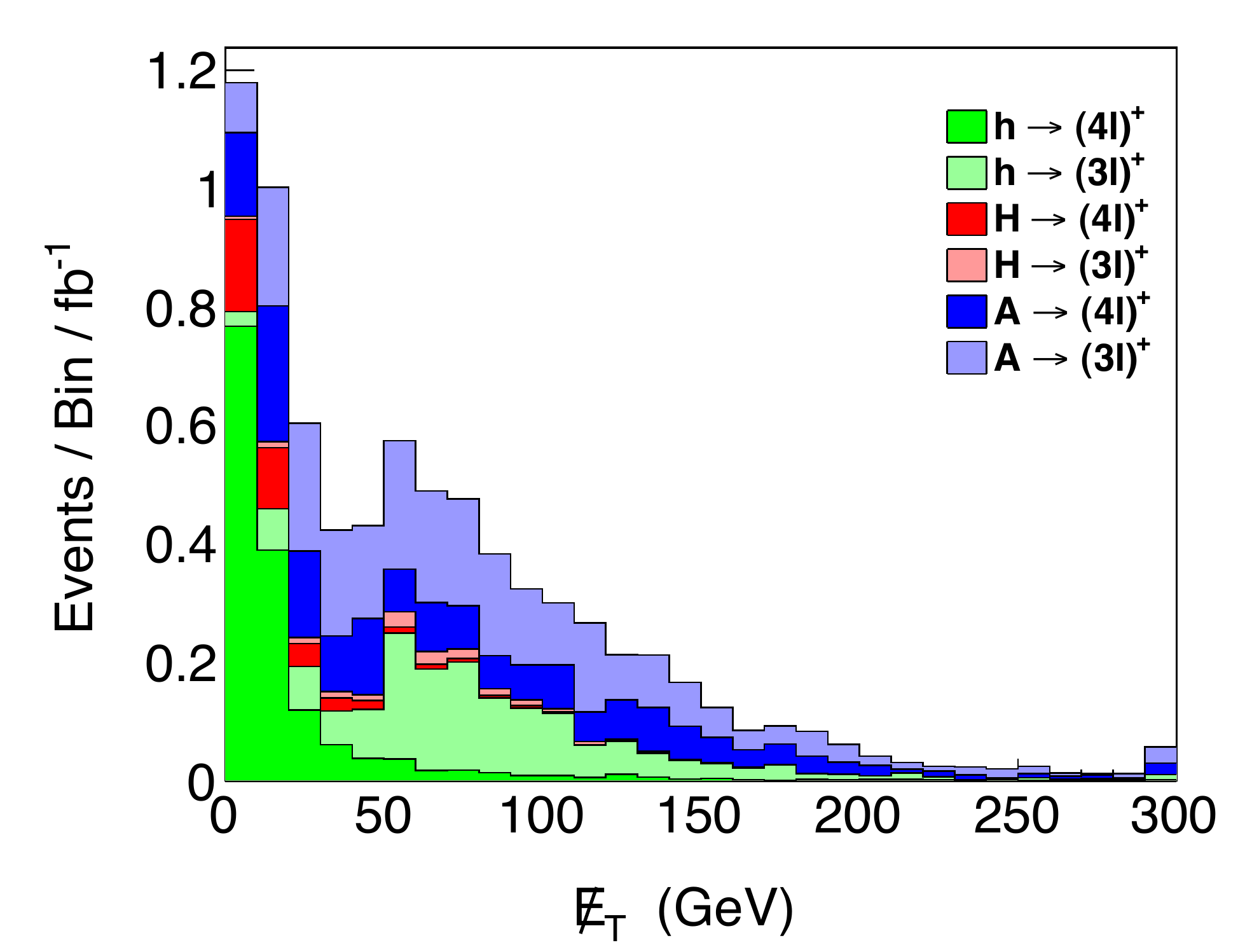}
   \caption{
    The 2HDM signal transverse
   hadronic energy distribution (left) and missing transverse energy distribution (right) 
   after acceptance and efficiency for 7 TeV proton-proton collisions arising from the production 
   and decay topologies of Benchmark Spectrum 1
   given in Table \ref{spec1} with $m_h = 125$ GeV, $m_H = 300$ GeV, 
   $m_{H^\pm} =  m_A = 500$ GeV,
   for Type II 2HDM couplings with  $\sin \alpha = -0.3$ and 
   $ \tan \beta = 1.0$. 
   Signal events correspond 
   to those falling in the exclusive three- or four-lepton channels
   labelled with 
   a dagger in Table \ref{tab:SM} that have moderate to good sensitivity.
   The colors indicate the initial type of Higgs boson produced.  
   For each color, the lighter shade corresponds to three-lepton channels, while the darker shade corresponds to
   four-lepton channels. 
   The bin size is 40 GeV for $H_T$ and 10 GeV for $\MET$, and in both cases the highest bin includes overflow.      
    }
   \label{fig:s1t2}
\end{figure}

At low $\tan \beta$, the direct multi-lepton decays of $h$ are still significant, as are the added contributions from $H \to hh, A \to Zh,$ and $A \to ZH$. The multi-lepton limits on the first benchmark spectrum for a Type II 2HDM  are strongest at low $\tan \beta$, where $h$ decays and inter-scalar decays to multi-lepton final states are enhanced; limits at $\sin \alpha \to -1$ come predominantly from direct decays of $H$, while those at $\sin \alpha \to 0$ come from direct decays of $h$. The contributions of the pseudoscalar in this limit are exemplified by Figure \ref{fig:s1t2}, which illustrates the $H_T$ and MET distributions for the sum of multi-lepton events at the point $(\sin \alpha = -0.3, \tan \beta = 1.0)$, for which there is a large contribution from $A \to Zh, ZH$.

The multi-lepton signals in the Type IV, or ``flipped,'' model are similar to that of the Type II model, since the couplings of the Higgs scalars to quarks and vectors are identical for these 2HDM types.  The notable exception are the reduced limits in the region of moderate $\sin \alpha$ and large $\tan \beta$. This reduction in sensitivity is due to the fact that in a Type IV 2HDM the partial width, $\Gamma(h \to \tau \tau)$, no longer scales with $\Gamma(h \to b \bar b)$, and so in the region where $\Gamma(h \to b \bar b)$ is particularly large there are no longer meaningful contributions to multi-lepton limits from $h \to \tau \tau$ with leptonically decaying $\tau$s. In particular, this removes possible multi-lepton signals from associated production of $h$ in this region, both through SM associated production and scalar cascades.

\subsection{Spectrum 2}

\begin{figure}[h]
   \centering
   \includegraphics[width=2.8in]{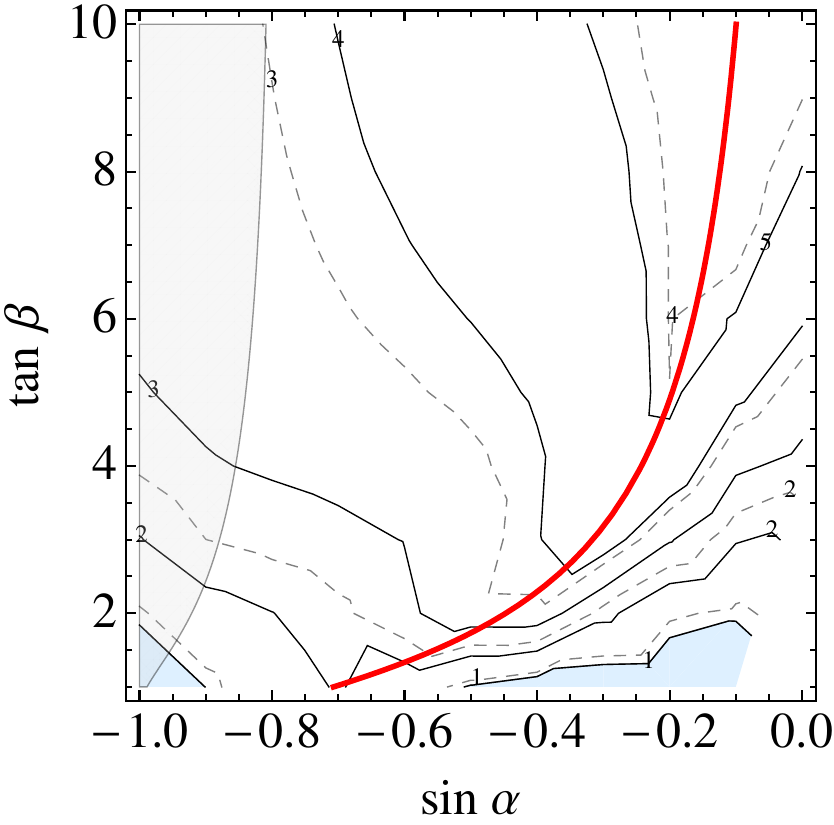}
   \includegraphics[width=2.8in]{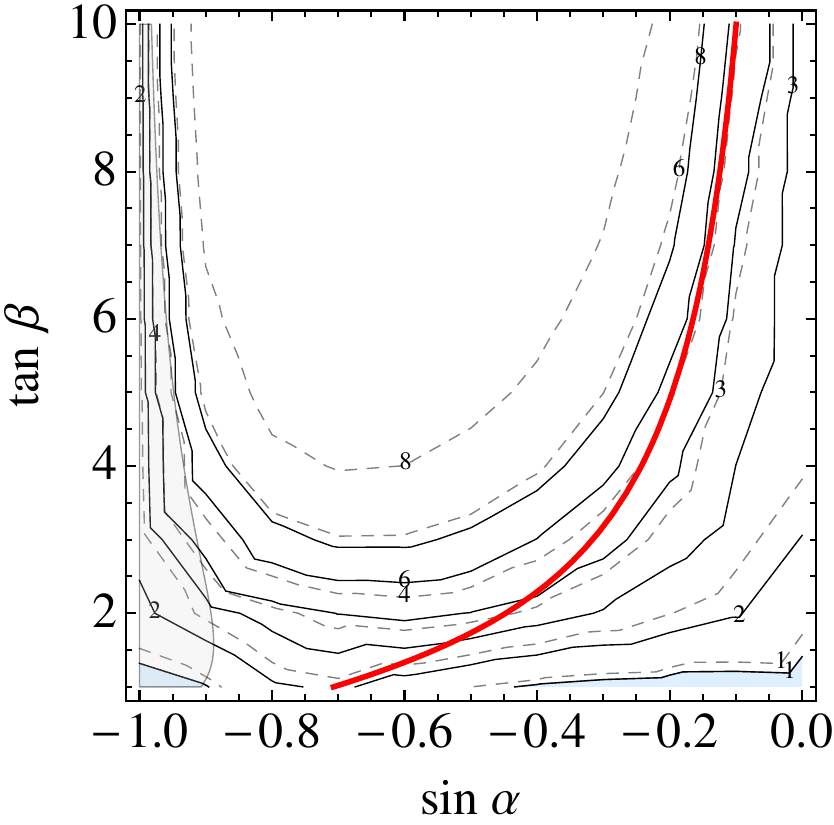}
   \includegraphics[width=2.8in]{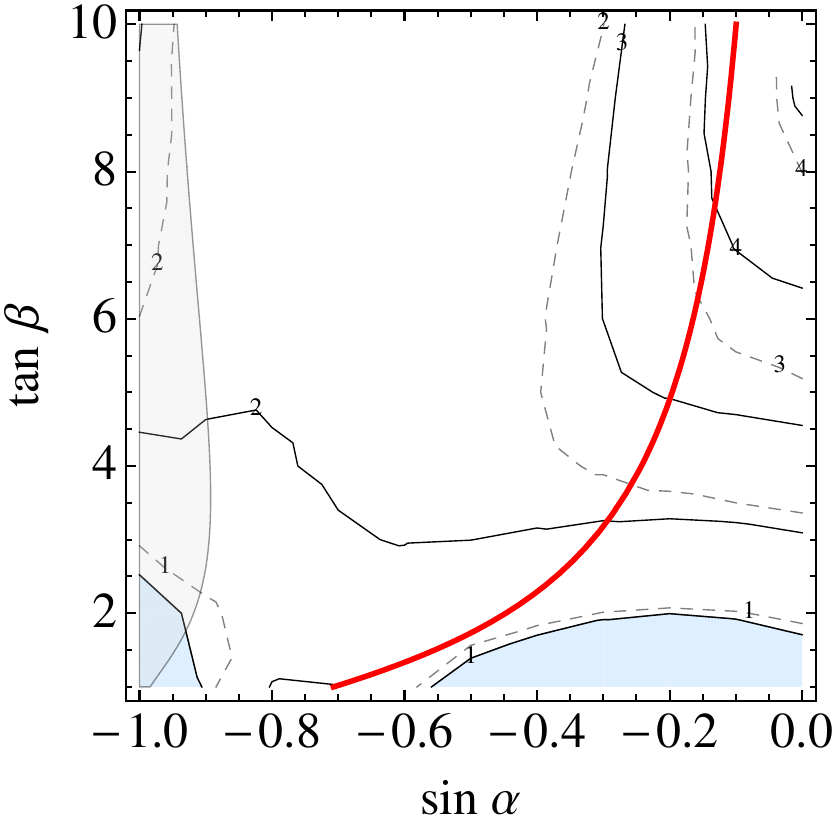}
   \includegraphics[width=2.8in]{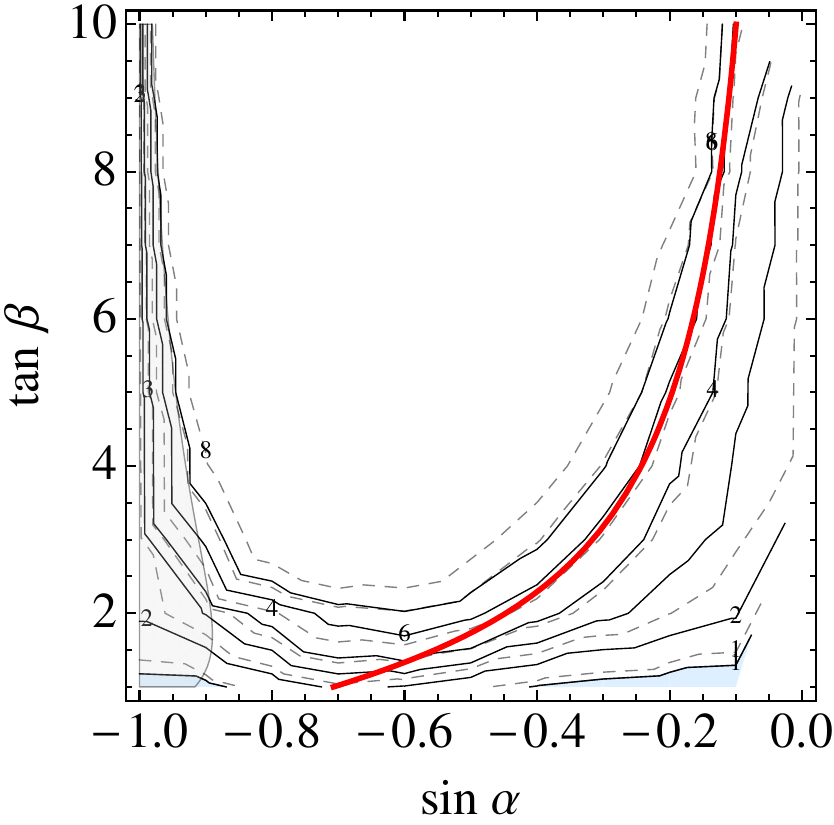}
   \caption{
%
%
   Multi-lepton limits 
   from the CMS multi-lepton search with 5 fb$^{-1}$ of 7 TeV proton-proton collisions \cite{CMSMulti5}   
   for the production and decay topologies of 
    Benchmark Spectrum 2 given in Table \ref{spec1}, 
    for Type I (top left), Type II (top right), Type III (bottom left), and Type IV (bottom right) couplings as a function 
    of $\sin \alpha$ and $\tan \beta$.  
        Limits were 
    obtained from an exclusive combination of the observed and expected number of events in all the 
    multi-lepton channels presented in Table \ref{tab:SM}.
        The solid and dashed lines correspond to the observed and expected 95\% CL limits 
    on the production cross section times branching ratio in multiples of the theory cross 
    section times branching ratio for the benchmark spectrum and 2HDM type. 
   The blue shaded regions denote excluded parameter space. 
   The solid red line denotes the alignment limit $\sin(\beta - \alpha) = 1$. 
   The gray shaded region corresponds to areas of parameter space where vector 
   decays of the heavy CP-even Higgs, $H \to WW^*$, are excluded at 95\% CL by the 
   SM Higgs searches at 7 TeV \cite{125HiggsCMS}.       
   }
   \label{fig:s2ex}
\end{figure}

The multi-lepton limits on the second benchmark spectrum are shown in Figure \ref{fig:s2ex}.
Much like the first benchmark spectrum, this spectrum includes the scalar decays $A\to Zh$ and $A \to ZH$, albeit with greater cross sections since $m_A = 250$ GeV in this spectrum. However, the decay $H \to hh$  is now kinematically forbidden. Since the parametric behavior of the relevant partial widths and $\sigma \cdot {\rm Br}$'s is the same as in the first benchmark up to overall rescalings, we do not show them explicitly, but emphasize that the cross sections for production of $A$ and $H$ are substantially larger compared to the first benchmark since both $A$ and $H$ are lighter in this case. 

\subsubsection*{Types I \& III}

The multi-lepton limits for Type I 2HDM are similar to those of the Type I model for Spectrum 1, albeit without the contributions from $H \to hh$. Particularly, the stronger limits around $\sin \alpha \sim -0.85$ in Spectrum 1 are absent here, but otherwise the parametric contributions are similar. The limits for this spectrum are stronger at small $\tan \beta$ because the now lighter $A$ has a larger production cross section, $\sigma(gg \to A)$, than in Spectrum 1. Similarly, the limits are stronger as $\sin \alpha \to -1$ since here the direct production and multi-lepton decays of $H$ dominate the limit, and the production cross section for $H$ is effectively SM-like in this region since $m_H = 140$ GeV.

Likewise, the multi-lepton limits for Type III 2HDM are similar to those of the Type III model for Spectrum 1, although they again lack the contributions from $H \to hh$, meaning that there is no significant $4\tau$ contribution with this spectrum.

\subsubsection*{Types II \& IV}

Unsurprisingly, the limits for Type II \& Type IV 2HDM are similar to the analogous 
limits in Spectrum 1, although 
 somewhat stronger due to the enhanced production cross sections for $A$ and $H$. Note that there is no significant weakening of the limit at large $\tan \beta$ and moderate $\sin \alpha$ compared to Spectrum 1, despite the disappearance of the decay $H \to hh$. This exemplifies the fact that in Type II and Type IV 2HDM, the multi-lepton decays of $h$ are suppressed in this range, so the presence or absence of $H \to hh$ does not significantly alter the limit.

\subsection{Spectrum 3}

\begin{figure}[h]
   \centering
   \includegraphics[width=2.8in]{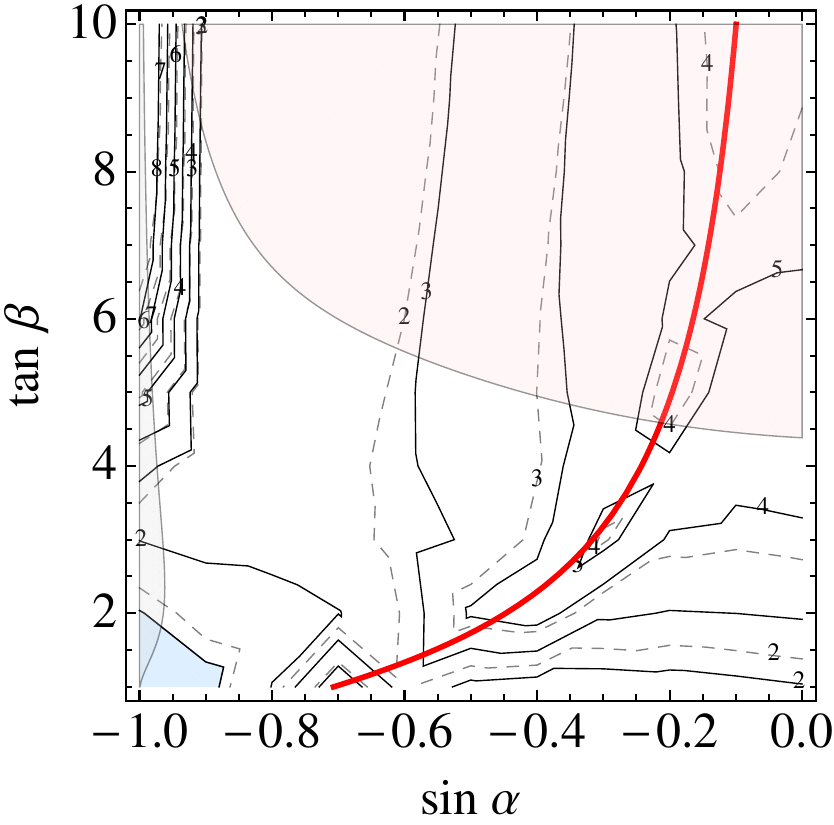}
   \includegraphics[width=2.8in]{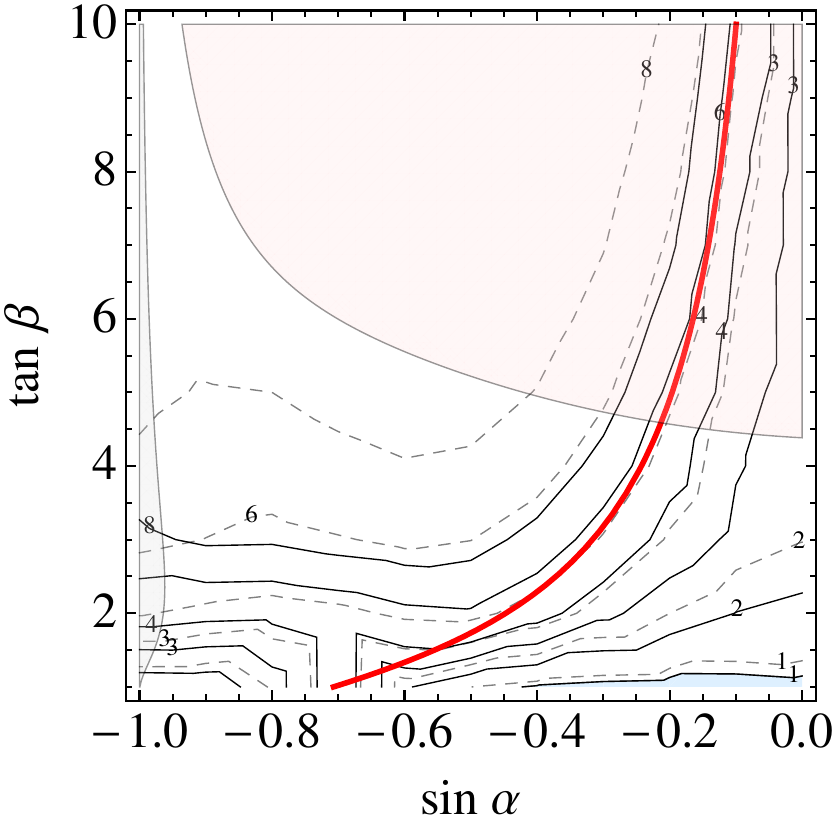}
   \includegraphics[width=2.8in]{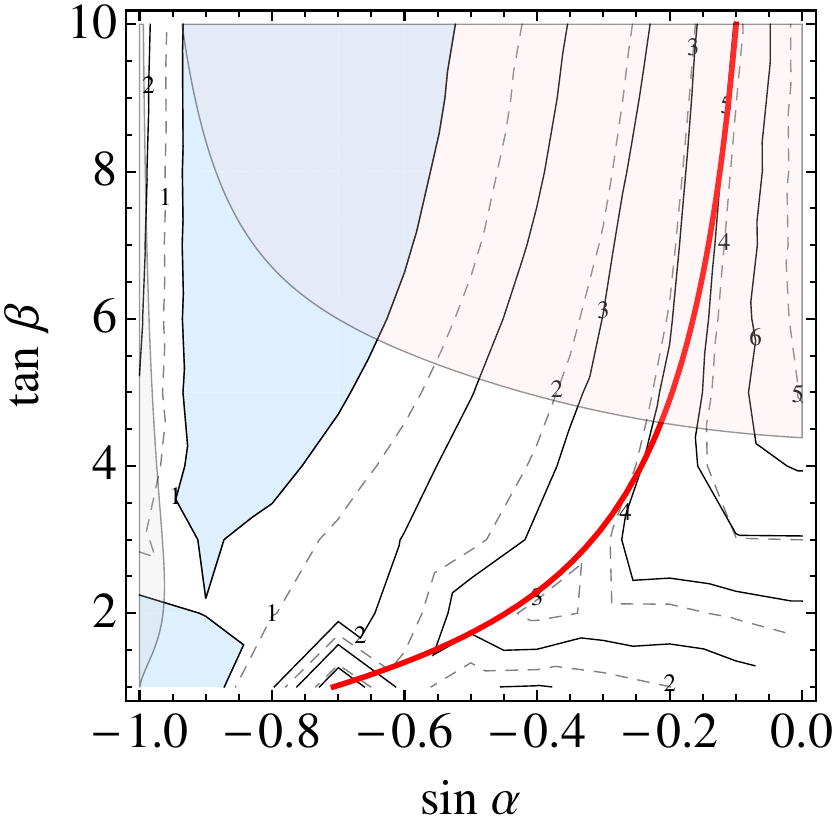}
   \includegraphics[width=2.8in]{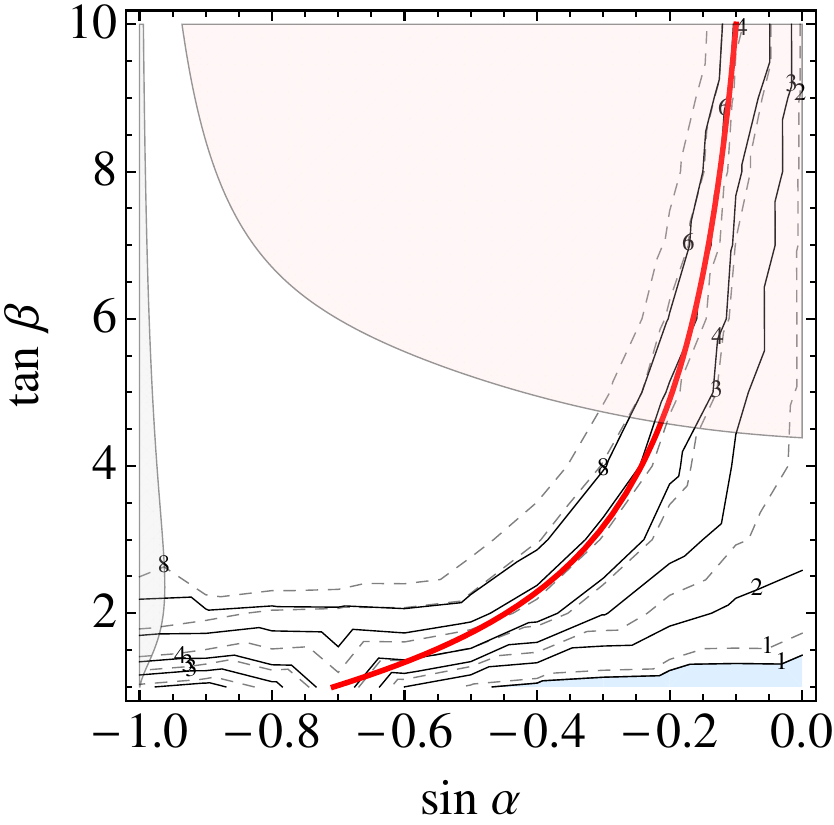}
   \caption{
%
%
   Multi-lepton limits 
   from the CMS multi-lepton search with 5 fb$^{-1}$ of 7 TeV proton-proton collisions \cite{CMSMulti5}   
   for the production and decay topologies of 
    Benchmark Spectrum 3 given in Table \ref{spec1}, 
    for Type I (top left), Type II (top right), Type III (bottom left), and Type IV (bottom right) couplings as a function 
    of $\sin \alpha$ and $\tan \beta$.  
        Limits were 
    obtained from an exclusive combination of the observed and expected number of events in all the 
    multi-lepton channels presented in Table \ref{tab:SM}.
        The solid and dashed lines correspond to the observed and expected 95\% CL limits 
    on the production cross section times branching ratio in multiples of the theory cross 
    section times branching ratio for the benchmark spectrum and 2HDM type. 
   The blue shaded regions denote excluded parameter space. 
   The solid red line denotes the alignment limit $\sin(\beta - \alpha) = 1$. 
   The gray shaded region corresponds to areas of parameter space where vector 
   decays of the heavy CP-even Higgs, $H \to VV^*$, are excluded at 95\% CL by the 
   SM Higgs searches at 7 TeV \cite{125HiggsCMS}.    
   In all cases, for $\tan \beta \gtrsim 5$ 
   and $\sin \alpha \gtrsim -0.8$ the total width of $H$ grows comparable to its mass 
   and the precise exclusion limit in this region is subject to large theoretical uncertainties, 
   these regions are highlighted in light red. 
   }
   \label{fig:s3ex}
\end{figure}

\begin{figure}[h]
\centering
\includegraphics[width=2in]{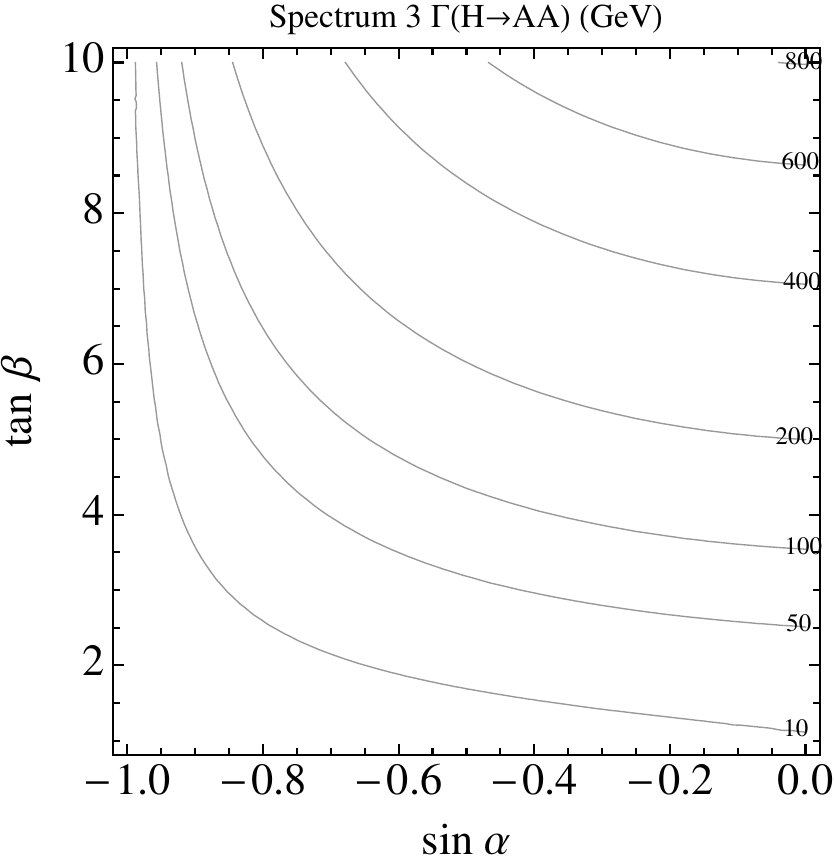}
\includegraphics[width=2in]{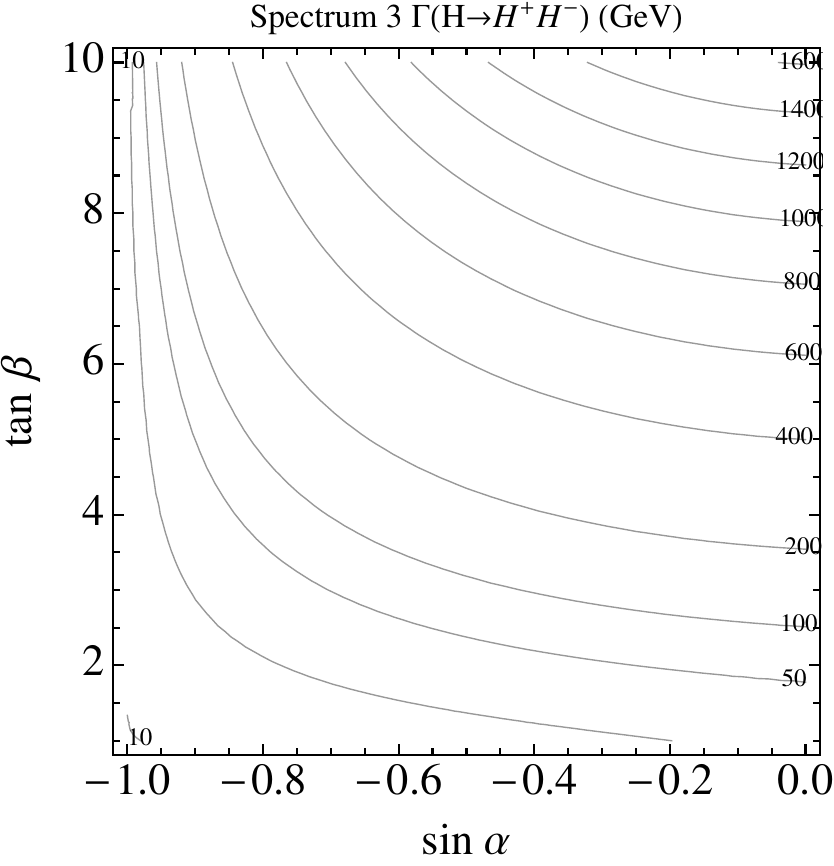}
\includegraphics[width=2in]{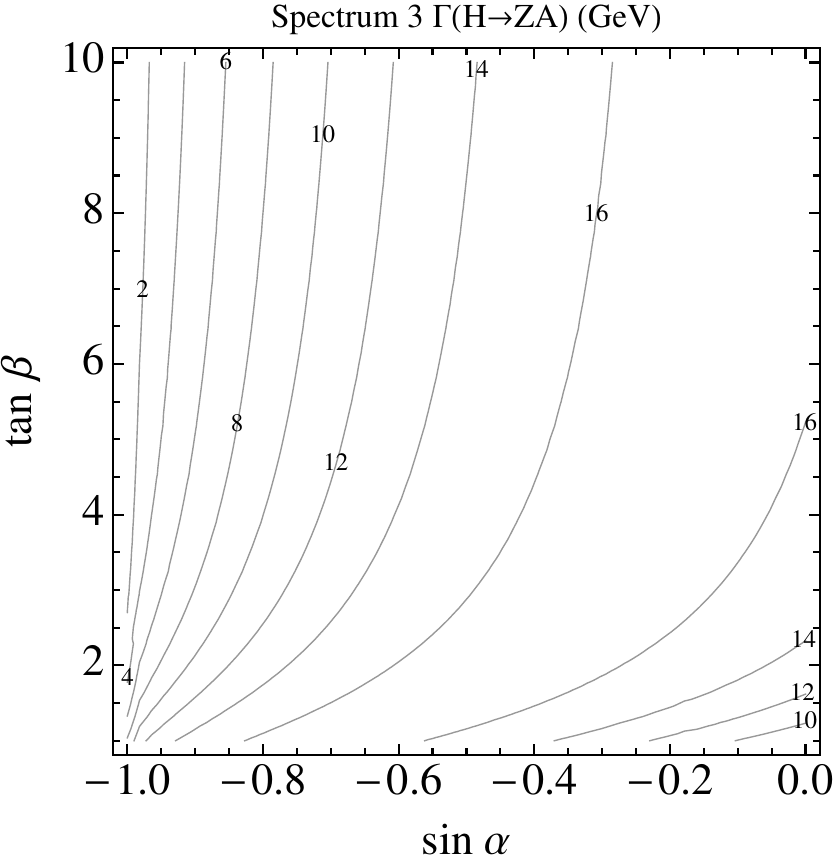}
\includegraphics[width=2in]{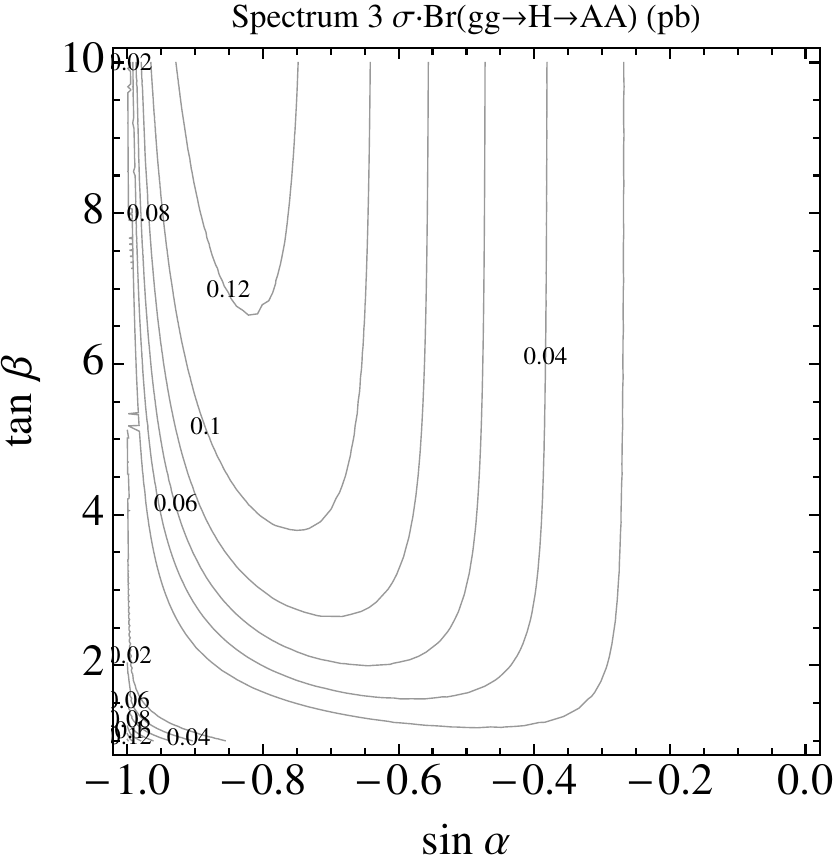}
\includegraphics[width=2in]{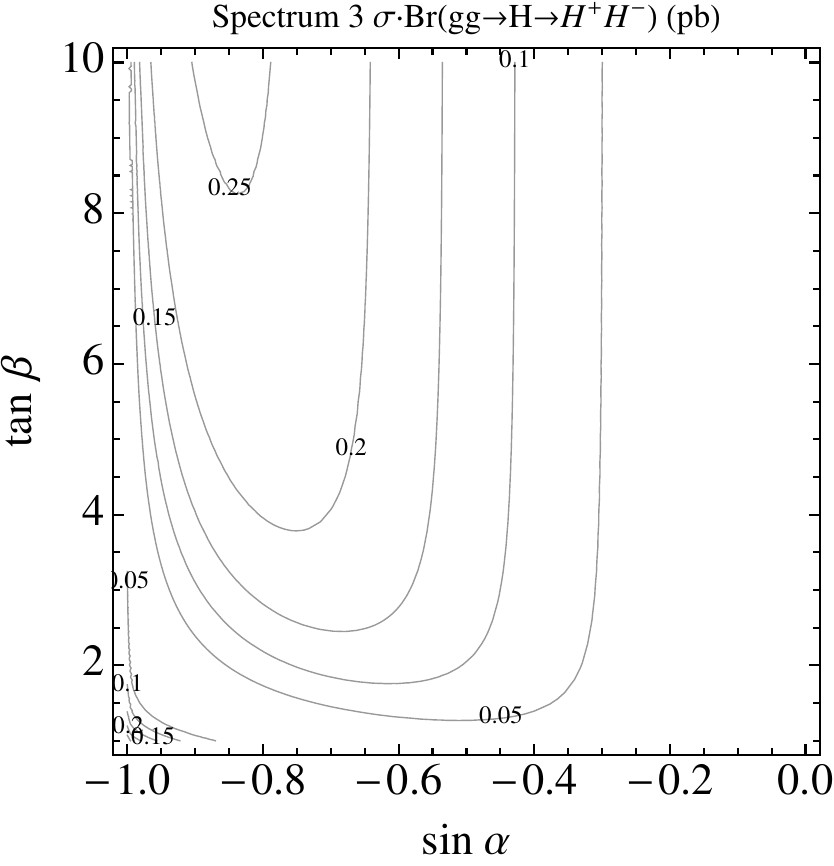}
\includegraphics[width=2in]{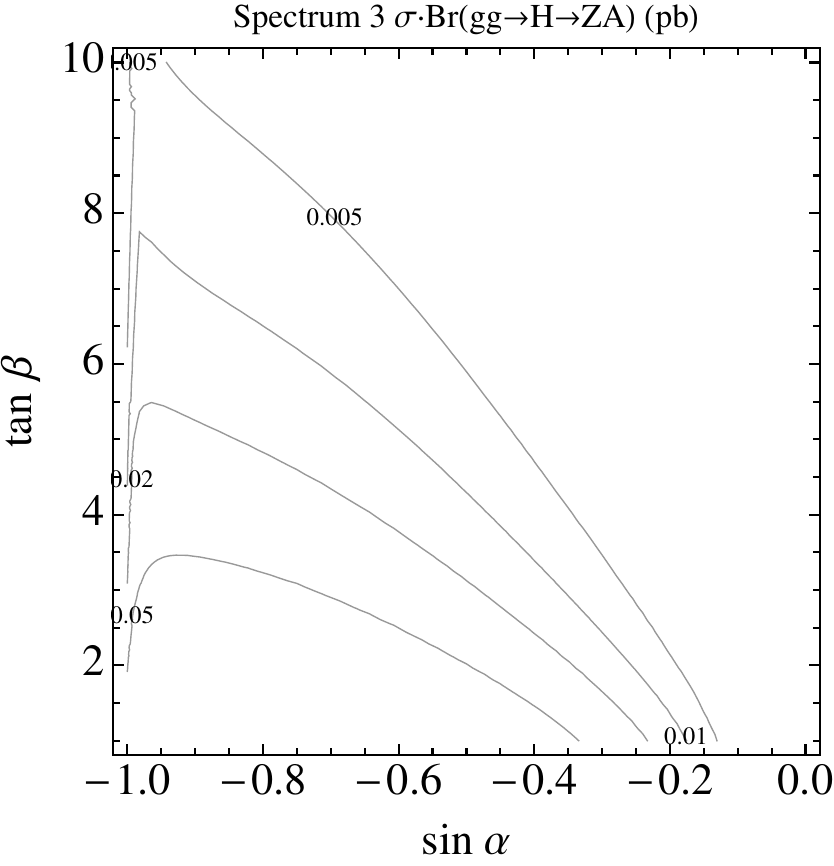}
\caption{2HDM Benchmark
Spectrum 3 partial widths $\Gamma(H \to AA)$, $\Gamma(H \to H^+ H^-)$, and $\Gamma(H \to ZA)$
in units of GeV,
and cross section times branching ratios 
$\sigma \cdot {\rm Br}(gg \to H \to AA)$, $\sigma \cdot {\rm Br}(gg \to H \to H^+ H^-)$, and 
$\sigma \cdot {\rm Br}(gg \to H \to ZA)$ in units of pb, all for Type I couplings. 
These partial widths and $\sigma \cdot {\rm Br}$s are qualitatively similar for the other types of 2HDM; 
the production cross section $\sigma(gg \to H)$ is moderately enhanced at large 
$\tan \beta$ for Type II and Type IV 2HDM due to the contribution from bottom loops.}
\label{fig:s3widths}
\end{figure}

The multi-lepton limits on the third benchmark spectrum for all four types of 2HDM are shown in Figure \ref{fig:s3ex}. 
The third benchmark spectrum enjoys a plethora of inter-scalar cascade decays. In particular, the important inter-scalar decays include $H \to hh$, $H \to AA$, $H \to H^+ H^-$, $H \to ZA$, $H^\pm \to W^\pm h$, and $A \to Zh$. The fact that $H\to H^+ H^-, AA, ZA$ and both $H^\pm \to W^\pm h$ and $A \to Zh$ are open allows for the possibility of multi-step cascades involving three Higgs scalars. Also note that the range of possible decays of $H$ means that the overlap of large $\Gamma(H \to hh)$ with multi-lepton decays of $h$ is not as important to limit-setting as it was in Spectrum 1, since, e.g., $H \to AA, ZA$ with $A \to \tau \tau$ may be important even when the multi-lepton decays of $h$ are small. However, since $H$ is relatively heavy in this benchmark ($m_H = 500$ GeV), the direct multi-lepton decays of $H$ are less important to limit-setting relative to other benchmarks due to the lower production cross section. The partial widths and $\sigma \cdot {\rm Br}$ for those processes unique to Spectrum 3 are shown in Figure \ref{fig:s3widths} (the parametric dependence of $H \to hh$ and  $A \to Zh$ were already shown in Figure \ref{fig:s1widths} and the dependence of $H^\pm \to W^\pm h$ will be shown in Figure \ref{fig:s4widths} when we discuss Spectrum 4).

The partial widths $\Gamma(H \to AA)$ and $\Gamma(H \to H^+ H^-)$ are complicated functions of $\alpha$ and $\beta$, but grow as $\tan \beta$ increases and $\sin \alpha$ goes to zero. The partial widths, $\Gamma(H \to ZA)$ and $\Gamma(H \to H^\pm W^\mp)$, scale simply as $\sin^2(\beta - \alpha)$, and so is largest in the alignment limit, while the partial widths, $\Gamma(A \to h Z)$ and $\Gamma(H^\pm \to W^\pm h)$, scale as $\cos^2(\beta - \alpha)$ and is largest away from the alignment limit.

Note in Figure \ref{fig:s3widths} the partial widths, $\Gamma(H \to AA)$ and $\Gamma(H \to H^+ H^-)$, grow quite large with increasing $\tan \beta$, such that the total width of $H$ exceeds its mass for $\tan \beta \gtrsim 5$ and $\sin \alpha \gtrsim -0.8$. In this regime, both the perturbative expansion in scalar couplings and the narrow width approximation break down, and the precise exclusion limit should be treated with caution.

On the production end, as noted earlier the dominant production mode for $H$, $\sigma(gg \to H)$, is largest at small $\tan \beta$ and $\sin \alpha\to -1$. The combination of this dependence and the partial widths implies that $\sigma \cdot {\rm Br}(gg \to H \to AA)$ and $\sigma \cdot {\rm Br}(gg \to H \to H^+ H^-)$ are largest at moderate $\sin \alpha$, peaking around $\sin \alpha \sim -0.8$ and increasing mildly with $\tan \beta$; both contribute over a somewhat wider range than $gg \to H \to hh$. In contrast, $\sigma \cdot {\rm Br}(gg \to H \to ZA)$ is largest at low $\tan \beta$ and $\sin \alpha \to -1$.

\subsubsection*{Types I \& III}

The signals of the Type I 2HDM for the third benchmark spectrum are similar to those of the first benchmark spectrum, to the extent that they are largely governed by the multi-lepton final states of $h$ combined with the scalar decays of $H$ and $A$. However, in contrast to Spectrum 1, here the direct multi-lepton decays of $H$ are less significant in limit-setting since the production cross section for $m_H = 500$ GeV is considerably smaller. Thus, the limits at large $\tan \beta$ and $\sin \alpha \to -1$ coming from direct multi-lepton decays of $H$ are noticeably weaker in this case. On the other hand, scalar decays of $H$ contribute meaningfully over a wide range in $\sin \alpha$ since $\sigma \cdot {\rm Br}(gg \to H \to AA)$ and $\sigma \cdot {\rm Br}(gg \to H \to H^+ H^-)$ change slowly as a function of $\sin \alpha$ compared to $\sigma \cdot {\rm Br}(gg \to H \to hh)$.

In the case of processes involving $H \to AA$, the multi-lepton limits are dominated by the decays $A \to Zh$ rather than $A \to \tau \tau$. This is because in a Type I model the $A \tau \tau$ coupling decreases with increasing $\tan \beta$, so that the branching ratio ${\rm Br}(A \to \tau \tau)$ is not large in the same region as $\sigma \cdot {\rm Br}(gg \to H \to AA)$. In contrast, the branching ratio ${\rm Br}(A \to Zh)$ is large precisely when ${\rm Br }(H \to AA)$ is large, hence $H \to AA \to Zh Zh$ contributes substantially to the limit at large $\tan \beta$ and $-0.9 \lesssim \sin \alpha \lesssim -0.4$, with $\sigma \cdot {\rm Br}(gg \to H \to AA \to Zh Zh)$ growing as large as $\sim 120$ fb in the region of study.

For processes involving $H \to H^+ H^-$, the multi-lepton limits always require at least one charged Higgs to decay via $H^\pm \to W^\pm h$, since the other decay modes such as e.g. $H^+ \to t \bar b, \tau^+ \nu$ give at most one lepton. In a Type I model, ${\rm Br} (H^\pm \to W^\pm h)$ is sizable when ${\rm Br}( H \to H^+ H^-)$ is large, so $H \to H^+ H^- \to W^+ h W^- h$ is important at large $\tan \beta$ in the range $-0.9 \lesssim \sin \alpha \lesssim -0.5$. Processes involving $H \to H^+ H^-$ with one decay to $t\bar{b}$ and $\tau \nu$ are also important at moderate $\tan \beta$.

\begin{figure}[h]
   \centering
   \includegraphics[width=3in]{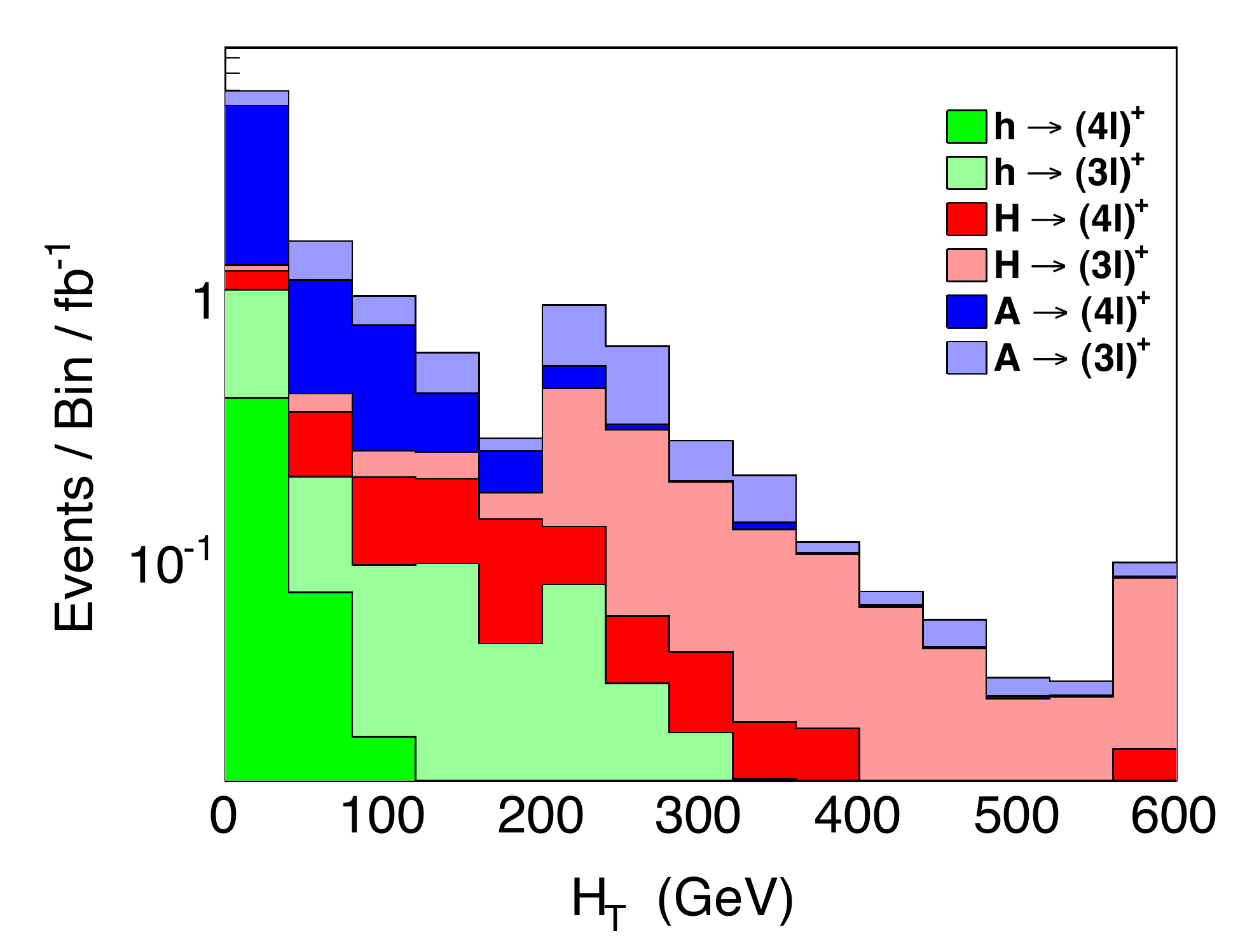}
   \includegraphics[width=3in]{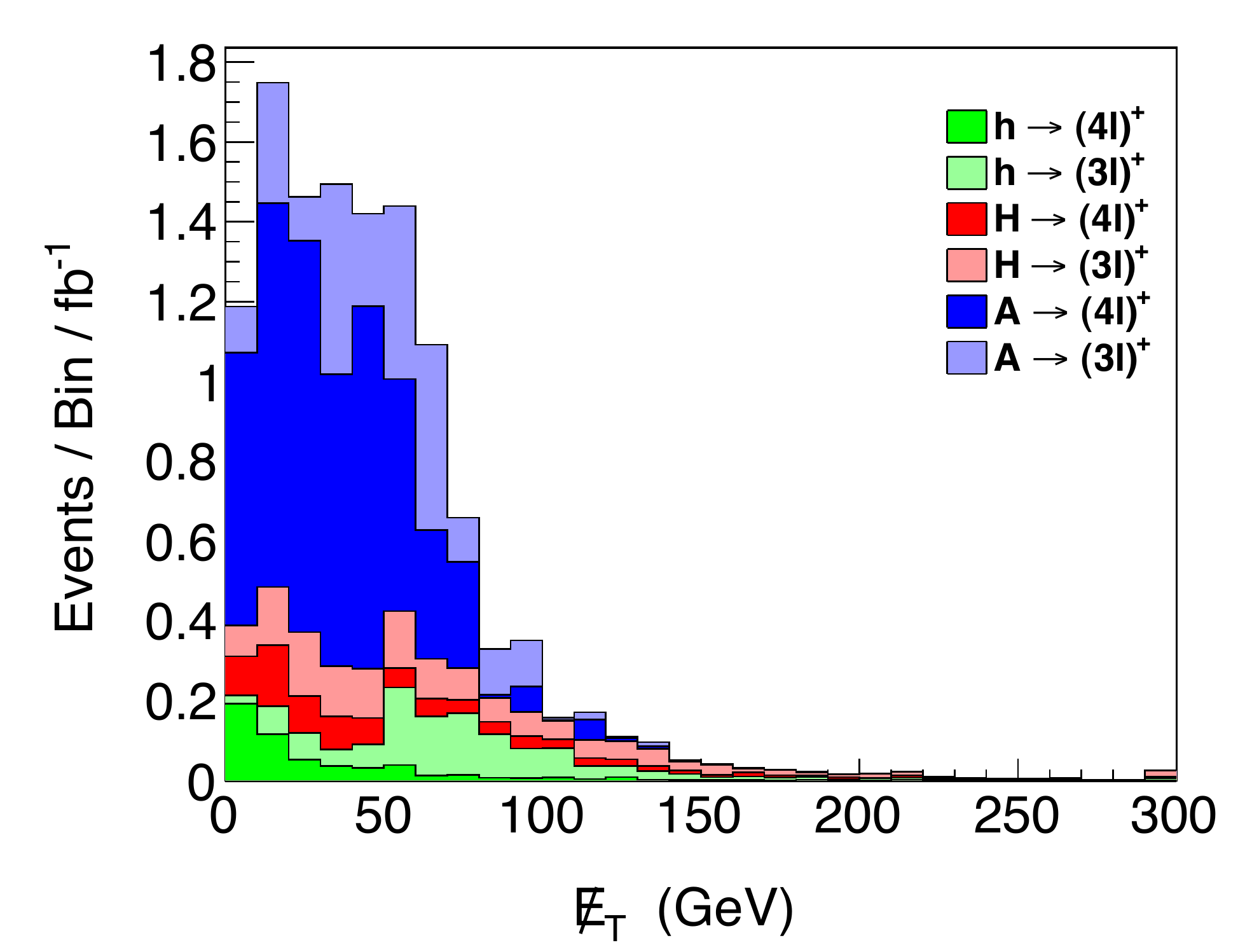}
   \caption{
      The 2HDM signal transverse
   hadronic energy distribution (left) and missing transverse energy distribution (right) 
   after acceptance and efficiency for 7 TeV proton-proton collisions arising from the production 
   and decay topologies of Benchmark Spectrum 3
   given in Table \ref{spec3} with $m_h = 125$ GeV, $m_H = 500$ GeV, 
   $m_{H^\pm} =  m_A = 230$ GeV,
   for Type I 2HDM couplings with  $\sin \alpha = -0.9$ and 
   $ \tan \beta = 1.0$. 
   Signal events correspond 
   to those falling in the exclusive three- or four-lepton channels
   labelled with 
   a dagger in Table \ref{tab:SM} that have moderate to good sensitivity.
   The colors indicate the initial type of Higgs boson produced.  
   For each color, the lighter shade corresponds to three-lepton channels, while the darker shade corresponds to
   four-lepton channels. 
   The bin size is 40 GeV for $H_T$ and 10 GeV for $\MET$, and in both cases the highest bin includes overflow.     
    }
   \label{fig:s3t1}
\end{figure}

As in previous cases, $gg \to A \to Zh$ is important at small $\tan \beta$, as is $gg \to H \to ZA$ with both $A \to \tau \tau$ and $A \to Zh$. Various exemplary features of the third benchmark spectrum with Type I 2HDM couplings are shown in Figure \ref{fig:s3t1}, which illustrates the $H_T$ and MET distributions for the sum of multi-lepton events at the point $(\sin \alpha = -0.9, \tan \beta = 1.0)$, distinguished by the initial scalar produced in each multi-lepton event.

The Type III 2HDM shares many of the qualitative features of the Type I 2HDM, albeit with additional contributions to multi-lepton signals coming from the fact that the partial widths $\Gamma(h \to \tau \tau)$ and $\Gamma(A \to \tau \tau)$ grow with $\tan \beta$.  So, in addition to the significant signals discussed earlier, both $H \to hh \to 4 \tau$ and $H \to AA \to 4 \tau$ are  important in the Type III 2HDM,  particularly at moderate $\sin \alpha$ and large $\tan \beta$ where ${\rm Br}(H \to hh, AA)$ are large and so too are ${\rm Br}(h,A \to \tau \tau)$. Taken together, these contributions are still not as great as in Spectrum 1 due to the reduced production cross section for $H$, but nonetheless lead to large regions already excluded using the 5 fb$^{-1}$ data.

\subsubsection*{Types II \& IV}

As in previous cases, the multi-lepton final states of $h$ decrease rapidly above the alignment limit, with the sole exception of $h \to \tau \tau$. Here, the reduced contribution from direct multi-lepton decays of $H$ is particularly noticeable, with a substantial weakening of the limit as $\sin \alpha \to -1$.

Much as in Spectrum 1 Type II, processes involving $H \to hh$ contribute little to the limit, since $h$ has suppressed multi-lepton final states when ${\rm Br}(H \to hh)$ is large. The decay, $H \to AA$, is somewhat more important, but, as with the Type III model, the contribution to multi-leptons comes primarily from $A \to \tau \tau$ as opposed to $A\to Zh$, especially at large $\tan\beta$. The $A \tau \tau$ coupling grows with $\tan \beta$ in a Type II 2HDM, but, as before, $A \to b \bar b$, with the same parametric scaling, still dominates the total width of $A$. Similarly, $H^\pm$ decays primarily to $tb$ and $\tau \nu$ at large $\tan \beta$, so $H^\pm \to W^\pm h$ is suppressed in this range and processes involving $H \to H^+ H^-$ do not contribute much to the multi-lepton limits. 

\begin{figure}[h]
   \centering
   \includegraphics[width=3in]{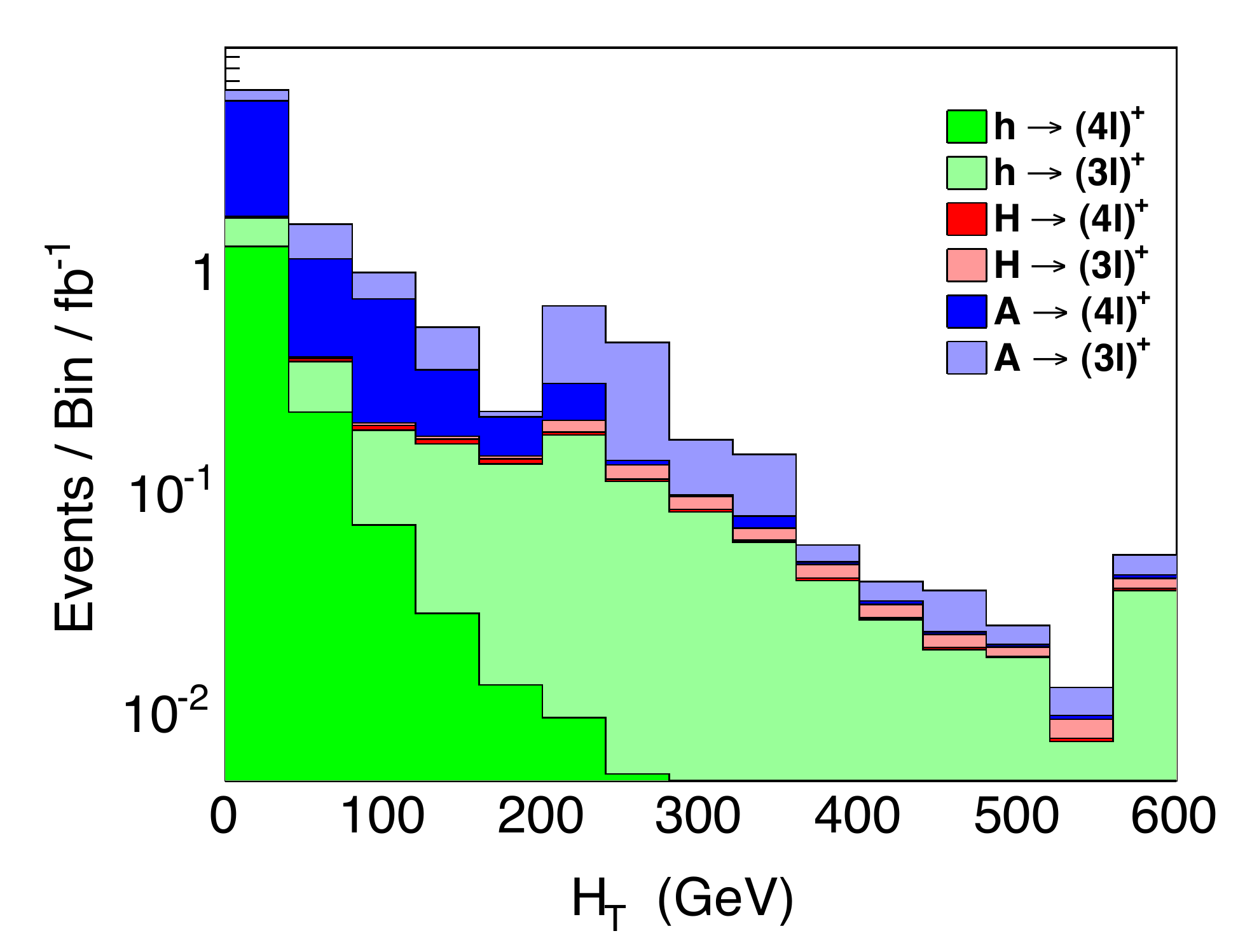}
   \includegraphics[width=3in]{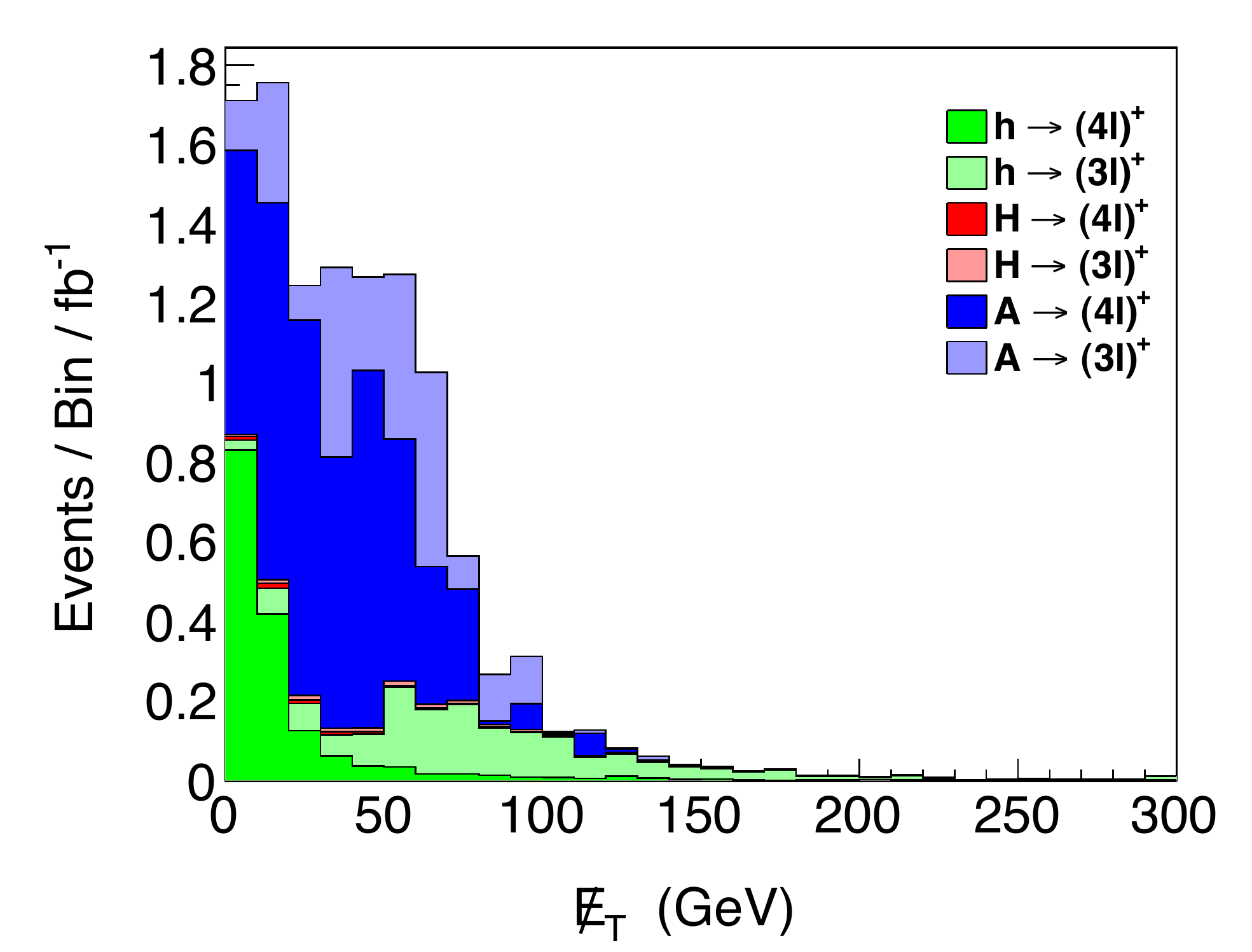}
   \caption{
   The 2HDM signal transverse
   hadronic energy distribution (left) and missing transverse energy distribution (right) 
   after acceptance and efficiency for 7 TeV proton-proton collisions arising from the production 
   and decay topologies of Benchmark Spectrum 3
   given in Table \ref{spec3} with $m_h = 125$ GeV, $m_H = 500$ GeV, 
   $m_{H^\pm} =  m_A = 230$ GeV,
   for Type II 2HDM couplings with  $\sin \alpha = -0.2$ and 
   $ \tan \beta = 1.0$. 
   Signal events correspond 
   to those falling in the exclusive three- or four-lepton channels
   labelled with 
   a dagger in Table \ref{tab:SM} that have moderate to good sensitivity.
   The colors indicate the initial type of Higgs boson produced.  
   For each color, the lighter shade corresponds to three-lepton channels, while the darker shade corresponds to
   four-lepton channels. 
   The bin size is 40 GeV for $H_T$ and 10 GeV for $\MET$, and in both cases the highest bin includes overflow.        
      }
   \label{fig:s3t2}
\end{figure}

The processes $gg \to A \to Zh$ and $gg \to H \to Z(A \to Zh)$ are important at small $\tan \beta$; here the multi-lepton decays of $h$ are enhanced below the alignment line, so that these processes contribute significantly to the limit through the direct multi-lepton decays of $h$. The contributions of the pseudoscalar  are exemplified by Figure \ref{fig:s3t2}, which illustrates the $H_T$ and MET distributions for the sum of multi-lepton events at the point $(\sin \alpha = -0.2, \tan \beta = 1.0)$, for which there is a large contribution from $A \to Zh$.

The Type IV 2HDM recapitulates many of the features of the Type II 2HDM, albeit without significant contributions from $h \to \tau \tau$ or $A \to \tau \tau$ at large $\tan \beta$. This eliminates contributions from, e.g., $H \to hh \to 4 \tau$ and $H \to AA \to 4 \tau$, so that the multi-lepton limits are particularly weak at moderate $\sin \alpha$ and large $\tan \beta$. As before, the multi-lepton decays of $h$ are important below the alignment line, and accumulate extra contributions from $gg \to A \to Zh$ and $gg \to H \to Z(A \to Zh)$ at low $\tan \beta$.

\subsection{Spectrum 4}

\begin{figure}[h]
   \centering
   \includegraphics[width=2.8in]{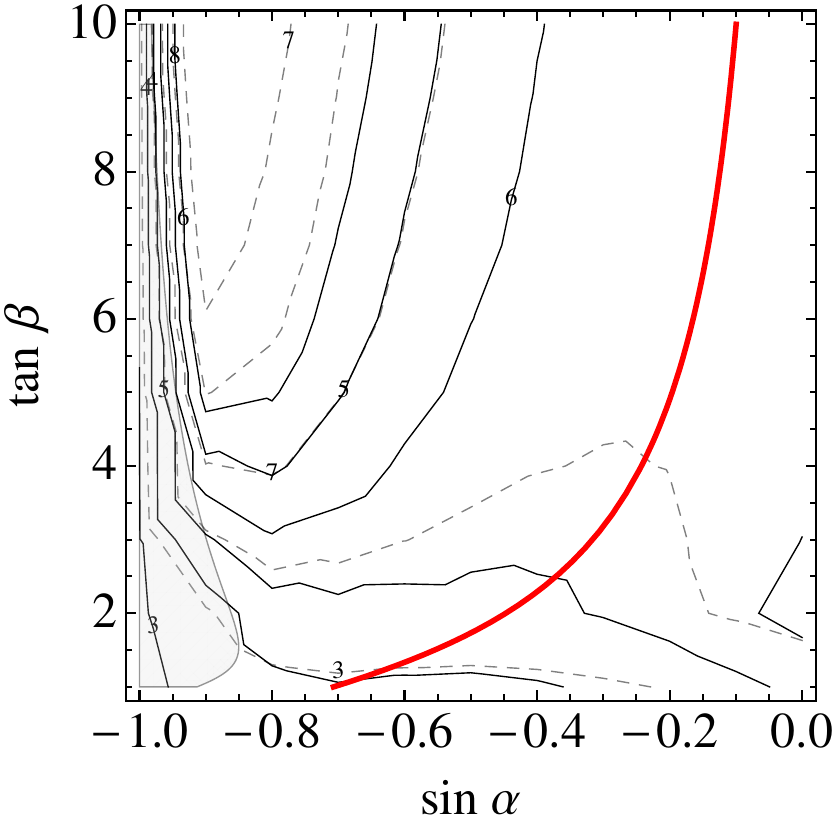}
   \includegraphics[width=2.8in]{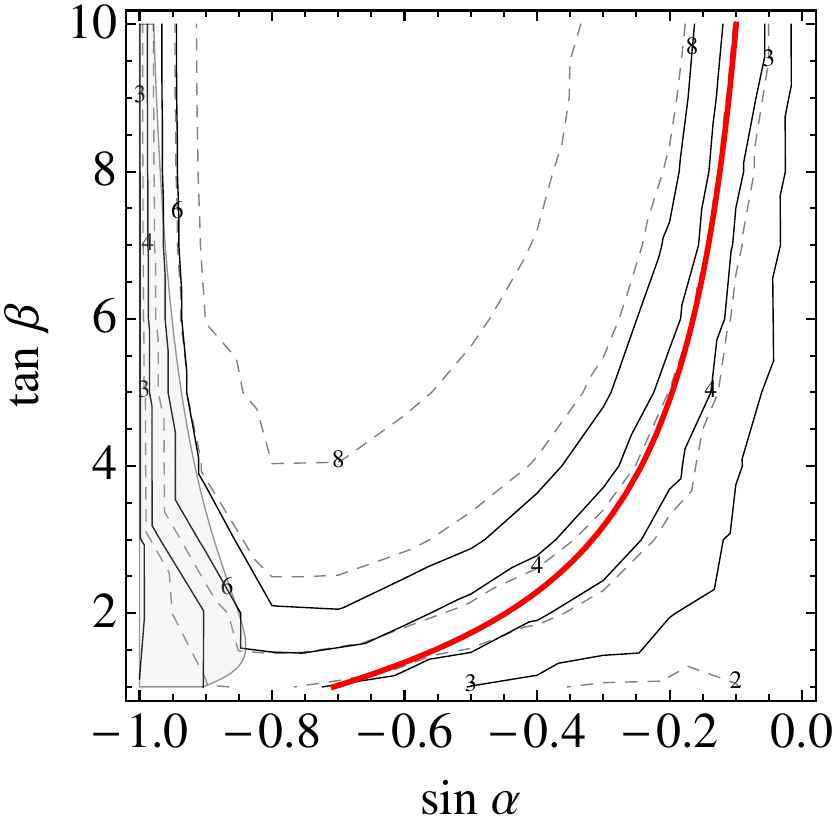}
   \includegraphics[width=2.8in]{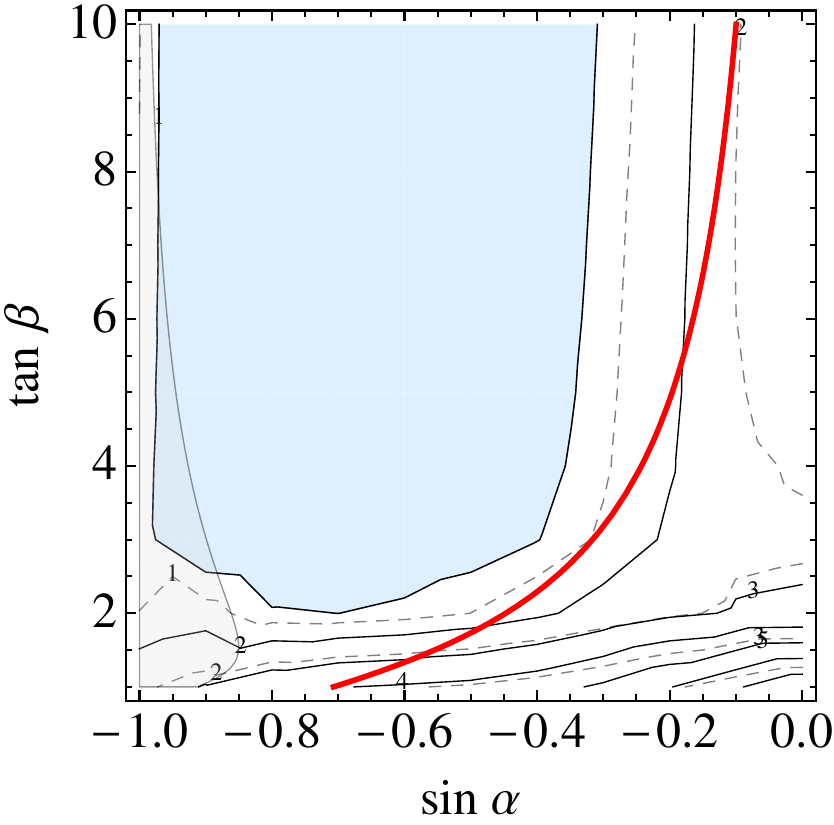}
   \includegraphics[width=2.8in]{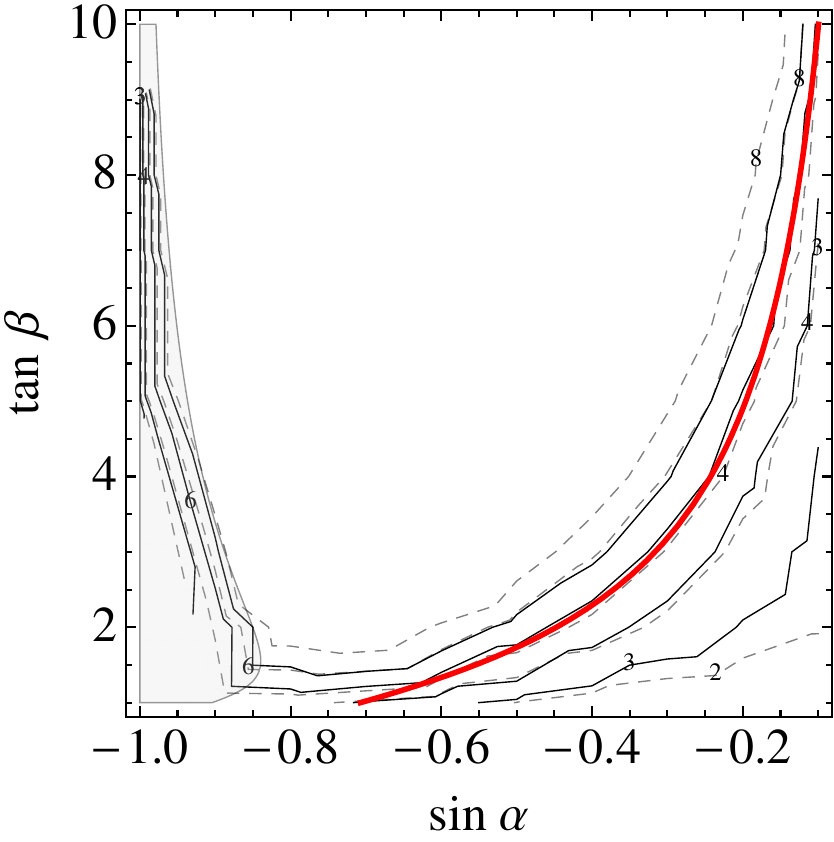}
   \caption{
%
%
      Multi-lepton limits 
   from the CMS multi-lepton search with 5 fb$^{-1}$ of 7 TeV proton-proton collisions \cite{CMSMulti5}   
   for the production and decay topologies of 
    Benchmark Spectrum 4 given in Table \ref{spec1}, 
    for Type I (top left), Type II (top right), Type III (bottom left), and Type IV (bottom right) couplings as a function 
    of $\sin \alpha$ and $\tan \beta$.  
        Limits were 
    obtained from an exclusive combination of the observed and expected number of events in all the 
    multi-lepton channels presented in Table \ref{tab:SM}.
        The solid and dashed lines correspond to the observed and expected 95\% CL limits 
    on the production cross section times branching ratio in multiples of the theory cross 
    section times branching ratio for the benchmark spectrum and 2HDM type. 
   The blue shaded regions denote excluded parameter space. 
   The solid red line denotes the alignment limit $\sin(\beta - \alpha) = 1$. 
   The gray shaded region corresponds to areas of parameter space where vector 
   decays of the heavy CP-even Higgs, $H \to VV^*$, are excluded at 95\% CL by the 
   SM Higgs searches at 7 TeV \cite{125HiggsCMS}.     
   }
   \label{fig:s4ex}
\end{figure}

\begin{figure}[h]
\centering
\includegraphics[width=2in]{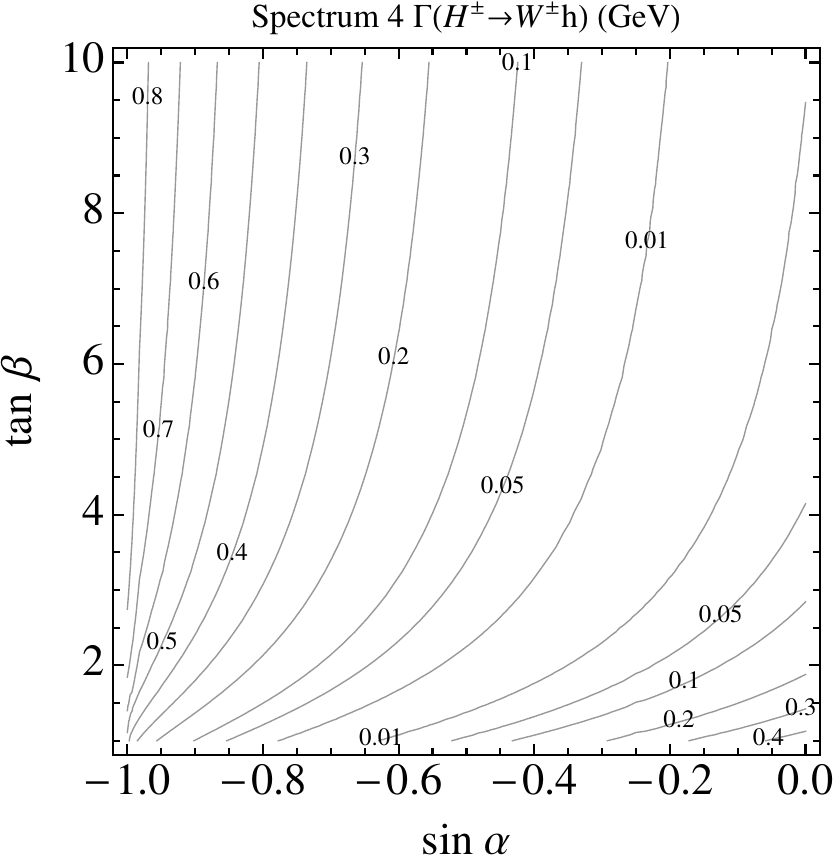}
\includegraphics[width=2in]{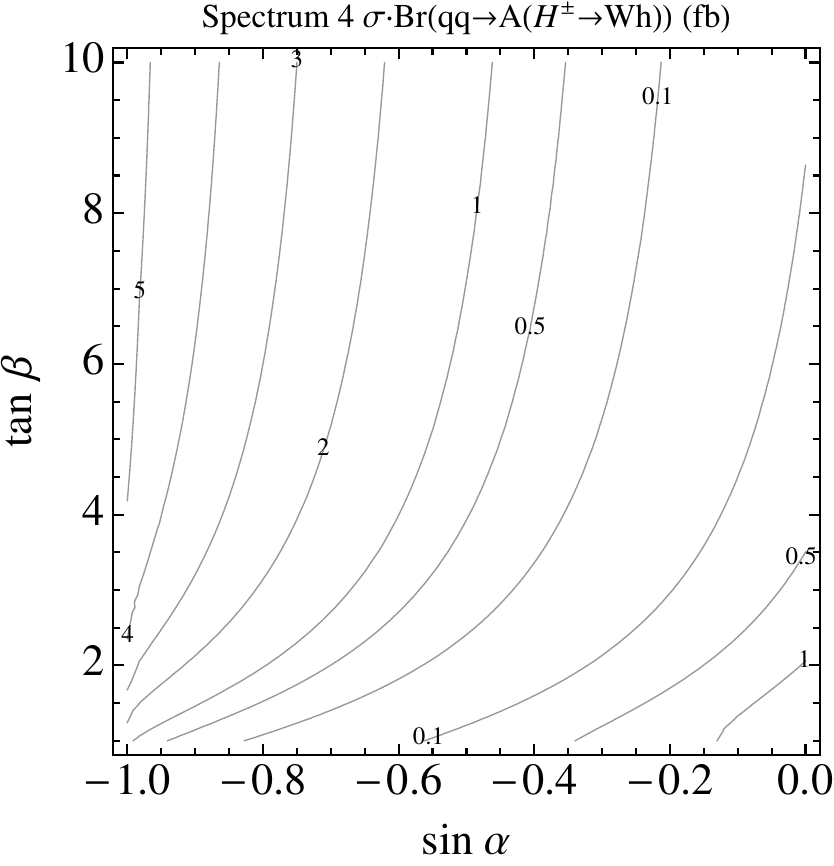}
\includegraphics[width=2in]{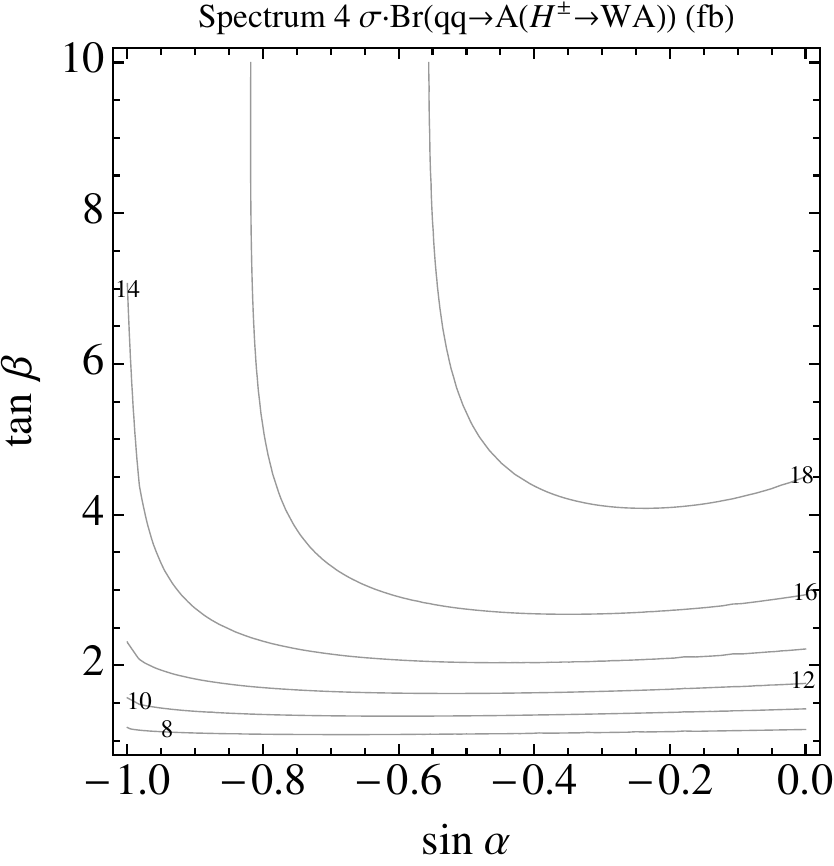}
\caption{2HDM Benchmark Spectrum 4 partial width $\Gamma(H^\pm \to W^\pm h)$ in units of GeV, 
and cross section times branching ratios $\sigma \cdot {\rm Br}(q \bar q \to A (H^\pm \to Wh))$ and $\sigma \cdot {\rm Br}(q \bar q \to A (H^\pm \to WA))$ in units of pb for Type I couplings. 
The partial width $\Gamma(H^\pm \to W^\pm A)$ is independent of $\alpha$ and $\beta$ and is not shown explicitly.}
\label{fig:s4widths}
\end{figure}

The multi-lepton limits on the first benchmark spectrum for all four types of 2HDM are shown in Figure \ref{fig:s4ex}. 
The fourth benchmark spectrum highlights the signals of a light pseudoscalar, 
both through decays of other scalars and through direct 
production in association with those scalars. Kinematically available 
inter-scalar decays include $H \to AA$, $H^\pm \to W^\pm h$, and $H^\pm \to W^\pm A$, 
while interesting associated production processes unique to this 
benchmark include $q \bar q \to H^\pm A,$  $q \bar q \to Ah$, and $q \bar q \to AH$ through off-shell $W$ and $Z$ bosons. 
The partial widths and $\sigma \cdot {\rm Br}$s for several of these processes are shown in Figure \ref{fig:s4widths}.

The partial width $\Gamma(H^\pm \to W^\pm h)$ scales as $\cos^2(\beta - \alpha)$ and hence grows away from the alignment limit. In contrast, $\Gamma(H^\pm \to W^\pm A)$ is entirely independent of the angles $\alpha, \beta$. On the production side, $\sigma(q \bar q \to Ah) \propto \cos^2(\beta - \alpha)$ grows away from the alignment limit, while $\sigma(q \bar q \to AH) \propto \sin^2(\beta - \alpha)$ grows as we approach the alignment limit. The production cross section $\sigma(q \bar q \to H^\pm A)$ is likewise independent of $\alpha, \beta$ since it scales as the square of the $H^\pm W^\mp A$ coupling.  However,  the partial widths of $H^\pm$ decays to SM states do depend on $\alpha$ and $\beta$, so the $\sigma \cdot {\rm Br}(q \bar {q} \to A (H^\pm \to W^\pm A))$ ultimately varies with $\sin \alpha$ and $\tan \beta$ due to the changing total width.   
As is apparent in Figure \ref{fig:s4widths}, the cross section for these processes is quite low, on the order of a few tens of femtobarns before further branching fractions are applied, so their inclusion is essentially for the sake of completeness; they contribute very little to the total multi-lepton limit.

Consequently, most qualitative features of this benchmark spectrum may be understood simply by the combination of the direct multi-lepton decays of $H$ and $h$ as well as the cascade decay $H \to AA$ with $A \to \tau \tau,$ which in this spectrum is the only source of multi-lepton signals from processes involving the pseudoscalar.

\subsubsection*{Types I \& III}

In a Type I 2HDM, the limit is largely governed by the direct multi-lepton decays of $h$ and $H$. In particular, the multi-lepton decays of $h$ are SM-like around the alignment limit and decrease slowly away from this limit. As $\sin \alpha \to -1$, the multi-lepton signals of $H$ become important and somewhat compensate for the vanishing signals of $h$. The branching ratio $H \to AA$ is large at moderate $\sin \alpha$ and large $\tan \beta$, but ${\rm Br}(A \to \tau \tau)$ does not grow exceptionally large in this regime, so the contribution to multi-lepton limits from $H \to AA$ is not great.

In the Type III 2HDM, the multi-lepton signals are much as in the Type I 2HDM with the exception of those involving $h \to \tau \tau$ and $A \to \tau \tau$. Thus, the process $gg \to H \to AA \to 4 \tau$ contributes significantly in this 2HDM type. Unsurprisingly, in the region excluded by 5 fb$^{-1}$ data, $\sigma \cdot {\rm Br}(gg \to H \to AA \to 4 \tau)$ is large, $\gtrsim 500$ fb, with the current exclusion contour tracking the contours of $\Gamma(H \to AA)$.

\subsubsection*{Types II \& IV}
In Type II, the multi-lepton signals of $h$ from decays to vectors decrease rapidly above the alignment limit and increase rapidly below it, again supplemented by the multi-lepton signals of $H$ as $\sin \alpha \to -1$. The multi-lepton signals of associated production with $h \to \tau \tau$ are somewhat important at large $\tan \beta$, but are not significantly enhanced over the SM rate since $h \to b \bar b$ grows equally quickly and controls the total width. Similarly, although the $A\tau \tau$ coupling grows with $\tan \beta$, so too does the coupling $A b \bar{b}$, so $H \to AA \to 4 \tau$ is not particularly important here.

For Type IV 2HDM the limits are much as in the Type II 2HDM, albeit with the loss of multi-lepton signals coming from $h \to \tau \tau$ and $A \to \tau \tau$ at large $\tan \beta$, leading to the weakest overall limits among 2HDM types.

\section{Conclusion}
\label{sec:conclusions}

In the wake of the discovery of a Standard Model-like Higgs, exploring and bounding extensions of the EWSB sector takes on paramount importance.  Models with two Higgs doublets are among the simplest and best motivated such extensions to the Higgs sector.  In this work, we have examined the reach of multi-lepton searches for probing the collective leptonic signatures resulting from the additional Higgs bosons in 2HDMs.  
In a study of 20 exclusive multi-lepton channels in four 
benchmark spectra with four discrete 
types of fermion couplings across 222 production and decay topologies, 
using a factorized mapping procedure \cite{sunil}
we determined regions of 2HDM parameter 
space probed by data from 
a recent CMS multi-lepton search \cite{CMSMulti5} with 5 fb$^{-1}$ of 7 TeV proton-proton 
collisions. 
These results provide new limits in some regions of 2HDM parameter space that have not 
been covered by other types of direct experimental investigations.  
Increased luminosity and production rates with 8 TeV proton-proton collisions and beyond 
will extend the 2HDM limits and 
discovery potential of multi-lepton searches.

Although the CMS multi-lepton searches \cite{CMSMulti, CMSMulti5} 
in their current incarnation are extremely powerful tools for probing 
new physics, with appropriate modifications the searches could be tailored in 
order to enhance sensitivity to 2HDM signals.  
Subdividing all exclusive multi-lepton channels by zero, one, or two or more 
$b$-tagged jets in an event should significantly increase sensitivity to 
2HDM final states with bottom quarks. 
Although many of 3- and 4- lepton events 
coming from production and decays of scalars in 2HDM populate the exclusive channels 
with relatively high backgrounds, most of the irreducible prompt 
background does not contain additional $b$-jets.  
For those backgrounds that do, 
very rarely, $b$-jets will provide isolated leptons, 
so two $b$-tags will substantially reduce major backgrounds 
(with the notable exception of $t \bar t$ plus a prompt fake lepton and  
$t\bar t V$), while leaving many 2HDM signal processes, 
such as $H \to hh \to ZZ bb$, $t\bar{t} A \to t \bar{t} Z h,$ $t \bar t A \to t \bar t \tau \tau$, 
$H \to A(A \to Zh) \to \tau \tau Z b b$, $H \to H^+ H^-\!\! \to t b Wh$, $H \to ZA \to ZZh \to ZZ b \bar b$, 
and, of course, $t\bar th$, relatively unaffected.   

Final states with multiple $\tau$-leptons 
are among the most promising for discovery or exclusion of various 2HDM.  In our study, we have focused solely on leptonically-decaying $\tau$s, since final states with hadronic $\tau$s will often have 
larger backgrounds. 
However, ignoring hadronic $\tau$s reduces sensitivity to, in particular, 
four-$\tau$ final states with low $\sigma \cdot {\rm Br}$.  
A further partitioning of the 4$\ell$, 2$\tau$ bins in a study optimized for four-$\tau$ signals 
may yield lower backgrounds in DY0 bins, 
e.g. $\tau_h^+\tau_h^+e^-\mu^-$, allowing for improved limits.  
As much of the energy in these events are going into leptons, defining signal regions either with harder $p_T$ cuts on leptons or with a cut on $\sum p_{T,\ell}$ could serve to significantly deplete the high SM backgrounds in some bins while leaving the signal largely unfazed.   
We have also restricted our focus to three- and four-lepton final states. 
Some additional sensitivity may be gained by adding exclusive channels with same-sign di-leptons 
subdivided by various combinations of $\MET$ and $H_T$. 
These channels would 
capture other decay modes of 
some of the production and decay topologies studied here, as well as 
bring in additional topologies that do not yield three or more leptons.  
Multiple Higgs bosons can also give rise to rare
five- or more lepton signatures; adding channels to 
separate out these signatures would also increase sensitivity, particularly 
at high luminosity. 

Finally, with a known Higgs mass, one can capitalize on partial or full 
kinematic constraints of its decays to help to isolate Higgs particles arising via 
new sources of associated production.  
Such kinematic tagging can serve to further reduce SM backgrounds.  
One example of this would be forward jet tagging to highlight VBF signals.
Another would be channel specific lepton kinematics focussed at specific decay topologies.  
One of the simplest and most effective ways to utilize kinematic tagging
to enhance sensitivity to certain multi-lepton 
signatures that include 
a SM-like Higgs boson would be to subdivide the 
DY2 four- or more lepton channels into an On Higgs category in which the invariant 
mass of the four leptons fall within a small window centered on the Higgs boson mass.  
Signals that include at least one SM-like Higgs boson that decays directly 
to four 
leptons fall in this sub-channel.   
The backgrounds in this special On Higgs sub-channel are very limited, 
thereby increasing sensitivity to such Higgs boson signals. 
Utilizing partial (rather than full) 
kinematic tagging could also increase sensitivity to other decay 
topologies that fall in other channels.  

While we have focused on 2HDMs, other extensions of the Higgs sector can lead to the production of new heavy, Higgs-like scalar resonances with decay topologies similar to those studied in this work.  Such new, Higgs-like particles generally lead to intermediate states composed of the heaviest SM particles, including $t$, $h$, $Z$, $W$, $b$ and $\tau$, whose final states contain multi-lepton signatures. If there exists an extended Higgs sector, multi-lepton searches optimized for the leptonic final states of Higgs scalars may prove an effective route for discovering new physics beyond the Standard Model.



\bigskip
\bigskip

\section*{Acknowledgments}
\noindent
We thank Spencer Chang, Amit Lath, Markus Luty, and Matthew Walker for useful conversations. The research of NC, JE, MP and ST was supported in part by DOE grant DE-FG02-96ER40959. The research of RG and SS  was supported in part by NSF grant PHY-0969282. The research of CK was supported in part by NSF grant PHY-0969020. NC gratefully acknowledges the support of the Institute for Advanced Study as well as the hospitality of the Weinberg Theory Group at the University of Texas, Austin during the inception of this work. NC and CK would like to thank the Aspen Center for Physics, supported by the NSF grant PHY-1066293, where part of this work was completed.


\end{document}